\documentclass[aps,pra,amsmath,amssymb,floatfix,twocolumn,amsmath,superscriptaddress,twocolumn,nofootinbib,tighten,letterpaper]{revtex4}
\usepackage{multirow}
\usepackage{bbold}
\usepackage{subfigure}
\usepackage{color}
\usepackage{mathrsfs}
\usepackage{hyperref}
\usepackage{bm}
\usepackage{stackrel}

\usepackage{amsfonts, relsize, color}
\usepackage{graphics}
\usepackage{graphicx}
\usepackage{subfigure}
\usepackage{color}

\hypersetup{pdfpagemode=UseNone}

\newcommand{\bsub}{\begin{subequations}}
\newcommand{\esub}{\end{subequations}}
\newcommand{\T}{\mathsf{T}}
\newcommand{\vex}[1]{\bm{\mathrm{#1}}}

\DeclareMathOperator{\sech}{sech}

\newcommand{\Mp}{M_{\mathsf{P}}}
\newcommand{\Mt}{M_{\mathsf{T}}}
\newcommand{\Ms}{M_{\mathsf{S}}}
\newcommand{\Mlp}{M_{\mathsf{LP}}}
\newcommand{\ys}{\mathsf{y}_{\mathsf{s}}}
\newcommand{\yb}{\mathsf{y}_{\mathsf{b}}}
\newcommand{\ts}[1]{{\textstyle{#1}}}

\begin{document}

\title{Quantum Multicriticality near the 
Dirac-Semimetal to Band-Insulator
Critical Point in Two Dimensions: A Controlled Ascent from One Dimension}

\author{Bitan Roy}
\affiliation{Department of Physics and Astronomy, Rice University, Houston, Texas 77005, USA}

\author{Matthew S. Foster}
\affiliation{Department of Physics and Astronomy, Rice University, Houston, Texas 77005, USA}
\affiliation{Rice Center for Quantum Materials, Rice University, Houston, Texas 77005, USA}

\date{\today}

\begin{abstract}
We compute the effects of generic short-range interactions on gapless electrons residing at the quantum critical point separating a two-dimensional Dirac semimetal 
and a symmetry-preserving band insulator. The electronic dispersion at this critical point is anisotropic ($E_{\mathbf k}=\pm \sqrt{v^2 k^2_x + b^2 k^{2n}_y}$ with $n=2$), 
which results in unconventional scaling of thermodynamic and transport quantities. Due to the vanishing density of states ($\varrho(E) \sim |E|^{1/n}$), this 
anisotropic semimetal (ASM) is stable against weak short-range interactions. However, for stronger interactions the direct 
Dirac-semimetal to band-insulator
transition can either $(i)$ become a fluctuation-driven first-order transition 
(although unlikely in a particular microscopic model considered here, the anisotropic honeycomb lattice extended Hubbard model), 
or $(ii)$ get avoided by an intervening broken-symmetry phase. 
We perform a controlled renormalization group analysis with the small parameter $\epsilon = 1/n$, augmented with a $1/n$ expansion 
(parametrically suppressing quantum fluctuations in the higher dimension) by perturbing away from the \emph{one-dimensional limit}, 
realized by setting $\epsilon=0$ and $n \to \infty$. 
We identify charge density wave (CDW), antiferromagnet (AFM) and singlet $s$-wave superconductivity as the three dominant candidates for 
broken symmetry. 
The onset of any such order at strong coupling $(\sim \epsilon)$ takes place through a continuous quantum phase transition across an 
interacting multicritical point, 
where the 
ordered phase, band insulator, Dirac and anisotropic semimetals
meet. We also present the phase diagram of an extended Hubbard model for the ASM, 
obtained via the controlled deformation of its counterpart in one dimension. 
The latter displays spin-charge separation and instabilities to CDW, spin density wave, and Luther-Emery liquid phases at arbitrarily weak coupling. 
The spin density wave and Luther-Emery liquid phases deform into pseudospin SU(2)-symmetric quantum critical points separating the ASM from the 
AFM and superconducting orders, respectively. Our phase diagram shows an intriguing interplay among CDW, AFM and $s$-wave paired states that can 
be germane for a uniaxially strained optical honeycomb lattice for ultracold fermion atoms, or the organic compound $\alpha$-(BEDT-TTF)$_2\text{I}_3$.
\end{abstract}

\maketitle

\section{Introduction}

A Dirac semimetal stands as a paradigmatic representative of a symmetry-protected gapless topological phase of matter that, for example, in two spatial dimensions can be realized in pristine monolayer graphene~\cite{graphene-1, graphene-2, graphene-review}. In a planar system, such a phase can be envisioned as a bound state of an equal number of vortices and anti-vortices (with unit vorticity) in reciprocal space, such that the net vorticity is zero. The difference of the vorticities, coined as the \emph{axial vorticity} (${\mathcal N}_a$), is however finite and given by ${\mathcal N}_a=2 n$, where $n$ is an integer ($n=1$ for graphene), that (modulo 2) in turn also defines the integer topological invariant/charge of a two-dimensional Dirac semimetal. Still it is conceivable to tune some suitable band parameter to the drive system through a continuous topological quantum phase transition where vortex and antivortex annihilate at a high symmetry point in the Brillouin zone, beyond which the system becomes a trivial band insulator. In the band insulator phase ${\mathcal N}_a=0$. While a Dirac semimetal features linearly dispersing quasiparticles in all directions down to arbitrarily low energy, the critical fermions residing at the 
Dirac-semimetal to band-insulator
quantum critical point (QCP) possess linear and quadratic dispersions along orthogonal directions in momentum space.

\begin{figure}[t!]
\subfigure[]{
\includegraphics[width=0.3\textwidth]{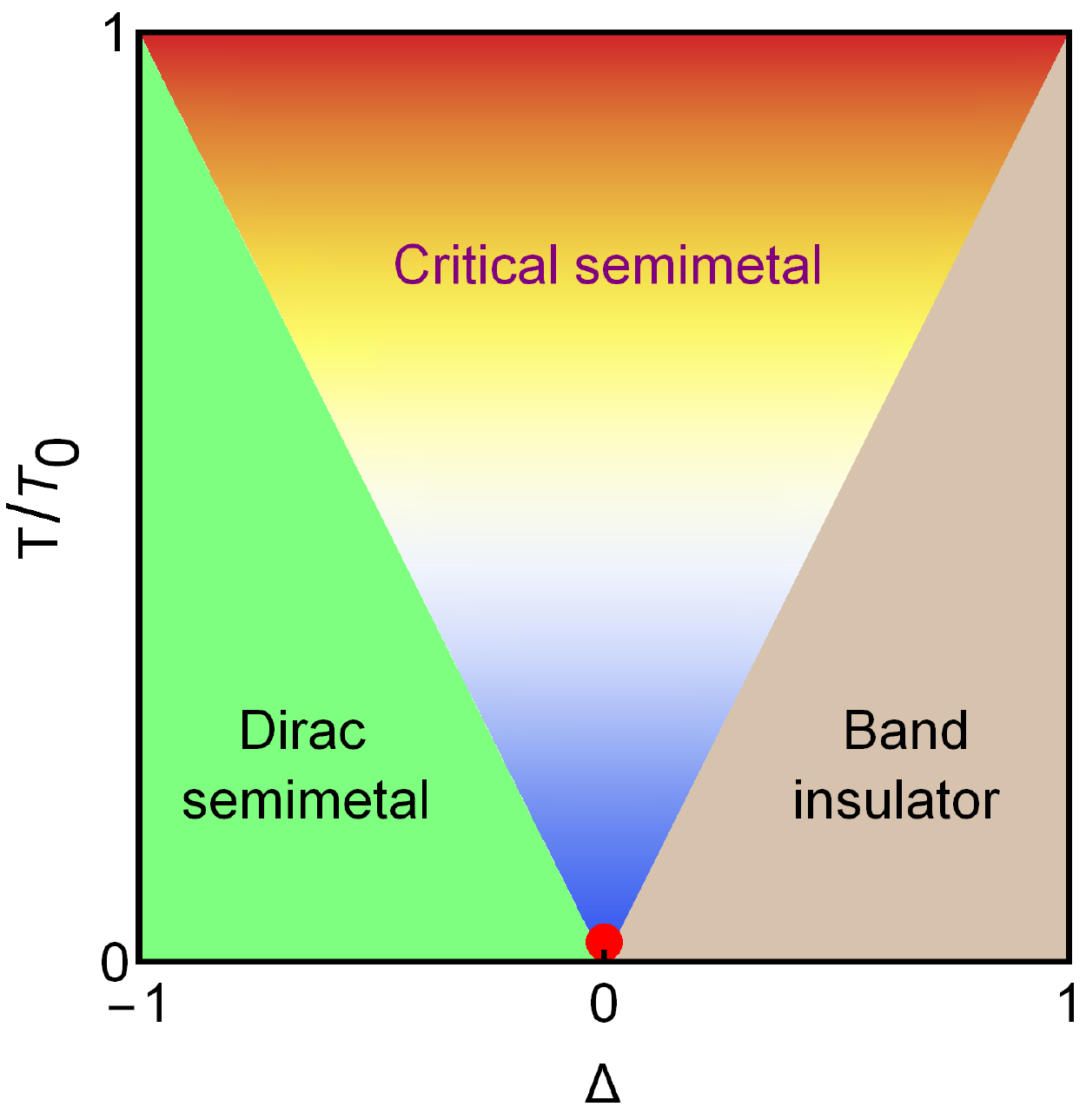}
~\label{critical-fan}
}
\\
\subfigure[]{
\includegraphics[width=0.21\textwidth]{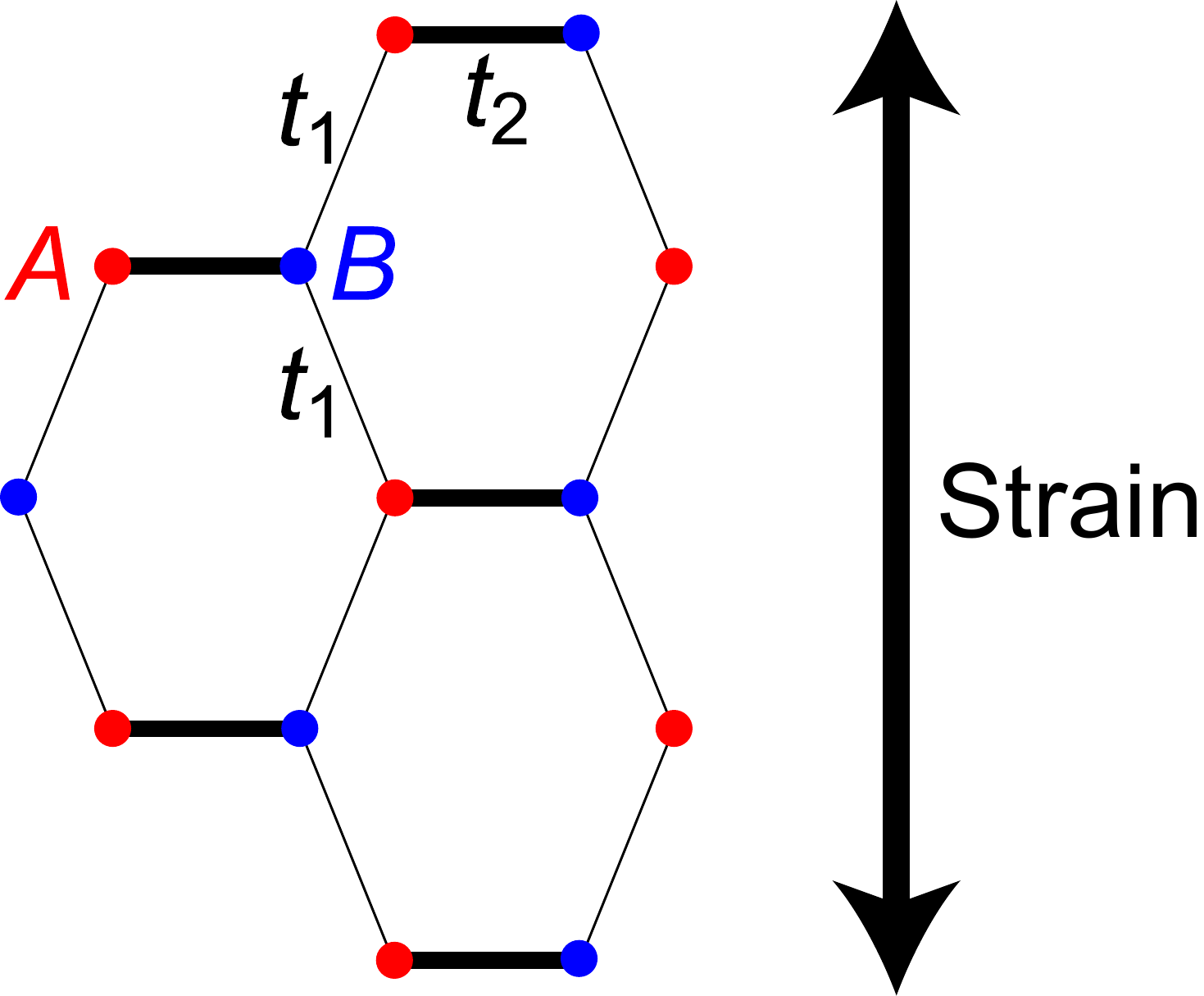}
~\label{lattice_hopping}
}
\caption{(a) Schematic representation of the quantum critical fan at finite temperature associated with the direct transition between a two-dimensional Dirac semimetal and a symmetry-preserving band insulator, with $T_0 \sim v^2/(k_B b)$. The signatures of the critical excitations, described by an anisotropic semimetal [$H({\mathbf k},0)$ from Eq.~(\ref{hamil-nonint:2D})], get gradually washed out as the temperature is increased and lattice effects set in at high temperature ($T \sim T_0$). The zero temperature quantum critical point in the noninteracting system, located at $\Delta=0$, is represented by the red dot. The critical regime can also be exposed by frequency and magnetic field (see text). The scaling of physical observables inside the critical fan, within the Dirac semimetal, or within the band insulator side of the transition is reported in Table~\ref{Table:scaling-noninteracting}. (b) Hopping pattern in a uniaxially strained honeycomb lattice. 
Strong (weak) hopping amplitudes are represented by thick (thin) lines. The sites on the two sublattices are shown in red (A) and blue (B). 
The anisotropic semimetal is realized when $t_2 = 2 t_1$.
}
\end{figure}

The quintessential properties of such a transition can be captured by a simple single-particle Hamiltonian 
\begin{equation}~\label{hamil-nonint:2D}
H(\mathbf k, \Delta)= \sigma_0 \left[ v k_x \, \tau_1 + \left( b k^2_y + \Delta \right) \tau_2 \right],
\end{equation}    
where two sets of Pauli matrices $\sigma_\mu =\left\{ \sigma_0, \sigma_1, \sigma_2,\sigma_3 \right\}$ and $\tau_\mu=\left\{ \tau_0, \tau_1, \tau_2,\tau_3 \right\}$ respectively operate on the spin and orbital space (and $\sigma_0 = \tau_0$ is the identity), and we set $\hbar = 1$.
Throughout this paper we assume that spin is a good quantum number (neglecting weak spin-orbit coupling). In Eq.~(\ref{hamil-nonint:2D})
$v$ and $b$ bear the dimensions of Fermi velocity and inverse mass, respectively, and $\Delta$ has the dimension of energy. The above Hamiltonian represents $(i)$ a Dirac semimetal for $\Delta < 0$, and $(ii)$ a band insulator for $\Delta>0$, as shown in Fig.~\ref{critical-fan}. In the Dirac semimetal phase the Dirac points are located at $k_x=0, k_y=\pm \sqrt{-\Delta/b}$.~\footnote{For $\Delta>0$ there is no real solution of $k_y$, implying that the phase is an insulator.} The 
Dirac-semimetal to band-insulator
QCP is located at $\Delta=0$,~\footnote{There is no symmetry distinction between a Dirac semimetal and band insulator.} where the quasiparticle spectra, given by $E_{\mathbf k}=\pm \sqrt{v^2 k^2_x + b^2 k^4_y}$, display both linear (along $k_x$) and quadratic (along $k_y$) dependence on the two different components of momenta. The density of states of the anisotropic semimetal (ASM) vanishes as $\varrho(E) \sim \sqrt{E}$. We compute key thermodynamic and transport properties of the noninteracting ASM [see Table~\ref{Table:scaling-noninteracting}].

\begin{figure}[t!]
\subfigure[]{
\includegraphics[width=4.0cm, height=4.0cm]{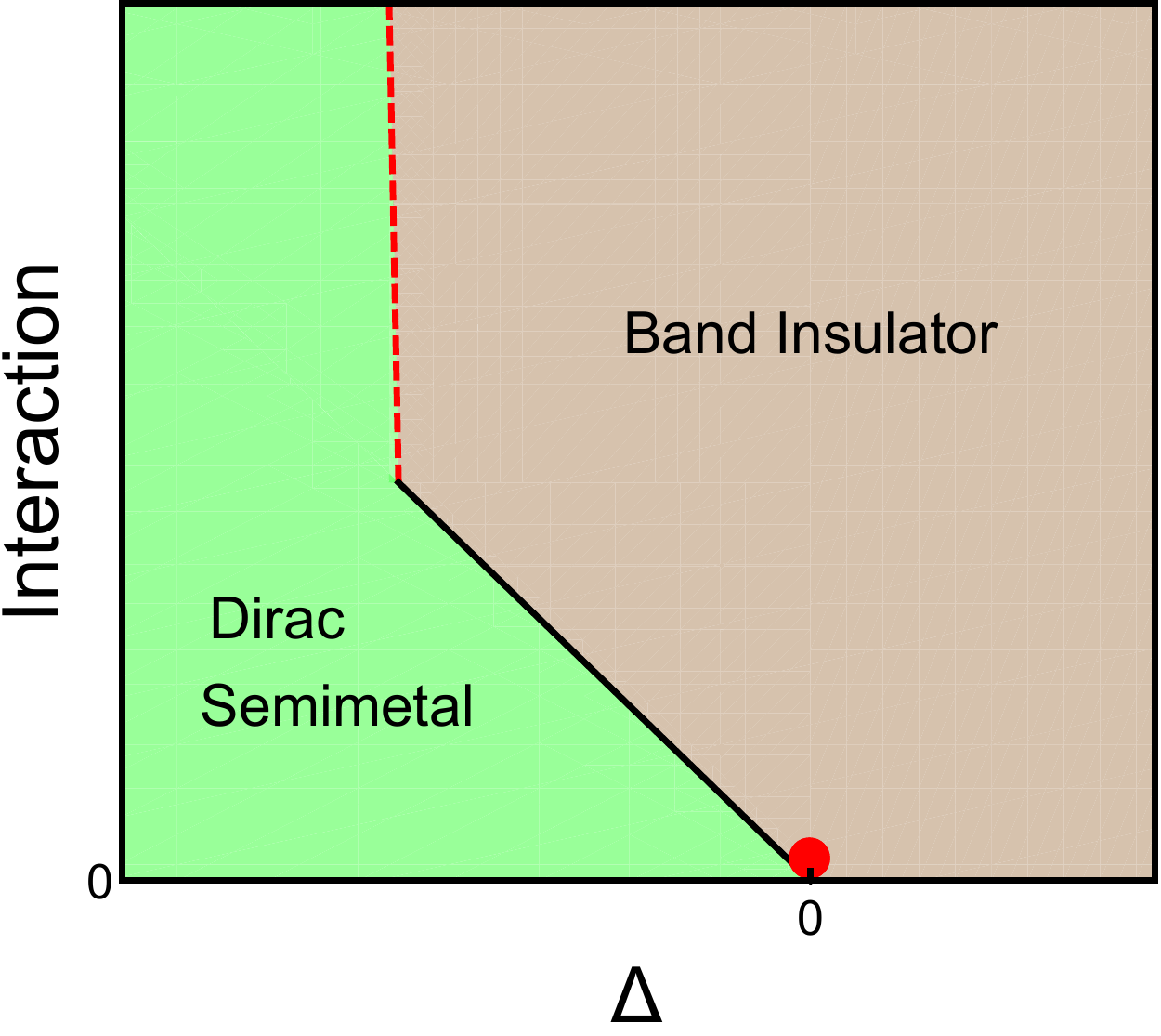}
\label{firstorder-schematic}
}
\subfigure[]{
\includegraphics[width=4.0cm, height=4.0cm]{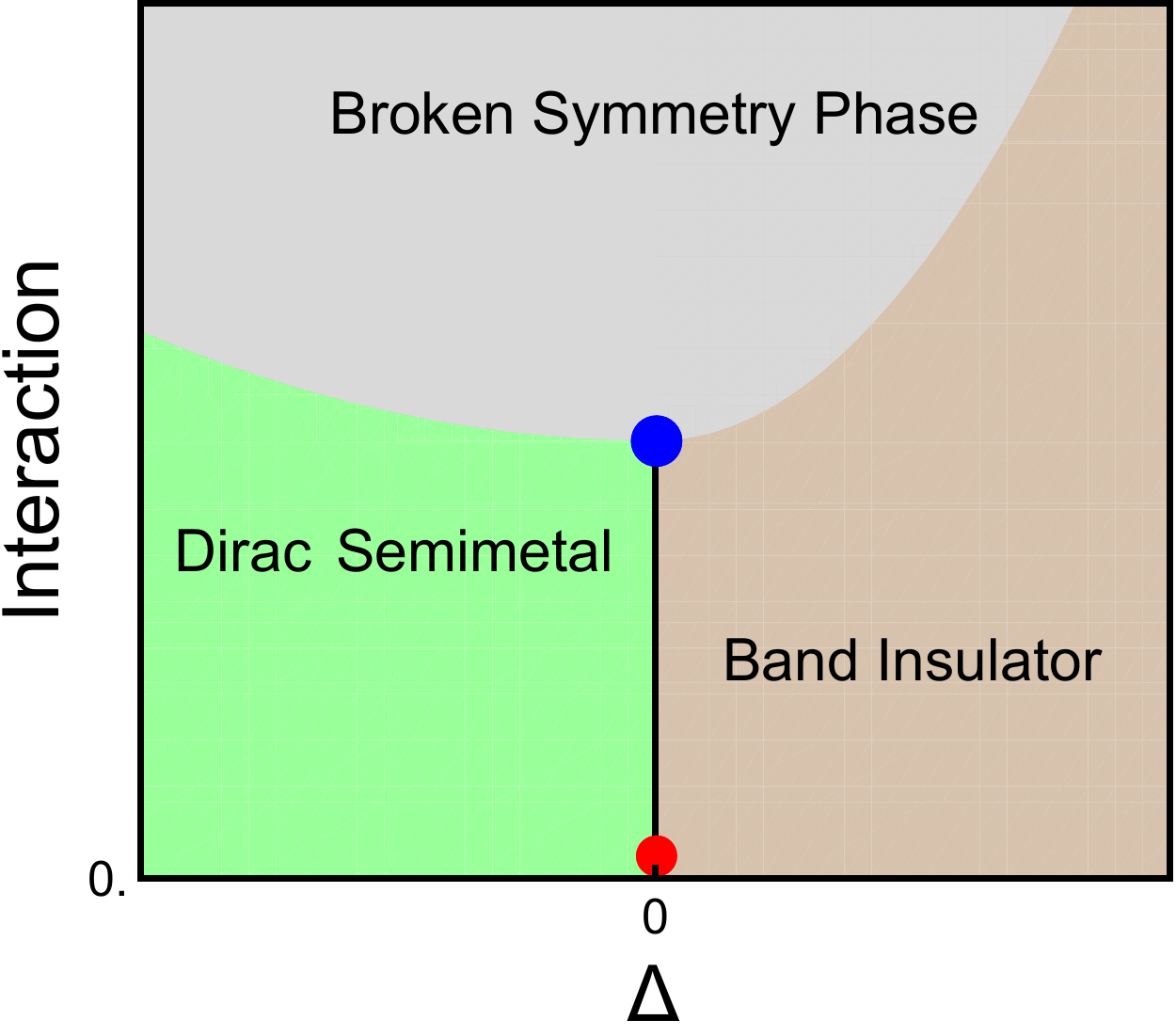}
\label{continuous-schematic}
}
\caption{Schematic phase diagrams of a two-dimensional Dirac material, residing in close proximity to the 
Dirac-semimetal to band-insulator
quantum critical point (the red dot), in the presence of generic short-range interactions. Critical fermions possessing anisotropic dispersion are found along the black line in either subfigure. This direct transition can be avoided in two separate scenarios. 
Subfigure (a) depicts a strong-coupling scenario in which the critical anisotropic semimetal (ASM) is replaced by a line of first-order transitions (red dashed line). Subfigure (b) depicts an alternative scenario in which the ASM becomes unstable to the formation of a broken-symmetry phase. In this case, the blue dot represents a multicritical point separating the ASM from spontaneous ordering. Charge density wave, antiferromagnet (N\'{e}el) and $s$-wave superconductivity are the prominent candidates for the broken-symmetry phase 
[see also Fig.~\ref{UV_PD} and Fig.~\ref{Phasediagram-finiten}]. For weak coupling in (a), the only effect of the interactions is to shift the 
phase boundary separating the band insulator and Dirac semimetal.
In (b), we assume that any such shift is compensated by the bare anisotropy ($\Delta$). 
We do not here discuss instabilities of the Dirac semimetal 
(see text for discussion on this issue~\cite{hou-chamon-mudry}).}
~\label{PD:schematic}
\end{figure}    

In this paper, we study the effects of generic short-range electronic interactions on such an ASM. We assume that long-range Coulomb interactions~\cite{longrange-1,longrange-2} are screened, e.g.\ via a proximate gate. Due to the vanishing density of states, the ASM is stable against sufficiently weak short-range interactions. By contrast, we show that the ASM can undergo a continuous quantum phase transition at strong interaction coupling through a multicritical point (tuned via interaction and anisotropy strengths) and enter into various broken-symmetry phases. At the 
multicritical points
the Dirac semimetal, band insulator, ASM and a broken symmetry phase meet [see the blue dot in Fig.~\ref{continuous-schematic}]. We identify charge density wave (CDW), antiferromagnet (AFM) and spin-singlet $s$-wave superconductivity as leading candidates for the broken symmetry phase. These conclusions obtain via the renormalization group (RG) controlled by an expansion about the \emph{one-dimensional limit}, described below. 
We also consider an alternative scenario that could occur at strong coupling, wherein
the Dirac semimetal can be separated from the band insulator by a \emph{fluctuation-driven first-order transition}. 
These two possibilities are schematically displayed in Fig.~\ref{PD:schematic}.  
We do not discuss in this paper quantum phase transitions between the Dirac semimetal and broken symmetry phases. 
In addition to CDW, AFM and $s$-wave pairing, a Dirac semimetal can accommodate additional 
fully gapped orders, 
such as quantum anomalous/spin-Hall insulator 
or  
Kekule valence-bond solid \cite{hou-chamon-mudry}.
Both of these arise due to the valley degree of freedom in the Dirac system,
which is annihilated at the 
Dirac-semimetal to band-insulator
(anisotropic semimetal) transition.  
Since the density of states in a Dirac semimetal (ASM) vanishes as $\varrho(E) \sim |E|$ ($\sqrt{E})$, 
any ordering in a 
Dirac semimetal tuned close to the transition into the band insulator
is expected to be preempted by those in the ASM, 
allowing us to focus solely on interaction effects in the latter.


In a half-filled honeycomb lattice the ASM can be realized by applying uniaxial strain~\cite{kohmoto, montambauz-2,Pereira2009,montambauz-1}
[see Fig.~\ref{lattice_hopping}], which in real graphene would require an extremely large (and possibly unrealistic) distortion of the lattice. Nonetheless, in an optical honeycomb lattice for ultracold fermion atoms, the ASM can be achieved by tuning the depth of the laser trap, as has been reported in recent 
experiments~\cite{coldatom-experiment-1,coldatom-experiment-2,coldatom-experiment-3}. 
In addition, the pressured organic compound $\alpha$-(BEDT-TTF)${}_2\text{I}_3$~\cite{materials-1, materials-2, organic-1, organic-2} as well as black phosphorous (a system with a few layers of phosphorene)~\cite{black-phosphorus, black-phosphorus-ARPES, doh-choi-black-phosphorus} may reside very close to the ASM QCP. A recent ARPES measurement is suggestive of anisotropic dispersion in black phosphorus~\cite{black-phosphorus-ARPES}. Furthermore, the quasiparticles residing at the interface of TiO$_2$/VO$_2$ are also believed to possess such an anisotropic dispersion~\cite{interface-1, interface-2}. 
The 
QCP separating the Dirac semimetal and band insulator
can also be accessed in uniaxially strained artificial/molecular graphene, as the hopping pattern in this system can be tuned quite efficiently~\cite{manoharan}.

A recent experiment has observed a first-order transition in $\alpha$-(BEDT-TTF)${}_2\text{I}_3$~\cite{firstorder-experiment}, also 
found in our theoretical analysis. Additionally, a semimetal-Mott (most likely AFM state) crossover in an optical honeycomb lattice driven by onsite repulsion has been reported in an experiment~\cite{coldatom-experiment-2}. Therefore, our results are germane to a plethora of condensed matter systems, but also directly applicable for a strained optical honeycomb lattice (populated by neutral atoms). At the same time, our theoretical approach to the interacting ASM smoothly connects it to strong correlation physics in one dimension, and might serve as a platform to explore how 1D physics such as spin-charge separation and fractionalization get modified in higher dimensions. We now highlight the central results of our study.

\subsection{Noninteracting system}

We begin by noting some hallmark signatures of the noninteracting ASM, the scaling behavior of thermodynamic and transport quantities that can directly be observed in experiments. Even though the ASM can only be found at $\Delta=0$ when $T=0$, its imprint can be realized over the wide quantum critical regime shown in Fig.~\ref{critical-fan} at finite temperature (up to a temperature $T_0 \sim v^2 / (k_B b)$ beyond which details of the lattice become important). The power-law scaling of various thermodynamic quantities, shown in Table~\ref{Table:scaling-noninteracting} (center column), is essentially governed by the scaling of density of states, which in the ASM vanishes as $\varrho(E) \sim \sqrt{E}$. Consequently the compressibility scales as $\kappa \sim \sqrt{T}$ and the specific heat vanishes as $C_v \sim T^{3/2}$ (at charge neutrality or for $T \gg |\mu|$, where $\mu$ denotes the chemical potential). If $\Delta<0$, the scaling of thermodynamic quantities displays a smooth 
crossover from the ones in an ASM to those in a 
Dirac semimetal
(displayed in the right column of Table~\ref{Table:scaling-noninteracting}) around a crossover temperature $T_\ast \sim |\Delta|/k_B$~\cite{sheehy}. On the other hand, for $\Delta>0$, the thermodynamic quantities display activated scaling for $T < T_\ast$.

\begin{table}
\begin{tabular}{|c||c|c|}
\hline
{\bf Physical observable} & {\bf ASM} & {\bf DSM} \\
\hline \hline
Compressibility ($\kappa$) & $T^{1/2}$ & $T$ \\
\hline
Specific heat ($C_v$) & $T^{3/2}$ & $T^2$ \\
\hline
Gr$\ddot{\mbox{u}}$neisen ratio ($\Gamma$) & $T^{-5/2}$ & $T^{-3}$ \\
\hline
Magnetic Gr$\ddot{\mbox{u}}$neisen ratio ($\Gamma_H$) & $T^{-3} \left( T^{-2} \right)$ & $T^{-4} \left(T^{-2}\right)$ \\
\hline
Dynamic conductivity ($\sigma_{xx, yy}$) & $\omega^{-1/2}, \omega^{1/2}$ & $\omega^0$ (constant) \\
\hline
Diamagnetic susceptibility ($\chi_0$) & $B^{-1/3}$ & $B^{-1/2}$ \\
\hline
Diamagnetic susceptibility [$\chi(T)$] & $T^{-1/2}$ & $T^{-1}$ \\
\hline
\end{tabular}
\caption{Power-law scaling of various physical observables for the anisotropic semimetal (ASM) and Dirac semimetal (DSM). The Gr$\ddot{\mbox{u}}$neisen ratio is defined as $\Gamma=\alpha/C_P$, where $\alpha$ is the thermal expansion parameter, and $C_P$ is the specific heat, measured at constant pressure. The magnetic Gr$\ddot{\mbox{u}}$neisen ratio is defined as $\Gamma_H = \left( \partial M/\partial T\right)_H/C_H$, where $M\propto H$ is the magnetization, $C_H$ is the molar specific heat, and $H$ is the magnetic field strength. When the Zeeman coupling dominates over the orbital one, $\Gamma_H$ has the same power-law dependence in the ASM and DSM (shown inside the parentheses in the 4$^{th}$ row). By contrast, when orbital coupling of the magnetic field dominates over Zeeman coupling $\Gamma_H$ has distinct scaling behaviors in the ASM and DSM (displayed outside the parentheses in the 4$^{th}$ row). The explicit form of the dynamic optical conductivity ($\sigma_{jj}$) and diamagnetic susceptibility ($\chi_0$) in the ASM are respectively shown in Eq.~(\ref{OC:interband}) and Eq.~(\ref{DMS:ASM}). The scaling of the diamagnetic susceptibility at finite temperature and low magnetic field (no Landau quantization) is shown in the last row. The crossover behavior between the ASM and the DSM at finite temperature, frequency and magnetic field is discussed in the text. All physical observables display activated behavior in the band insulator.   
 }~\label{Table:scaling-noninteracting} 
\end{table}

The quantum critical regime shown in Fig.~\ref{critical-fan} can also be exposed by frequency ($\omega$) as long as $\omega \lesssim \omega_0 \sim v^2/b$ and $\omega > |\Delta|$. For example, the Drude conductivity for noninteracting electrons in the ASM scales as
\begin{align}~\label{drude}
	\!\!\!\!\!
	\sigma_{jj} \left( \omega \right) \Big|_{\omega \to 0} = \frac{a_j e^2}{h} \left[\frac{T_0}{T} \right]^{\frac{3}{2}-j} \!	F_j \left( \frac{  \mu}{2 k_B T}\right) \delta\left( \frac{\omega}{k_B T}\right)\!,\! 
\end{align} 
while the interband component of the optical conductivity goes as
\begin{eqnarray}~\label{OC:interband}
	\sigma_{jj}(\omega,T)	= \frac{a_j e^2}{2 h} \left[\frac{\omega_0}{\omega} \right]^{\frac{3}{2}-j} 
	\sum_{\alpha=\pm} \tanh \left[\frac{\omega + 2 \alpha \mu}{4 k_B T}\right], 
\end{eqnarray}   
where $a_x\approx 0.06262$, $a_y \approx 0.20602$, $k_B T_0 = \omega_0 = v^2/b$. In the last two equations $j=1,2$ respectively corresponds to $x$ and $y$. The scaling of the two universal functions $F_1(x)$ and $F_2(x)$ is shown in Fig.~\ref{conductivity}, and the detailed calculation of the optical conductivity is presented in Appendix~\ref{append_conductivity}. Note that as frequency is lowered the interband component of the optical conductivity displays a smooth crossover from the scaling highlighted in Eq.~(\ref{OC:interband}) to the one for massless Dirac fermions ($\sigma_{xx, yy} \sim e^2/h$ without any leading power-law dependence on $\omega$) for $\omega< |\Delta|$ 
and $\Delta<0$. By contrast, the optical conductivity displays activated behavior at low frequency ($\omega<\Delta$) when $\Delta>0$.

The diamagnetic susceptibility of the ASM at $T=0$ is given by 
\begin{equation}~\label{DMS:ASM}
\chi_0=-{\mathcal A} \; (e/h) \; \left( e v \sqrt{b/2} \right)^{2/3} \; B^{-1/3},  
\end{equation}     
which diverges as $B^{-1/3}$, where ${\mathcal A} \approx 0.075$. A detailed calculation 
is presented in Appendix~\ref{diamagnetic}. 
By contrast, the 
diamagnetic susceptibility
of a two-dimensional 
Dirac semimetal
diverges as $\chi_0 \sim B^{-1/2}$~\cite{goswami-ghosal}, while that for three-dimensional Dirac~\cite{goswami-chakravarty} and Weyl~\cite{roy-sau} semimetals scales as $\chi_0 \sim \log(B/B_0)$, where $B_0 \sim 1/a^2$ and $a$ is the lattice spacing. The power-law dependence of the 
diamagnetic susceptibility, shown in Eq.~(\ref{DMS:ASM}), can be observed in experiments only when $B > \Delta^2$, so that we expose the critical regime associated with the noninteracting QCP [Fig.~\ref{critical-fan}] by an external magnetic field.~\footnote{In other words, we require $\ell_B<\xi_\Delta$, where $\xi_\Delta \sim 1/\Delta$ is the correlation length associated with the 
Dirac-semimetal to band-insulator QCP
and $\ell_B \sim B^{-1/2}$ is the magnetic length, so that the magnetic length provides the infrared cut-off. This is required to observe the scaling
reported in Eq.~(\ref{DMS:ASM}).} Otherwise, with decreasing strength of the magnetic field the 
diamagnetic susceptibility
displays a smooth crossover from $B^{-1/3}$ to $B^{-1/2}$ scaling when $B<\Delta^2$ and $\Delta<0$.

\begin{figure}
\subfigure[]{
\includegraphics[width=4.0cm, height=4.25cm]{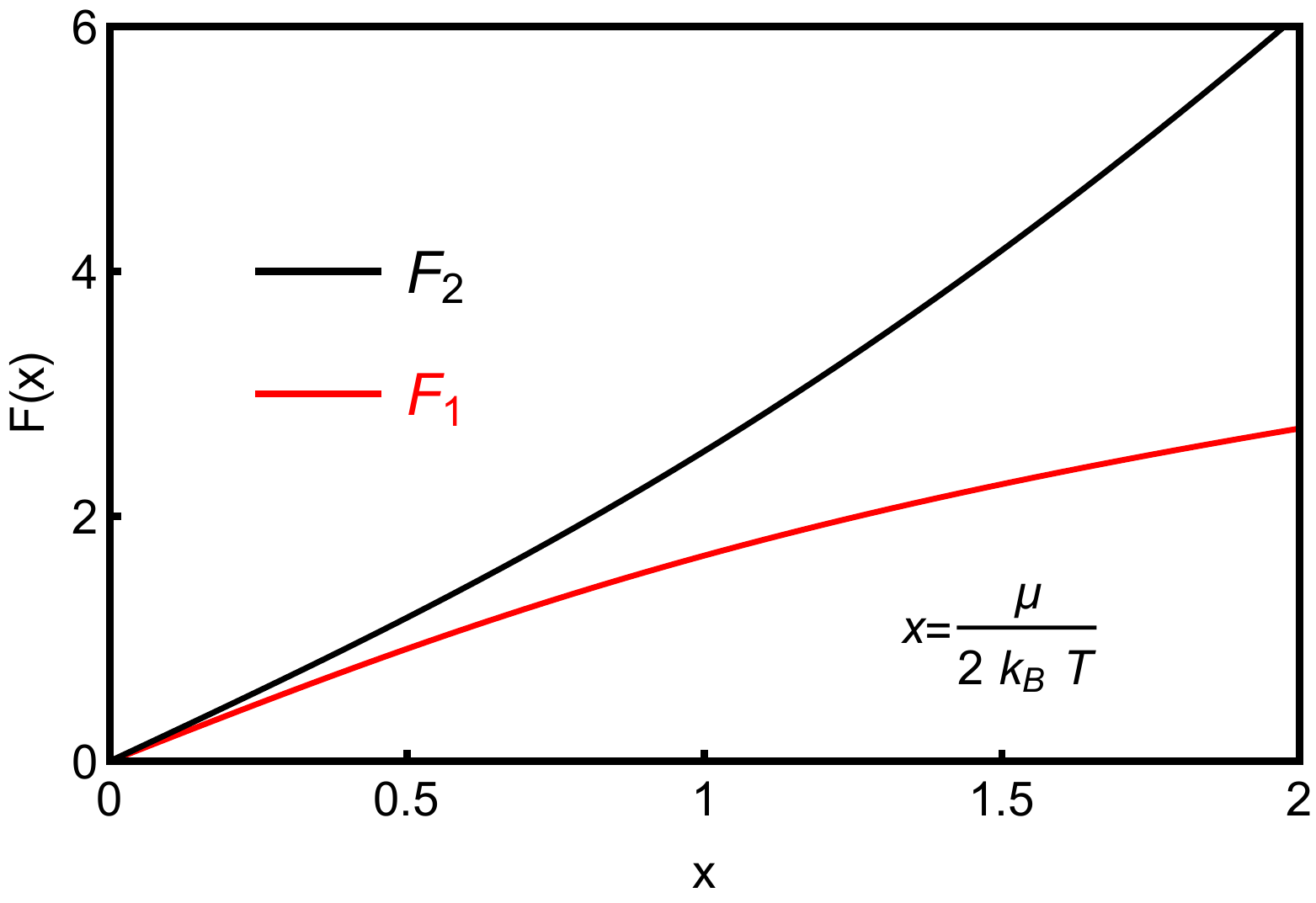}
\label{conductivity}
}
\subfigure[]{
\includegraphics[width=4.0cm, height=4.15cm]{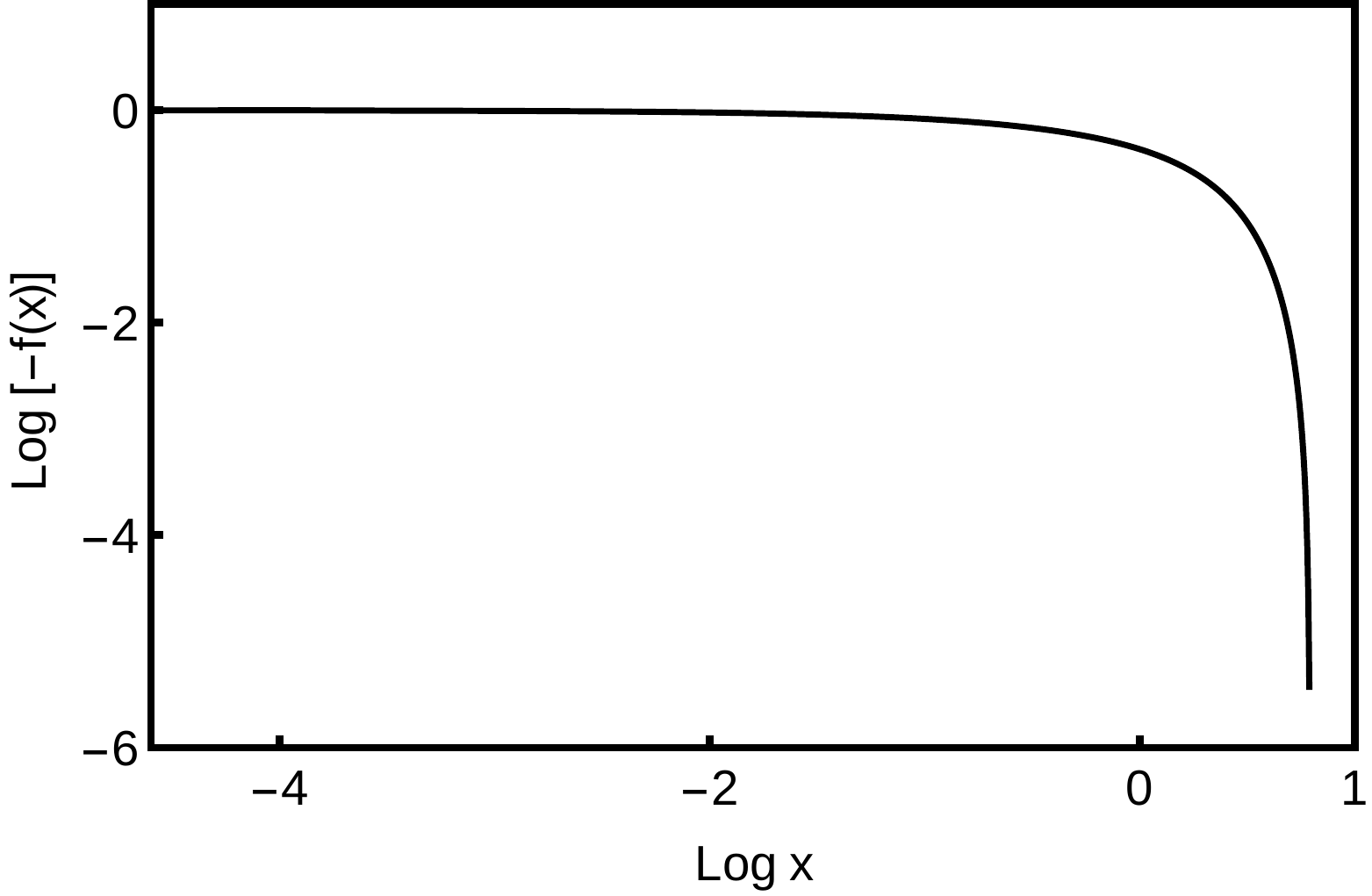}
\label{diamagnetism}
}
\caption{(a) Scaling of two functions appearing in the expression for the Drude conductivity of the anisotropic semimetal in the $x$ (given by $F_1$) and $y$ (given by $F_2$) directions [see Eq.~(\ref{drude})] versus the dimensionless argument $X = \mu/(2 k_B T)$. Here $\mu$ is the chemical potential and $T$ is the temperature. (b) Scaling of the function $f(x)$ versus the dimensionless argument $x$ (the ratio of thermal de Broglie wavelength to magnetic length) that controls the diamagnetic susceptibility of an anisotropic semimetal at finite temperature [see Eq.~(\ref{DMS:finiteT})].
}
\end{figure}

At finite temperature the
diamagnetic susceptibility
assumes the following universal scaling form (see Appendix~\ref{diamagnetic} for details)
\begin{equation}~\label{DMS:finiteT}
\chi_T=\chi_0 \left[ 1+ f\left( \lambda_{th}/ \ell_B \right) \right],
\end{equation}
where $\lambda_{Th}=\hbar \sqrt{v} (b/2)^{1/4}/\left( k_B T\right)^{3/4}$ is the \emph{thermal de Broglie wavelength} for the critical fermions.~\footnote{By contrast, the thermal de Broglie wavelength for relativistic and nonrelativistic fermions scales as $\lambda_{Th} \sim T^{-1}$ and $T^{-1/2}$, respectively.} The scaling of the universal function $f(x)$ is shown in Fig.~\ref{diamagnetism}. Our proposed scaling 
at finite temperature ($\chi_T$) is valid as long as $\ell_B<\lambda_{Th}<\xi_\Delta \sim 1/\Delta$.~\footnote{The scaling of the 
diamagnetic susceptibility
for the anisotropic semimetal at finite temperature and weak magnetic field (no Landau quantization) can be estimated as follows: $\chi(T) \sim B^{-1/3} \sim \ell^{2/3}_B \sim \lambda^{2/3}_{Th}$, yielding $\chi(T) \sim T^{-1/2}$. On the other hand, for a two-dimensional Dirac semimetal the 
diamagnetic susceptibility
at finite-$T$ scales as $\chi(T) \sim B^{-1/2} \sim \ell_B \sim \lambda_{th}$, yielding $\chi(T) \sim T^{-1}$.}

\begin{figure*}
\subfigure[]{
\includegraphics[width=6.5cm,height=5.6cm]{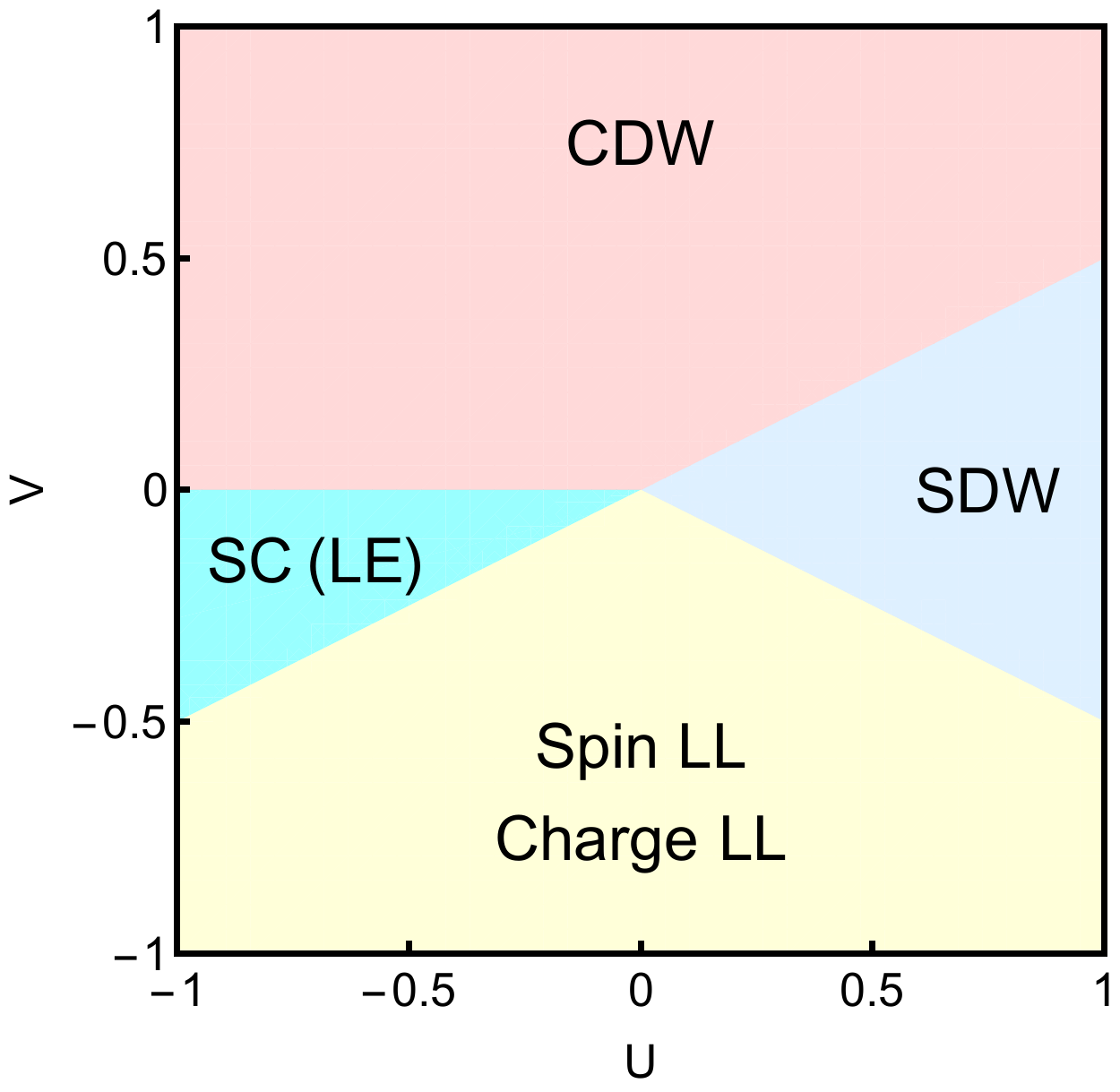}
\label{UV_PD_1D}
}
\subfigure[]{
\includegraphics[width=6.5cm,height=5.5cm]{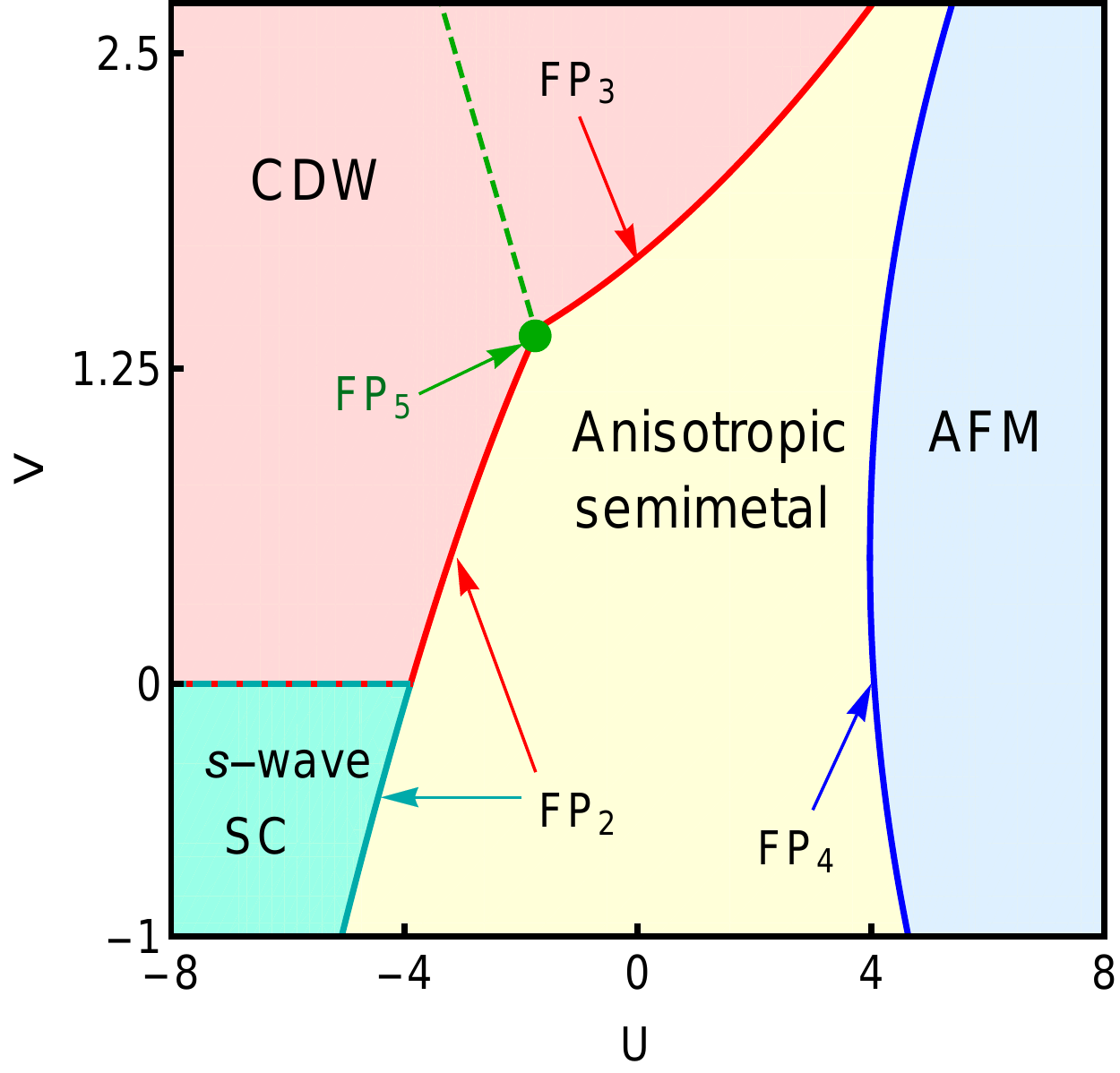}
\label{UV_PD}
}
\caption{The phase diagram of the extended Hubbard model for spin-1/2 electrons in (a) one dimension and (b) the two-dimensional anisotropic semimetal (ASM), obtained in this work. Here, $U$ and $V$ respectively correspond to onsite and nearest-neighbor interactions, see Eq.~(\ref{hubbard}). The 1D extended Hubbard model has been extensively studied, see e.g.~\cite{1DHub1,1DHub2,1DHub3,1DHub4}; here we limit attention to the well-known predictions that obtain from bosonization \cite{giamarchi,tsvelik}, expected to hold for sufficiently weak coupling. In 1D, electronic quasiparticles are ill-defined for arbitrarily weak interactions. The interacting phases are shown in panel (a) and consist of 
(i) spin density wave (SDW), (ii) charge density wave (CDW), (iii) Luther-Emery liquid (LE), and (iv) coexisting spin and charge Luttinger liquids (LLs). In the continuum limit these phases exhibit complete spin-charge separation \cite{tsvelik}; the charge is gapped in the CDW and SDW phases, while the spin is gapped in the CDW and LE phases. The latter is a 1D precursor to superconductivity. We obtain a qualitatively similar phase diagram for the ASM, shown in (b). The key differences are that the quasi-long-range-ordered SDW and LE phases are respectively replaced with true long-range-ordered antiferromagnetism (AFM) and s-wave superconductivity (SC) (at zero temperature), and that the CDW, AFM, and SC orders appear at finite coupling due to the vanishing density of states. In (b) we set $n=10$ [Eq.~(\ref{hamil-nonint})] to parametrically suppress higher dimensional quantum fluctuations (see Sec.~\ref{spinless-RG} for a detailed discussion); 
here $U$ and $V$ are dimensionless couplings measured in units of $\epsilon$. The phase diagram (b) is obtained by simultaneously solving the RG flow equations for four-fermion coupling constants and order parameter source terms (see Sec.~\ref{susceptibility}). For $V=0$ and sufficiently attractive $U < 0$, simultaneous nucleation of CDW and s-wave pairing occurs (red/cyan dashed line). Along this line pseudospin SU(2) symmetry \cite{SCZhang90,Auerbach94,Hermele07} gets spontaneously broken. Various quantum critical fixed points FP$_j$ (see Table~\ref{table-1overn}) control different segments of the phase boundaries, indicated in panel (b). Fixed points FP${}_{2,4}$ exhibit emergent pseudospin SU(2) symmetry, while this is broken at FP${}_{3}$. In the CDW phase (red region) in panel (b), the green dashed line is a crossover boundary. Above the line of coexistence with the CDW, the $s$-wave superconductivity exhibits short-range ordering that smoothly vanishes at the green dotted line. This is governed by the bicritical point FP$_5$ (see Table~\ref{table-1overn}), 
where the ASM-CDW phase boundary displays a kink (green dot).    
 }
\end{figure*}

\subsection{Electron-electron interactions}

Next we turn to the effects of generic short-range electronic interactions on the ASM. We focus on the particle-hole symmetric system at charge neutrality with $\mu = 0$, although our results will also strongly influence the finite temperature physics when $T \gg |\mu|$.
Scaling of the density of states yields a \emph{negative} scaling dimension for any four-fermion interaction coupling $g$, namely $\left[ g \right]=-1/2$. 
Consequently, the noninteracting ASM QCP separating the 
Dirac semimetal and the band insulator
remains stable against sufficiently weak short-range interactions. However, at stronger interaction coupling we predict that one of two possible scenarios occurs: 
Either $(i)$ the 
Dirac-semimetal to band-insulator
transition becomes a \emph{fluctuation-driven first-order transition} 
(see Sec.~\ref{1storder}), or 
$(ii)$ the direct 
transition gets masked by an intervening broken-symmetry phase (see Sec.~\ref{spinful-RG}). These two scenarios are schematically depicted in Fig.~\ref{firstorder-schematic} and Fig.~\ref{continuous-schematic}, respectively. Demonstrating the first-order transition requires a nonperturbative approach, which we here implement using large-$N$ mean field theory (where $N$ is the number of flavors of critical fermions; $N = 2$ for spin-1/2 electrons in the strained honeycomb model). The onset of 
broken symmetry
takes place through a continuous 
quantum phase transition. 
The nucleation of 
broken symmetry
through a continuous 
transition
at intermediate coupling can be established by an RG analysis.

To control the RG calculation, we deform the Hamiltonian from Eq.~(\ref{hamil-nonint:2D}) to 
\begin{equation}~\label{hamil-nonint}
	\!	H(\mathbf k, \Delta) \to H_n(\mathbf k, \Delta) = \sigma_0 \left[ v k_x \tau_1 + \left( b k^n_y + \Delta \right) \tau_2 \right]\!,\!
\end{equation}    
where $n$ is an \emph{even integer}, so that this transformation does not alter the symmetry of the system (discussed in Sec.~\ref{symmetries}). The density of states in the deformed system scales as $\varrho(E) \sim E^{1/n}$, and consequently the scaling dimension for any four-fermion interaction coupling $g$ becomes $[g]=-1/n$. In the limit $n \to \infty$, the density of states becomes finite. With purely local interactions, this limit corresponds to a decoupled collection of \emph{one-dimensional systems} composed of massless Dirac fermions with spectra $\varepsilon_{\mathbf k}=\pm v k_x$. All local interactions have vanishing engineering dimension in this one-dimensional limit. Hence, $n \to \infty$ sets the \emph{marginality} condition for generic short-range interactions in the ASM, and the perturbative RG calculation can be controlled in terms of a small parameter $\epsilon=1/n$, following the spirit of the $\epsilon$-expansion~\cite{zinn-justin} with $[g]=-\epsilon$. 
Thus our RG calculation is performed in an \emph{effective dimension} $d_\ast=1+\epsilon$, 
and precisely at $\epsilon = 0$ local four-fermion interactions are marginal.

The above construction allows us to further control quantum fluctuations arising from the finite quasiparticle dispersion along the $k_y$-direction by a parameter $1/n$, entering through loop corrections in the perturbative RG calculation. Thus, our RG analysis is simultaneously controlled by \emph{two} small parameters, $\epsilon$ (which sets the engineering dimension of the interaction coupling $g$) 
and $1/n$ (which controls the quantum fluctuations). In this framework, the RG flow equations take the form 
\begin{align}~\label{beta-function}
	\!\!\!\!\!\!
	\frac{d g_\mu}{d l}	\! \equiv \! \beta_{g_\mu} \!	= -	\epsilon g_\mu + \sum_{\mu, \nu} A_{\mu, \nu} \, g_\mu g_\nu + \frac{1}{n} \sum_{\mu, \nu} B_{\mu, \nu} \, g_\mu g_\nu,
\end{align}
where $l$ is the logarithm of the RG length scale, and the summation over $\mu,\nu$ runs over all symmetry-allowed, linearly independent coupling constants $\{g_\mu\}$. To leading order, $A_{\mu,\nu}$ and $B_{\mu,\nu}$ are numerical ($n$-independent) matrices that we compute.

\emph{We emphasize that we treat $\epsilon$ and $1/n$ as \emph{independent parameters} in our RG scheme, although
we should set these equal at the end.}
There are physical and technical reasons for this. 
Physically, $\epsilon$ enters as the engineering dimension for interaction couplings $[g_\mu]=-\epsilon$ 
[Eq.~(\ref{beta-function})]. This is due entirely to dimensional analysis. 
While $\epsilon > 0$ makes all such interactions irrelevant at weak coupling, it preserves key aspects of the physics special to one dimension, such as spin-charge separation. By contrast, the explicit $1/n$ corrections in Eq.~(\ref{beta-function}) obtain from loop integrations involving the $k_y$-dispersion, and thus encode quantum fluctuations beyond 1D. The Hubbard model phase diagram in Fig.~\ref{UV_PD} 
(discussed below) obtains only after quantum corrections are included, since these select between competing orders that are degenerate when $\epsilon > 0$ but $1/n = 0$
[see Table~\ref{anomalous-dim-table} and Sec.~\ref{extended-hubbard} for details]. 
Technically, our calculation works like an $\epsilon$-expansion,~\footnote{
An additional advantage of our $\epsilon$-scheme is that it is performed
in fixed $d = 2$ spatial dimensions; no analytic continuation of the Clifford algebra is required.}
since we expand 
order-by-order in the interaction strengths. The expansion of the one-loop corrections (see Appendix~\ref{Append-loop-integral}) 
in $1/n$ is \emph{not} technically required, but is performed for physical 
clarity.~\footnote{A comparison can be drawn between our combined $\epsilon$- and $1/n$-expansion with 
a conventional double-expansion in $\epsilon$ and $1/N$, where $N$ is the flavor number of the original problem~\cite{moshe-moshe}. 
Typically $\epsilon$ captures the deviation from the marginal dimension.
However, the parameter $1/N$ only controls the quantum loop corrections and does not enter in the engineering dimension of
the coupling $[g]=-\epsilon$. By contrast, we here gain control over loop corrections by tuning the band curvature in the 
$k_y$-direction (through the parameter $1/n$, which does not enter the scaling dimension of $g$, namely $[g]=-\epsilon$, in $d_\ast=1+\epsilon$) 
instead of artificially increasing the flavor number. }

As a benchmark, we will show that if we set $\epsilon=0$ and send $n \to \infty$ in Eq.~(\ref{beta-function}), we recover well-established results for interacting fermions in one dimension. For example: $(i)$ the $\beta$-function vanishes for spinless fermions (described by a two-component Dirac fermion and a single Luttinger current-current interaction coupling $g$), and $(ii)$ the above set of flow equations displays spin-charge separation, with independent charge and spin Kosterlitz-Thouless transitions for spin-$1/2$ electrons~\cite{giamarchi, tsvelik}.

The central outcome of our analysis can be summarized in the form of a phase diagram for an extended Hubbard model appropriate for uniaxially strained graphene, 
tuned to the 
Dirac-semimetal to band-insulator
QCP. The interactions are defined via 
\begin{eqnarray}~\label{hubbard}
	H_{UV} &=& 	\frac{U}{2} \sum_{\vec{X}=\vec{A},\vec{B}} n_\uparrow (\vec{X}) n_\downarrow (\vec{X}) 
\nonumber \\
	&+& \frac{V}{2} \sum_{\vec{A}, i, \sigma=\uparrow ,\downarrow} n_{\sigma} (\vec{A}) n_{\sigma} (\vec{A} + \vec{b}_i).
\end{eqnarray}
Here, $U$ and $V$ respectively denote the onsite and nearest-neighbor (NN) interaction strengths, and $n_\sigma(\vec{X})$ is the fermion number operator at position $\vec{r}=\vec{X}$ with spin projection 
$\sigma=\uparrow, \downarrow$. For the sake of simplicity, we assume that the strength of NN repulsion is the same among all three NN sites in uniaxially strained graphene.~\footnote{The 
Dirac-semimetal to band-insulator
QCP in the uniaxially strained honeycomb lattice is achieved at the cost of $C_{3v}$ symmetry (present in the absence of strain), see Fig.~\ref{lattice_hopping}. Thus, the coupling $V$ is expected to be different for the three NN sites. Such modification can only lead to a non-universal shift of the phase boundaries in Fig.~\ref{UV_PD}, without altering its qualitative and universal features.} The extended Hubbard model in one dimension has been extensively studied by analytical \cite{giamarchi,tsvelik} and numerical \cite{1DHub1,1DHub2,1DHub3,1DHub4}
methods (see also Sec.~\ref{Hubbard:1D}). The phase diagrams for one and two dimensional systems are respectively shown in Fig.~\ref{UV_PD_1D} and Fig.~\ref{UV_PD}. A detailed analysis of the extended Hubbard model is presented in Sec.~\ref{extended-hubbard}.

The relevance of Eq.~(\ref{hubbard}) and Fig.~\ref{UV_PD} to experiments is as follows. 
A strained optical honeycomb lattice for ultracold atoms 
\cite{coldatom-experiment-1,coldatom-experiment-2,coldatom-experiment-3}
constitutes an ideal platform to realize the entire $V=0$ axis of the phase diagram, as the strength of repulsive and attractive Hubbard interaction can be tuned quite efficiently, while NN interactions are negligible~\cite{Esslinger:Review}.

In pristine (unstrained) graphene $U \sim 9$ eV, while $V \sim 5$ eV for $\kappa=2.5$, where $\kappa$ is the effective dielectric constant of the medium~\cite{katsnelson:Interaction};
the hopping strength $t \sim 2.8$ eV. 
When uniaxially strained, the strength of onsite repulsion in graphene 
is likely 
unchanged, while the 
NN interaction gets slightly weaker, thus allowing access to the onsite-repulsion--dominated regime of the phase diagram.
A very large strain of order 20\% is required to tune to the 
Dirac-semimetal to band-insulator
transition 
point \cite{Pereira2009}.
A strain around 13\% was recently achieved using MEMS
(albeit in three-layer graphene \cite{Staufer2014}),
corresponding to $t_2 \sim 1.5 t_1$ \cite{Pereira2009} (see Fig.~\ref{lattice_hopping}).  
This is still relatively far from the transition at $t_2 = 2 t_1$, but 
even larger values of strain may be possible. 
As emphasized above, it is not necessary to tune precisely to the 
transition point to potentially realize the Hubbard phase diagram 
in Fig.~\ref{UV_PD}, since transitions out of the ASM are expected to 
dominate over those out of the 
Dirac semimetal
due to the larger density of states 
of the former. 
The parabolic curvature responsible for density of states enhancement 
turns on at 
intermediate
single-particle energies even when there is still some splitting between
degenerate
Dirac semimetal
valleys at zero energy (see the wide quantum critical regime for ASM in Fig.~\ref{critical-fan}), and intermediate energies are more relevant for $U,V > t$.

Another method to achieve the ASM in graphene could exploit a Moir\'e pattern.
Such a pattern can occur in twisted bilayers \cite{Moire}, where it generates 
satellite Dirac points in the vicinity of the $K$ and $K'$ points. 
A smaller (but still substantial) strain could be used to further merge two
satellites along a high-symmetry direction.

The strength of $U$ and $V$ in black phosphorus, 
at a TiO$_2$/VO$_2$ interface, 
and in $\alpha$-(BEDT-TTF)$_2\text{I}_3$ 
are 
not precisely known. 
Nonetheless, by changing the substrate on which these quasi-two-dimensional systems
reside, one can efficiently tune the strength of NN repulsion, with suspended samples 
endowed with strongest NN repulsion. 
Furthermore, by changing the distance ($\xi$) between a gate and the 
sample 
one can also tune the strength of NN repulsion~\cite{vafek-throckmorton}. 
The screened Coulomb interaction is 
\bsub
\begin{align}
	V(r) 
	\sim&\, 
	U_0 \, \xi \left[1/r -1/\sqrt{r^2+\xi^2}\right], 
	\\
	\sim&\,
	U_0 \, \sqrt{\xi/r} \, e^{-\pi r/\xi}, 
\end{align}
\esub
where the first (second) result applies to a single
(double)-gated sample. 
In this equation $U_0 \sim e^2/(\kappa \xi)$; the nearest neighbor 
repulsion is $V = V(a)$, with $a$ the lattice spacing.
Thus different gate configurations can tune the relative strength of $U$ and $V$, 
and experimentally access various repulsive regimes and 
quantum phase transitions
of the phase diagram in Fig.~\ref{UV_PD}.  
Black phosphorous (already residing close to the ASM fixed point) in particular can be tuned to the ASM band structure via
impurity doping and the giant Stark effect~\cite{black-phosphorus-ARPES},
but this also shifts the chemical potential away from charge neutrality. 
The effects of finite chemical potential on the phase diagram shown 
in Fig.~\ref{UV_PD} are an important avenue for future work. 
We hope that the above discussion will motivate first principle calculations in other 
materials (similar to Ref.~\cite{katsnelson:Interaction}) to estimate interaction strengths 
in various 2D ASMs, and future experiments to search for various broken symmetry phases. 

Superconductivity in an ASM can also be induced via proximity effect. 
Proximity-induced superconductivity has been recently achieved in monolayer graphene, 
when deposited on Rhenium (an $s$-wave superconductor), and Pr$_{2-x}$Ce$_x$CuO$_4$ (a $d$-wave superconductor, with coherence length $\xi_S \sim 30$nm), with respective superconducting transition temperatures $T_c \sim 2$K~\cite{SC:graphene-1} and $T_c \sim 4.2$K~\cite{SC:graphene-2}. Therefore, it is conceivable to induce superconductivity in an ASM by controlling the 
distance ($\xi$) between strained graphene or black phosphorus and bulk superconducting materials 
(such as Rh, Pr$_{2-x}$Ce$_x$CuO$_4$).

We now discuss salient features of the two phase diagrams shown in Figs.~\ref{UV_PD_1D} and ~\ref{UV_PD}. In 1D, electronic quasiparticles are ill-defined for arbitrarily weak interactions. The 1D extended Hubbard phase diagram reviewed in Fig.~\ref{UV_PD_1D} is that obtained via bosonization, which predicts spin-charge separation \cite{giamarchi,tsvelik}. The spin and charge sectors can each separately exhibit gapless Luttinger liquid or gapped Mott insulating phases. This leads to the four composite phases shown in Fig.~\ref{UV_PD_1D}: (i) spin density wave (SDW), (ii) charge density wave (CDW), (iii) Luther-Emery liquid (LE), and (iv) coexisting spin and charge Luttinger liquids. The charge (spin) is gapped in the CDW and SDW (CDW and LE) phases. The SDW and Luther-Emery phases exhibit quasi-long-range order in the spin and charge sectors, respectively; these are precursors to antiferromagnetism and superconductivity in higher dimensions. The CDW phase exhibits true long-range, sublattice-staggered charge order at zero temperature.

The phase diagram of the extended Hubbard model in the two-dimensional ASM is shown in Fig.~\ref{UV_PD}, as obtained via our RG analysis. 
The 2D ASM (yellow shaded region) remains stable against sufficiently weak local interactions. The onset of ordered phases takes place at a finite strength coupling ($g_\mu \sim \epsilon$) in the ASM. In terms of the interacting phases, 1D [Fig.~\ref{UV_PD_1D}] and 2D [Fig.~\ref{UV_PD}] are qualitatively similar. The key difference is that the quasi-long-range-ordered SDW and LE phases in 1D are respectively replaced with true long-range-ordered N\'{e}el antiferromagnetism (AFM) and s-wave superconductivity (SC) in 2D (at zero temperature).

Sufficiently strong repulsive $U > 0$ drives the ASM through a continuous 
quantum phase transition
and places it into the N\'{e}el AFM phase, which breaks spin SU(2) and time-reversal symmetry on the honeycomb lattice. Strong onsite attraction $U < 0$ with $V = 0$ induces the 
simultaneous nucleation of $s$-wave pairing and CDW orders, stemming from the spontaneous breaking of the exact pseudospin SU(2) symmetry.~\footnote{The half-filled Hubbard model on a bipartite lattice (with $V = 0$) exhibits exact pseudospin SU(2) symmetry \cite{SCZhang90,Auerbach94,Hermele07}.} 
Nonzero $V$ breaks the pseudospin symmetry, preferring CDW ($s$-wave superconductivity) for repulsive $V > 0$ (attractive $V < 0$) coupling. The CDW and AFM phases respectively exhibit sublattice-staggered charge and collinear spin orders; they preserve translational invariance on the bipartite honeycomb lattice, but break reflection invariance across any line of bonds.

Thus the phase diagram of the extended Hubbard model in the ASM can be considered a controlled deformation (by two small parameters $\epsilon$ and $1/n$) of its counterpart in one spatial dimension. In addition, the results obtained from our leading order RG analysis are consistent with pseudospin SU(2) symmetry~\cite{SCZhang90,Auerbach94,Hermele07} for $V = 0$. The fixed points FP${}_2$ and FP${}_4$ that govern the transitions along this line exhibit degenerate scaling dimensions for all operators residing in each irreducible representation 
of this symmetry (see Table~\ref{anomalous-dim-table} and Fig.~\ref{pseudospinrotation}). The proposed phase diagram in Fig.~\ref{UV_PD} should therefore qualitatively describe the strongly interacting ASM, residing at the phase boundary between the Dirac semimetal and band insulator, for which $\epsilon=\frac{1}{2}$ and $n=2$.

Approaching from within the $s$-wave SC and AFM 
broken symmetry phases, 
the pairing amplitude and N\'eel order parameters vanish continuously at the quantum critical points FP${}_{2}$ and FP${}_{4}$, respectively. 
These 
transitions
can therefore be regarded as the two-dimensional deformations of the 1D quasi-long-range-ordered LE and SDW phases. To the leading order in the $\epsilon$- and $1/n$-expansion the \emph{correlation length exponent} 
$\nu^{-1}=\epsilon$ is the same across all continuous 
transitions
to
broken symmetry phases.~\footnote{Thus for the physically relevant system (strained honeycomb lattice) 
$\nu=2$ to the leading order, distinct from the result for the 
Dirac-semimetal to broken-symmetry-phase transition
characterized by $\nu=1$ (to the leading order).} This is, however, an artifact of the leading order calculation. 
Setting $\epsilon = 1/2$, the transition or crossover temperature ($T_c$) 
between
broken symmetry phases
and the ASM
scales as $T_c \sim \delta^2$, while that 
to the
Dirac semimetal
scales as $T_c \sim \delta$, where 
$\delta=\left( X-X_c\right)/X_c$ is the reduced control parameter for the zero-temperature 
quantum phase transition
from ASM or 
Dirac semimetal into
the ordered phase, with $X=U, V$ [Eq.~(\ref{hubbard})] for example, and $X_c$ as the critical strength for ordering. 
Thus, even though the ASM lives adjacent to a 
Dirac semimetal, 
the scaling of transition/crossover temperature to these 
scale distinctly. The 
quantum phase transition
in the former system should precede the one in the 
Dirac semimetal
if the latter is tuned close to the transition
into the band insulator,
since the ASM density of states vanishes more slowly with energy.

Each of the interacting QCPs (discussed in Sec.~\ref{spinful-RG}), controlling various  
continuous transitions into broken symmetry phases
are expected to 
accommodate strongly interacting \emph{non-Fermi liquids} that lacks sharp quasiparticle excitations. It would be extremely interesting to look for remnants of spin-charge separation in the two-loop self-energy, which will give an anomalous dimension and lifetime to the fermion field. Our proposed phase diagram and associated quantum critical phenomena can be directly tested by quantum Monte Carlo simulations on the honeycomb lattice with onsite ($U$) and nearest-neighbor ($V$) interaction, as the extended Hubbard model can now be simulated without encountering the infamous \emph{sign problem}~\cite{sorella-1, herbut-assaad-1, herbut-assaad-2, sorella-2, kaul, Gremaud, troyer-honeycomb, hong-yao-NN-honeycomb}. The phase diagram (or at least some part of it) can also be exposed in future in optical honeycomb lattice experiments with ultracold fermion atoms, wherein the strength of interactions can possibly be efficiently tuned. Our methodology can also be adopted to investigate the effects of electronic interactions in various other itinerant systems, possessing anisotropic fermionic dispersion, such as general Weyl semimetals~\cite{goswami-roy-juricic}, which we discuss qualitatively in Sec.~\ref{conclusion}.

\subsection{Outline}

The rest of the paper is organized as follows. In Sec.~\ref{hamiltonian} we establish the field theory description of the noninteracting 
Dirac-semimetal to band-insulator quantum phase transition
and discuss its symmetries. Possible broken-symmetry phases (both excitonic and superconducting) and the associated quasiparticle spectra are discussed in Sec.~\ref{BSP-classification}. In Sec.~\ref{interaction} we describe the minimal models for spinless and spin-1/2 
versions of the interacting ASM. In Sec.~\ref{spinless-RG} we determine the effects of electronic interactions on the ASM for spinless fermions. The RG analysis for spin-$1/2$ fermions, connection to 1D spin-charge separation, existence of various interacting QCPs and associated quantum critical phenomena are discussed in Sec.~\ref{spinful-RG}. In Sec.~\ref{susceptibility} we determine the nature of the broken-symmetry phases at strong coupling across various continuous
transitions
for the interacting spin-1/2 model. Sec.~\ref{extended-hubbard} provides a more detailed discussion of the extended Hubbard model [Eq.~(\ref{hubbard})]. We summarize our findings, discuss applications of our methodology in other correlated systems and highlight prospects for future work in Sec.~\ref{conclusion}. Additional technical details are relegated to the Appendices.


\section{Noninteracting system: Field theory and symmetries}~\label{hamiltonian}

In this section we construct the field theoretic description of the QCP separating a two-dimensional 
Dirac semimetal and a band insulator
in a noninteracting system, and discuss the symmetries of the ASM separating these two phases.

\subsection{Hamiltonian, Lagrangian, and scaling \label{sec:HamLagAct}}

The universality class of the 
Dirac-semimetal to band-insulator quantum phase transition
is captured by the quadratic Hamiltonian from Eq.~(\ref{hamil-nonint:2D}) for $\Delta=0$. The four-component spinor basis can be chosen as 
$\Psi^\top= \left[ c_{A, {\mathbf k}, \uparrow}, c_{B, {\mathbf k}, \uparrow}, c_{A, {\mathbf k}, \downarrow}, c_{B, {\mathbf k}, \downarrow} \right]$, where $c_{X, {\mathbf k}, \sigma}$ is the fermion annihilation operator with momentum ${\mathbf k}$, spin projection $\sigma=\uparrow, \downarrow$ and on sublattice $X=A/B$ of the honeycomb lattice. In other compounds, such as black phosphorus or $\alpha$-(BEDT-TTF)${}_2\text{I}_3$, $X$ represents orbital degrees of freedom. 

The imaginary time ($t$) action associated with the noninteracting Hamiltonian reads 
\begin{eqnarray}\label{S0} 
	{\mathcal S}_0= \int d^2 {\mathbf r} \, dt \; \Psi^\dagger \left( {\mathbf r}, t \right) {\mathcal L}_0 \Psi \left( {\mathbf r}, t \right),
\end{eqnarray} 
where ${\mathbf r} \equiv \left( x,y \right)$ and the kinetic operator ${\mathcal L}_0$ is given by
\begin{equation}
	{\mathcal L}_0 = \partial_t +	H_n \left( {\mathbf k} \to -i {\bm{\nabla}}, \Delta \right).	 
\end{equation}
Here we define ${\mathcal L}_0$ via the deformation $H({\mathbf k}, \Delta) \to H_n({\mathbf k}, \Delta)$ [see Eq.~(\ref{hamil-nonint})], which leaves all the symmetries of the Hamiltonian unaffected (since $n$ is here an even integer).

We now introduce the notion of coarse-graining for this model, under which we send $t \to e^{l} t$, $x \to e^{l} x$, and $y \to e^{l/n} y$; the field transforms as $\Psi \to e^{-l(1+1/n)/2} \Psi$. Here $l$ denotes the logarithm of the renormalization group length scale. The imaginary time action ${\mathcal S}_0$ remains invariant, so that the Fermi velocity ($v$) and inverse mass ($b$) are effectively dimensionless (carry zero scaling dimension $[v]=[b]=0$). On the other hand, the scaling dimension of $\Delta$ is $[\Delta]=1$. This is a relevant perturbation at the 
Dirac-semimetal to band-insulator
QCP that controls the transition between these two phases that possess the same symmetry. The scaling dimension of $\Delta$ yields the \emph{correlation length exponent} for this noninteracting QCP: $\nu^{-1}=1$.

\subsection{Symmetries}~\label{symmetries}

The Hamiltonian describing the ASM possesses a discrete chiral symmetry [also known as sublattice symmetry (SLS)], generated by a unitary operator $\sigma_0 \tau_3$ and encoded in the condition
\begin{align}\label{SLS}
	- \sigma_0 \tau_3 \, H_n ({\mathbf k},\Delta) \, \sigma_0 \tau_3 = H_n ({\mathbf k},\Delta).
\end{align} 
Time-reversal symmetry is generated by the antiunitary operator ${\mathcal T}=\sigma_2 \tau_3 K$, where $K$ is complex conjugation, and ${\mathcal T}^2=-1$. The Hamiltonian $H_n ({\mathbf k},\Delta)$ also possesses spin SU(2) symmetry generated by $\vec{S}=\vec{\sigma} \tau_0$, and $x$-reflection symmetry defined via 
\begin{align}\label{x-refl}
	\sigma_0 \tau_2 \, H_n (-k_x,k_y,\Delta) \, \sigma_0 \tau_2 = H_n (k_x,k_y,\Delta).
\end{align}
This symmetry operation encodes invariance of the system under the exchange of two sublattices or orbitals, and will be denoted by ${\mathcal R}_\pi$ [see Table~\ref{order-parameters}]~\cite{HJR, foster-aleiner}. We relegate a detailed discussion on the symmetry properties of the ASM to Appendix~\ref{symmetry-append}. In the next section we discuss various possible ordered phases that break at least one of the above symmetries of the noninteracting system.

\begin{table}
\begin{tabular}{|c|c|c|c|c|c|c|c|}
\hline
Bilinear & description & SLS & ${\mathcal T}$ & ${\mathcal R}_\pi$ &  $\protect{\stackrel{\textstyle{\text{spin}}}{\text{SU(2)}}}$ & gap ($?$) & OP \\
\hline \hline
$\Psi^\dagger \sigma_0 \tau_0 \Psi$ & density & 		$\times$	& $\checkmark$ 	& $\checkmark$ 	& 0 	& No & $\Delta^s_0$ \\
\hline 
$\Psi^\dagger \sigma_0 \tau_1 \Psi$ & $x$-current & 		$\checkmark$	& $\times$ 	& $\times$ 	& 0 	& No & $\Delta^s_1$ \\
\hline
$\Psi^\dagger \sigma_0 \tau_2 \Psi$ & AP & 			$\checkmark$	& $\checkmark$	& $\checkmark$ 	& 0 	& Yes/No & $\Delta^s_2$ \\
\hline
$\Psi^\dagger \sigma_0 \tau_3 \Psi$ & CDW & 			$\times$	& $\checkmark$ 	& $\times$ 	& 0 	& Yes & $\Delta^s_3$ \\
\hline
$\Psi^\dagger \vec{\sigma} \tau_0 \Psi$ & ferromagnet & 	$\times$	& $\times$ 	& $\checkmark$	& 1 	& No & $\Delta^t_0$ \\
\hline 
$\Psi^\dagger \vec{\sigma} \tau_1 \Psi$ & $x$-spin-current & 	$\checkmark$	& $\checkmark$ 	& $\times$ 	& 1 	& No & $\Delta^t_1$ \\
\hline
$\Psi^\dagger \vec{\sigma} \tau_2 \Psi$ & spin-BDW & 		$\checkmark$	& $\times$	& $\checkmark$ 	& 1 	& No & $\Delta^t_2$ \\
\hline
$\Psi^\dagger \vec{\sigma}\tau_3 \Psi$ & AFM & 			$\times$	& $\times$ 	& $\times$ 	& 1 	& Yes & $\Delta^t_3$ \\
\hline \hline
$\Psi \sigma_2 \tau_3 \Psi$ & $s$-wave SC & 			$\checkmark$	& $\checkmark$ 	& $\checkmark$ 	& 0 	& Yes & $\Delta_s$ \\
\hline
$\Psi \sigma_2 \tau_1 \Psi$ & chiral SC$_1$ & 			$\times$	& $\checkmark$ 	& $\checkmark$ 	& 0 	& No & $\Delta^1_{ch}$ \\
\hline
$\Psi \sigma_2 \tau_0 \Psi$ & chiral SC$_2$ & 			$\checkmark$	& $\checkmark$ 	& $\times$ 	& 0	& No & $\Delta^2_{ch}$ \\
\hline
$\Psi \sigma_{(0,1,3)} \tau_2 \Psi$ & triplet SC & 		$\times$	& $\checkmark$	 & $\times$ 	& 1 	& No & $\Delta_t$ \\
\hline
\end{tabular}
\caption{ All possible local (momentum independent or intra unit-cell) order parameters in the particle-hole or excitonic channel (first 8 rows) and particle-particle or superconducting channel (last 4 rows), and their transformation properties under various discrete and continuous symmetry operations, discussed in Sec.~\ref{symmetries}. Under spin SU(2), bilinears transform in either the singlet (0) or triplet (1) representations. Note that the anisotropy parameter (AP) preserves all microscopic symmetries, and respectively gives rise to gapless and gapped spectra for $\Delta<0$ and $\Delta>0$ on the corresponding sides of the 
Dirac-semimetal to band-insulator transition.
The second to the last column shows whether the quasiparticle spectrum inside a given broken-symmetry phase is fully gapped or not. Notice that only three ordered phases, namely the charge density wave (CDW), N\'{e}el antiferromagnet (AFM), and $s$-wave superconductor (SC) yield a fully gapped spectrum. These three orders are energetically most favored, at least within the framework of the weak coupling renormalization group analysis [see, for example, the phase diagram of the extended Hubbard model in Fig.~\ref{UV_PD}]. The last column labels the corresponding order-parameter (OP) source term [see also Sec.~\ref{susceptibility}]. The density, $x$-current, and $x$-spin-current are included here for completeness, although these are not true order parameters.  
 We note that due to the lack of valley degrees of freedom a two-dimensional anisotropic semimetal does not allow any topological order such as quantum anomalous/spin Hall insulator, \emph{which are odd under the exchange of two valleys}, unlike the situation in a Dirac semimetal~\cite{hou-chamon-mudry}.
}~\label{order-parameters}
\end{table}

\section{Broken-symmetry phases}~\label{BSP-classification}

Since the DOS of the 2D ASM vanishes as $\varrho(E) \sim \sqrt{E}$, the noninteracting QCP separating a 
Dirac semimetal and a band insulator
remains stable against sufficiently weak short-range interactions. Nevertheless, beyond a critical strength of the interactions the ASM can become unstable toward the formation of various broken-symmetry phases. Here we consider all 
possible phases,
including both particle-hole or excitonic as well as particle-particle or superconducting orders. However, we restrict ourselves to the momentum-independent (local or intra-unit-cell) orderings. Then the ASM all together supports eight excitonic and four pairing orders, enumerated in Table~\ref{order-parameters}. We distinguish these with the labels in the 
order parameter (``OP'') 
column of this table. The transformation of these fermion bilinears under the symmetry generators of the noninteracting system are also highlighted in Table~\ref{order-parameters}.

One can add eight particle-hole channel fermion bilinears to the noninteracting Hamiltonian. Of these, only the anisotropy parameter $\Delta^s_2$ that tunes through the 
Dirac-semimetal to band-insulator
transition preserves all discrete and continuous symmetries. The charge density $\Delta^s_0$, $x$-current $\Delta^s_1$, and $x$-spin-current $\Delta^t_1$ each break at least one discrete or continuous symmetry, but these conserved symmetry currents do not represent true order parameters. The remaining four bilinears are order parameters for the following 
broken symmetry phases:
sublattice-staggered charge density wave (CDW) $\Delta^s_3$, sublattice-staggered collinear (N\'eel) antiferromagnet (AFM) $\Delta^t_3$, uniform ferromagnet $\Delta^t_0$, and spin bond density wave (spin-BDW) $\Delta^t_2$. The last of these corresponds to spin-dependent hopping, since it preserves sublattice symmetry SLS [see Eq.~(\ref{SLS})]. 

Only the CDW and AFM 
order parameters
induce a fully gapped spectrum in the ordered phase. We therefore anticipate CDW and AFM to be the dominant orders for the ASM with strong repulsive interactions [see Figs.~\ref{UV_PD} and \ref{Phasediagram-finiten}], since they optimally minimize the free energy (at least at $T=0$). Nevertheless, it is worth understanding the effects of other fermionic bilinears in the ordered phase. For example, a nonzero vacuum expectation value for $\Delta^s_1$ renormalizes the Fermi velocity in the $x$-direction, while a ferromagnetic compensated ASM is realized for $\Delta^t_0 \neq 0$. In this phase, perfectly nested electron-like and hole-like Fermi surfaces arise 
for opposite projections of the electron spin. Thus, such a compensated semimetallic phase can undergo a subsequent BCS-like weak coupling 
instability towards the formation of an AFM phase~\cite{keldysh, rice, aleiner-kharzeev-tsvelik, roy-galitski}. However, the 
order parameter
in this AFM phase will be locked in the \emph{spin-easy-plane}, perpendicular to the ferromagnetic moment in the compensated semimetal. 
For $\Delta^t_1 \neq 0$, the Fermi velocity in the $x$-direction acquires opposite corrections for the two projections of electronic spin, while a nonzero value of $\Delta^t_2$ yields both the Dirac semimetal and band insulator, but for opposite spin projections.

Sufficiently strong attractive interactions could induce various superconducting phases. When we restrict ourselves to local (momentum independent or intra-unit-cell) pairings, all together the ASM supports only three spin-singlet and one spin-triplet pairings. Among them only the singlet $s$-wave pairing $\Delta_s$ gives rise to a fully gapped quasiparticle spectrum. Thus we expect rest of the pairings to be energetically inferior to the singlet $s$-wave superconductor. We will demonstrate that for strong attractive interactions the ASM becomes unstable towards the formation of this $s$-wave state [see Figs.~\ref{UV_PD} and \ref{Phasediagram-finiten}].


\section{Interacting theory: Minimal model}~\label{interaction}

We now consider the minimal model of electronic interactions in the 2D ASM separating the 
Dirac semimetal and band insulator, compatible with the symmetry of the system, discussed in Sec.~\ref{symmetries}. We discuss spinless and spin-$1/2$ fermions separately.~\footnote{For the clarity of presentation we here denote the spinor for spinless (two-component) and spinful (four-component) fermions as $\psi$ and $\Psi$, respectively.}

\subsection{Spinless fermions}

For spinless fermions, generic local four-fermion interactions are described by four quartic terms,
\begin{eqnarray}~\label{Hint_spinless}
	H^{spinless}_{int} 	&=& 	-	\int d^2{\mathbf r}	\,	\sum_{\mu = 0}^3 \left[ g_\mu \left( \psi^\dagger \tau_\mu \psi \right)^2 \right].
\end{eqnarray}       
In this notation, $g_\mu >0$ represents a repulsive electron-electron interaction that promotes condensation of the corresponding Hermitian fermion bilinear $\left\langle\psi^\dagger \tau_\mu \psi\right\rangle$. However, not all quartic terms are linearly independent. The Fierz identity mandates that generic interactions for spinless fermions can be captured by only \emph{one} local four-fermion interaction (see Appendix~\ref{fierz_spinless}). Nevertheless, the final outcome regarding the fate of the critical fermions in the ASM at strong coupling 
depends upon the bilinear used to decouple the interactions (in mean field theory), as discussed in Sec.~\ref{spinless-RG}.

\subsection{Spin-$1/2$ electrons}

Since we throughout assume spin to be a good quantum number, a collection of interacting spin-$1/2$ electrons in the ASM in the presence of generic short-range interactions is described in terms of eight quartic terms    
\begin{align}~\label{Hint_spinful_bare}
	H^{\prime}_{int} = -\int d^2{\mathbf r} \, \sum_{\mu = 0}^3
	\left[	
	\begin{aligned}
	&\,
	g^{s}_{\mu} \left( \Psi^\dagger \sigma_0 \tau_\mu \Psi \right)^2 
	\\&\,
	+
	g^{t}_{\mu} \left( \Psi^\dagger \vec{\sigma} \tau_\mu \Psi \right)^2
	\end{aligned}
	\right].
\end{align}
Again, not all quartic terms are independent. As shown in Appendix~\ref{fierz_spinful}, a maximum of four independent interactions 
are possible for spin-1/2 fermions. With a judicious choice (see Sec.~\ref{spinful-RG}) we define the interacting Hamiltonian as 
\begin{eqnarray}~\label{Hint_spinful}
	H_{int} &=& -	\int d^2{\mathbf r} \; \bigg[ g^{s}_{1} \left( \Psi^\dagger \sigma_0 \tau_1 \Psi \right)^2 + g^{s}_{3} \left( \Psi^\dagger \sigma_0 \tau_3 \Psi \right)^2 
\nonumber \\
		&+& g^{t}_{2} \left( \Psi^\dagger \vec{\sigma} \tau_2 \Psi \right)^2 + g^{t}_{3} \left( \Psi^\dagger \vec{\sigma} \tau_3 \Psi \right)^2 \bigg].
\end{eqnarray}
In Sec.~\ref{spinful-RG} we will perform an RG calculation with $H_{int}$ to understand the effect of electronic interactions in the spinful 2D ASM.

It is instructive to express the interacting theory in terms of a different set of four couplings that will allow us to connect with well-established results for spin-$1/2$ electrons in one dimension, which in our formalism is achieved by setting $\epsilon=0$ and taking $n \to \infty$ in the RG analysis. 
 We can cast the interacting Hamiltonian $H_{int}$ in terms of four-fermion interactions that capture the physics of spin-1/2 electrons in one dimension most efficiently, according to  
\begin{eqnarray}~\label{Hint_1D}
	H^{1D}_{int} &=& \int d^2{\mathbf r} \; \bigg[ U_N \, {\mathcal O}_N + U_A \, {\mathcal O}_A  \nonumber \\
	&+& W \, {\textstyle{\frac{1}{2}}}\left({\mathcal O}_W + \bar{{\mathcal O}}_W\right) + X \, {\mathcal O}_X \bigg],
\end{eqnarray} 
where 
$U_N$, $U_A$, $W$, and $X$ are coupling constants respectively for (i) an SU(2)${}_1$ spin current-current interaction (denoted by ${\mathcal O}_N$), (ii) a U(1) charge current-current interaction (denoted by ${\mathcal O}_A$), (iii) umklapp interactions [denoted by $\left( {\mathcal O}_W + \bar{{\mathcal O}}_W \right)/2 $], and (iv) U(1) stress tensor interactions (denoted by ${\mathcal O}_X$). The precise definition of these operators is relegated to Appendix~\ref{1D:Coupling-Append}, see Eq.~(\ref{1DOps--Def}). The correspondence between the four coupling constants, appearing in Eqs.~(\ref{Hint_1D}) and ~(\ref{Hint_spinful}) is 
\begin{eqnarray}~\label{U-g}
	U_N &=& g^s_{3}-g^t_{2}-g^t_{3},	\,\,	U_A=2 g^s_{1}+g^s_{3} +3 g^t_{2}+3 g^t_{3}, \nonumber \\
	W &=& -g^s_{3}-3 g^t_{2} + 3 g_{j3}, \,\,	X=-g^s_{1},
\end{eqnarray}
or 
\begin{gather}
	g^s_{1}=-X, \,\, g^t_{2} = \ts{\frac{1}{12}} \left[ -3U_N+U_A-2W+2X \right], \nonumber\\ 
	\!\!\!
	g^s_{3}= \ts{\frac{1}{4}} \left[ 3 U_N + U_A+2 X \right], \,\,	g^t_{3} = \ts{\frac{1}{6}} \left[ U_A + W+2X \right]. \!
\label{g-U}
\end{gather}
This mapping will prove useful to connect with the physics of interacting electrons in 1D, as discussed in Sec.~\ref{spinful-RG}.

\section{Interacting spinless fermions}~\label{spinless-RG}

\begin{figure}
\subfigure[]{
\includegraphics[width=4.0cm, height=3.75cm]{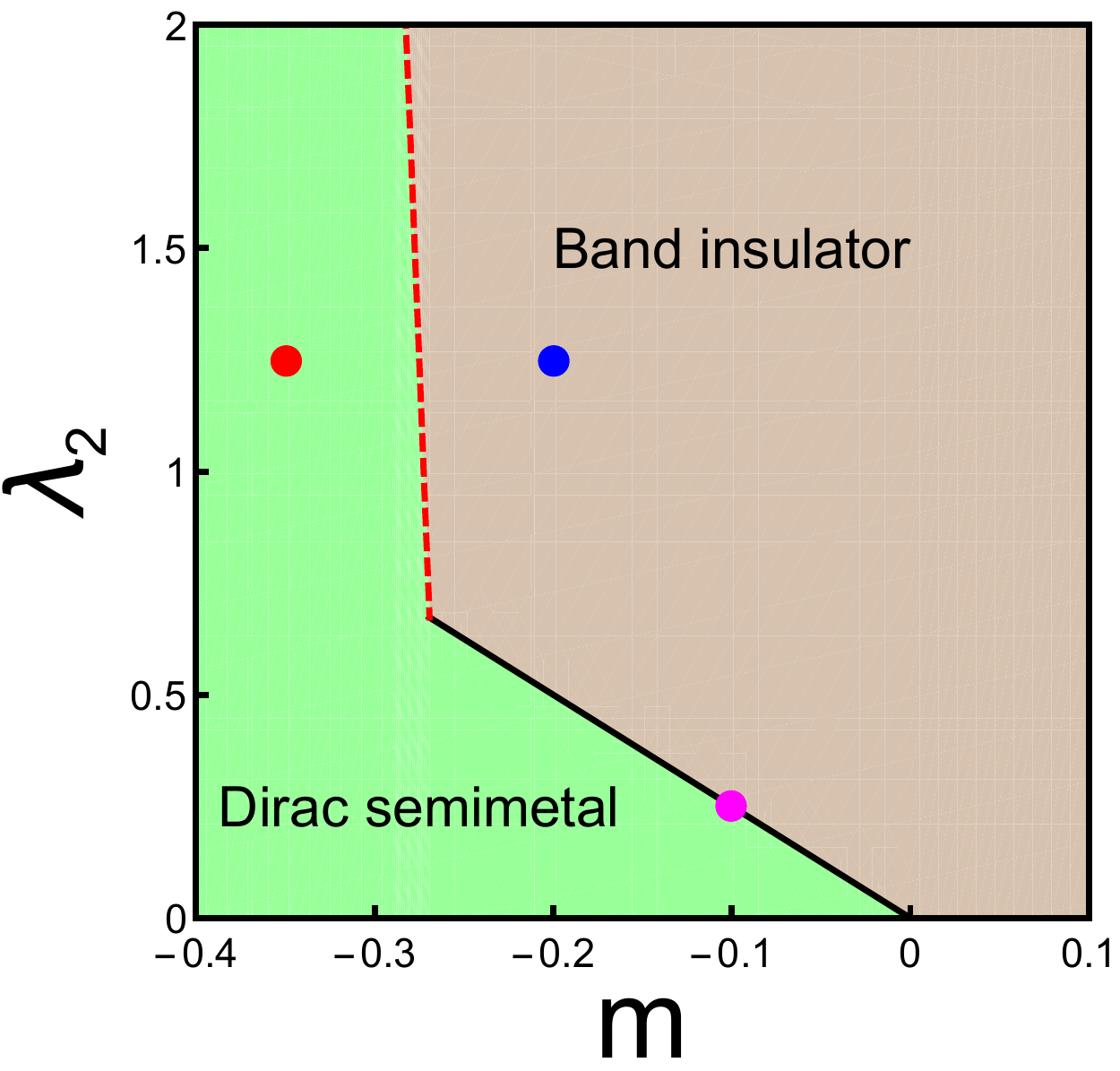}
\label{Firstorder}
}
\subfigure[]{
\includegraphics[width=4.0cm, height=3.75cm]{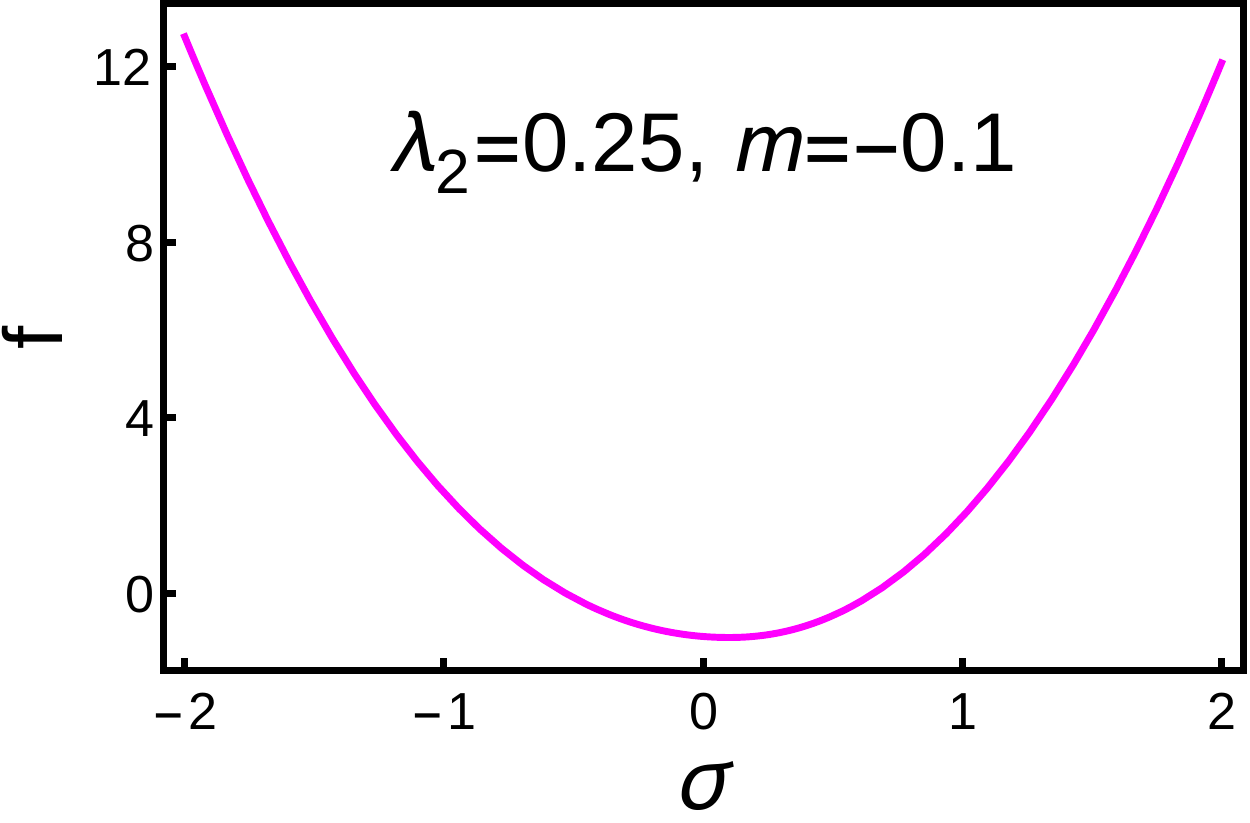}
\label{free-energy-continuous}
}
\subfigure[]{
\includegraphics[width=4.0cm, height=3.75cm]{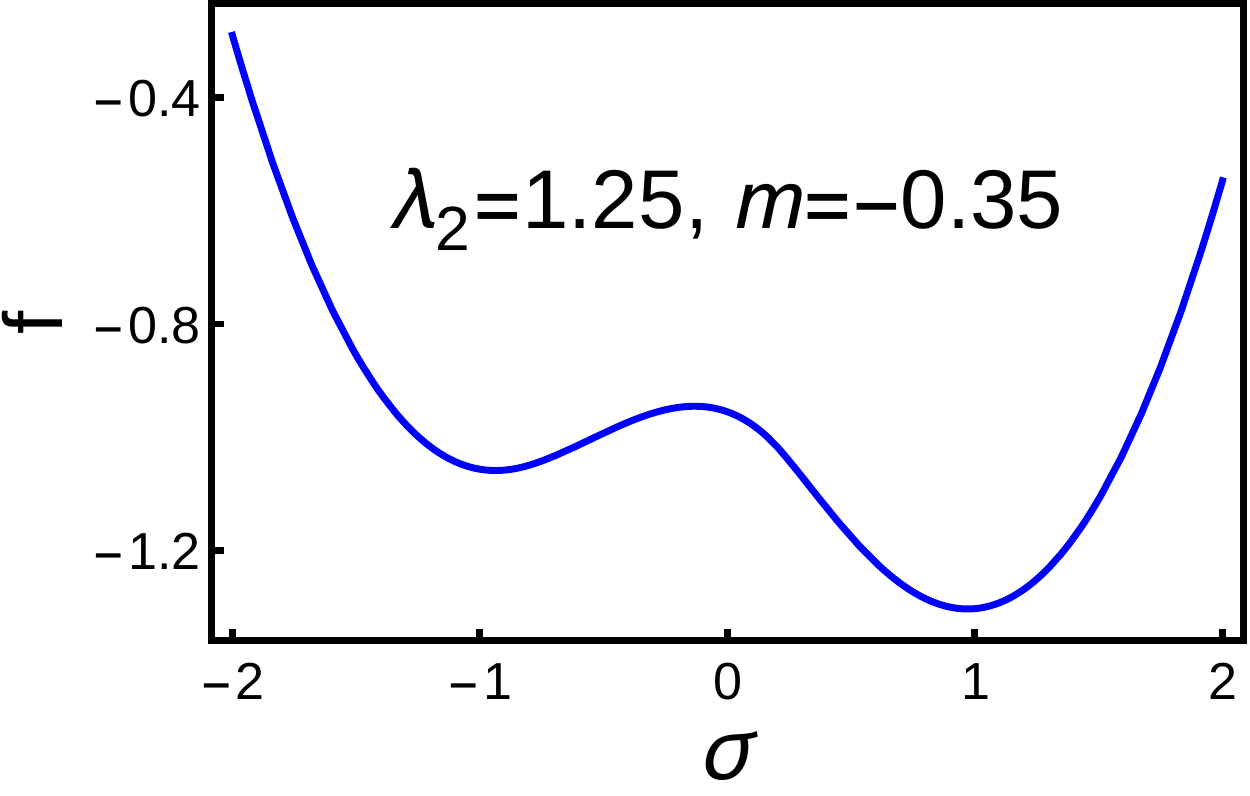}
\label{free-energy-insulator}
}
\subfigure[]{
\includegraphics[width=4.0cm, height=3.75cm]{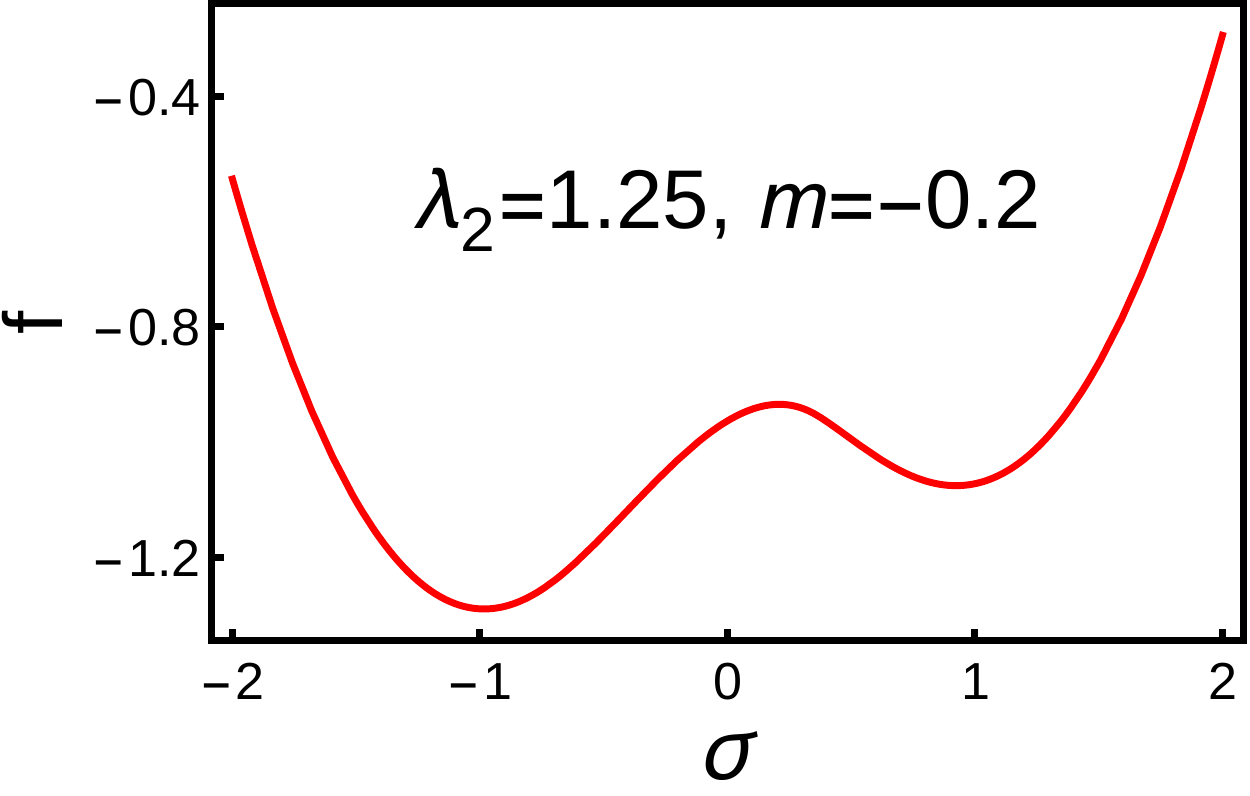}
\label{free-energy-DSM}
}
\caption{(a) Phase diagram of interacting spinless fermions obtained by minimizing the free energy, shown in Eq.~(\ref{free-energy-integral}). The profile of the (dimensionless) free energy  at three points in the phase diagram, marked in Fig.~\ref{Firstorder} is shown in panels (b)--(d). Values of various parameters are displayed in the corresponding figures. Here $\lambda_2$ and $m$ are the dimensionless interaction coupling and anisotropy parameter, and $\sigma$ is the interaction-induced condensate of the anisotropy parameter (also dimensionless) 
[see Eq.~(\ref{dimlesspara_1storder})]. (b) For weak enough interaction the free energy possesses only one global minimum, 
and thus the direct transition between Dirac semimetal and band insulator is continuous. The universality class of this transition is determined by the two-dimensional anisotropic semimetal. For stronger interaction the free-energy profile develops two inequivalent local minima, and respectively in panel (c) and (d) we display the free energy on the insulating and Dirac semimetal sides of the transition. Consequently the direct transition between these two phases at strong interaction becomes first order in nature [represented by the red dashed line in (a)].    }~\label{free-energy}
\end{figure}

We first discuss the effects of electronic interactions on spinless fermions. As shown in the previous section, generic local interactions for spinless fermions are described by only one quartic term. However, we will demonstrate that for stronger interaction the direct transition between the 
Dirac semimetal and band insulator can either 
$(i)$ get replaced by a first-order transition or 
$(ii)$ get avoided by an intervening broken-symmetry phase, depending on how we address the effects of strong electron-electron interactions.

\subsection{Topological first-order transition}~\label{1storder}

We first illustrate the possibility of the first-order transition between the 
Dirac semimetal and band insulator
at strong interaction, for which we consider the interacting Hamiltonian with local interaction $g_2$ [see Eq.~(\ref{Hint_spinless})]. This outcome can be demonstrated from the following mean-field or large-$N$ free energy 
\begin{equation}
	\frac{F}{N} =\frac{\Sigma^2}{2 g_2} -	\int \frac{d^2{\mathbf k}}{(2 \pi)^2} \; 
	\sqrt{v^2 k^2_x + \left( b k^2_y + \Delta + \Sigma \right)^2 },
\end{equation}
where $N$ is the number of two-component spinors, obtained after performing a Hubbard-Stratonovich transformation of the four-fermion interaction proportional to $g_2$ in favor of a bosonic field $\Sigma=\langle \psi^\dagger \tau_2 \psi \rangle$, and subsequently integrating out the critical fermions. The parameter $N$ (flavor number) should not be confused with $n$ [controlling the degree of anisotropy in ASM, Eq.~(\ref{hamil-nonint})]. To proceed with the calculation, let us now define a set of variables according to 
\begin{equation}
	v k_x=\rho \cos\theta, \: b k^2_y =\rho \sin \theta, \nonumber 
\end{equation}
with $0 \leq \theta \leq \pi$ and $0<\rho<E_\Lambda$, where $E_\Lambda$ is the high-energy cutoff up to which the quasiparticle dispersion is anisotropic (given by $E_{\mathbf k}$). In terms of dimensionless variables, defined as 
\begin{equation}~\label{dimlesspara_1storder}
	x		=\frac{\rho}{E_\Lambda}, \,	m		=\frac{\Delta}{E_\Lambda}, \,	\sigma=\frac{\Sigma}{E_\Lambda}, \,
	\lambda_2	= \frac{2 g_2 \sqrt{E_\Lambda}}{ (2 \pi)^2 v \sqrt{b} },
\end{equation}
(here $\sigma$ should not be confused with Pauli matrices) and the dimensionless free-energy density $f=F(2\pi)^2 2 v \sqrt{b}/(2 N E^{5/2}_\Lambda)$, we can express the above equation as 
\begin{eqnarray}~\label{free-energy-integral}
f &=& \frac{\sigma^2}{\lambda_2}-\int^1_0 \sqrt{x} dx \; \int^1_0 dy \; \frac{1}{\left( 1-y^2\right)^{3/4}} \nonumber \\
& \times & \sqrt{x^2 y^2 +\left( x \sqrt{1-y^2} + m +\sigma \right)^2}.
\end{eqnarray}   

We numerically minimize the free energy and the resulting phase diagram is shown in Fig.~\ref{Firstorder}. Nonetheless, the salient features of the phase diagram can be appreciated by expanding the above free energy in powers of $\sigma$, yielding 
\begin{equation}
	f	= \left[ \frac{m^2}{\lambda_2} +f_0 \right] -	\sigma \left[ \frac{m}{\lambda_2}-f_1 \right] + 
	\sum_{j \in \mbox{integer}} \sigma^{j} f_{j},
\end{equation}
after shifting the variable $\sigma+m \to \sigma$. It should, however, be noted that such expansion of the free-energy density 
in powers of $\sigma$ is not a well-defined procedure, and for $j>4$ the $f_j$s are non-analytic functions. But, most importantly, \emph{the free energy $f$ contains all odd powers of $\sigma$}. When the interaction is sufficiently weak the $\sigma$ field does not condense, and the profile of the free energy possesses a single global minimum, as shown in Fig.~\ref{free-energy-continuous}. Consequently, the nature of the direct, continuous transition between the 
Dirac semimetal and band insulator
remains unchanged at weak coupling. In this regime only the term linear in $\sigma$ 
is important and $\delta=\lambda_2 f_1$ defines the phase boundary between these two phases; this is represented by the black solid line in Fig.~\ref{Firstorder}.

On the other hand, beyond a critical strength of the interaction $\lambda_2 > \lambda^\ast_2 \sim 1$, the $\sigma$ field acquires a finite vacuum expectation value and all powers (including the odd ones) of $\sigma$ in $f$ become important. In this regime the profile of the free energy accommodates two inequivalent minima, as shown in Fig.~\ref{free-energy-insulator} and Fig.~\ref{free-energy-DSM}, respectively on the 
band-insulator and Dirac-semimetal sides of the phase diagram. As a result, the direct transition between these two phases at strong coupling becomes a \emph{fluctuation-driven first-order transition} along the red dashed line 
in Fig.~\ref{Firstorder}~\footnote{Even though we here do not account for any local or spatial fluctuation of the order parameter $\Sigma$ or $\sigma$ field, the terminology ``fluctuation driven first-order transition" is justified in this case since it can only be captured upon incorporating quantum corrections to the free-energy density ($f$) due to electronic interaction $g_2$ or $\lambda_2$. In particular, in the large-$N$ (flavor number) limit such corrections arises from the Feynman diagrams (b) and (d), shown in Fig.~\ref{Feynman}. See also Ref.~\cite{first-order:old} for discussion on similar issue in one spatial dimension.}. 
When the 
Dirac semimetal
is separated from the 
band insulator
by a first-order transition, there exists no ASM at the transition. A similar first-order transition at strong coupling has also been predicted in two- and three-dimensional topological insulators~\cite{roy-goswami-sau, amaricci-1, amaricci-2, juricic-abargel} and also in a three-dimensional Weyl material residing in proximity to the 
semimetal to band-insulator QCP~\cite{goswami-roy-juricic}.

It is worth mentioning that a first-order transition has recently been observed in the pressured organic compound $\alpha$-(BEDT-TTF)${}_2\text{I}_3$~\cite{firstorder-experiment}. Thus our theory provides a possible explanation of this experimental observation.

\subsection{Continuous quantum phase transition
}~\label{Sec:spinless-RG}

\begin{figure}
\includegraphics[width=8cm, height=5cm]{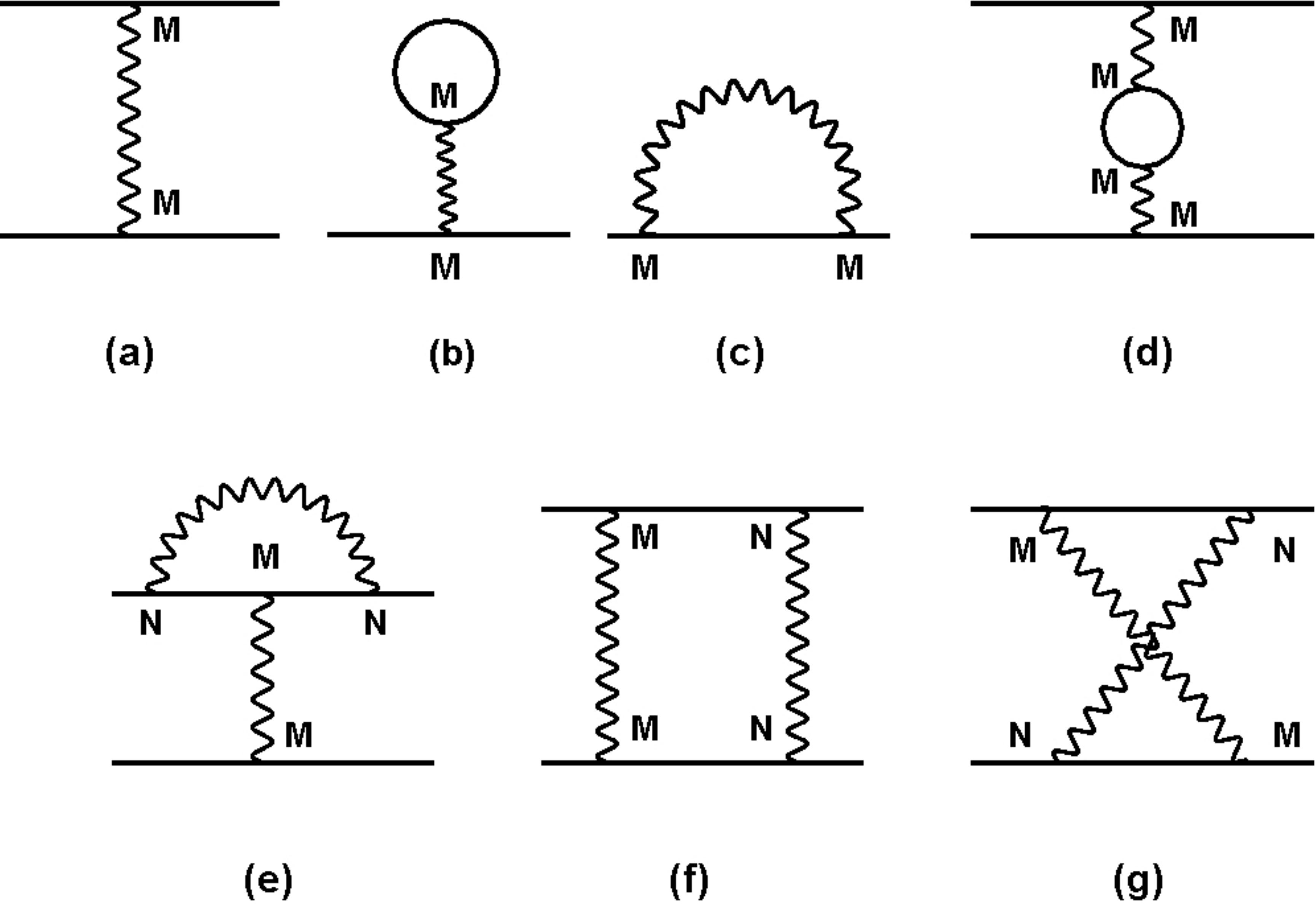}
\caption{(a) Bare four-fermion interaction vertex $\left( \psi^\dagger M \psi \right)^2$ (for spinless fermions) or $\left( \Psi^\dagger M \Psi \right)^2$ (for spin-$1/2$ electrons). The Feynman diagrams contributing to the renormalization group flow equations to the leading order in $\epsilon$ and $1/n$ expansion are shown as (b)--(g). Here $M$ and $N$ are $2 \, \times \, 2$ (for spinless fermions) and $4 \, \times \, 4$ (for spin-$1/2$ electrons) Hermitian matrices.  }~\label{Feynman}
\end{figure}

Next we will demonstrate the second possibility regarding the fate of the direct transition between the 
Dirac semimetal and band insulator, when strong interactions causes nucleation of a 
broken symmetry phase
that masks the direct transition between these two symmetry-preserving phases [see Fig.~\ref{continuous-schematic}]. For spinless fermions, only the CDW phase gives rise to a fully gapped spectrum; inside this ordered phase $\langle \psi^\dagger \tau_3 \psi \rangle \neq 0$. Thus CDW order is energetically advantageous at temperature $T=0$ for sufficiently large $g_3 > 0$ [Eq.~(\ref{Hint_spinless})]. Next we will establish the onset of such a CDW phase by carrying out a RG calculation on the interacting model 
with local interaction $g_3$. The imaginary-time action for such interacting system reads
\begin{eqnarray}\label{S--spinless}
	{\mathcal S} 
	&=& 
	{\mathcal S}_0 
	-
	g_3 \int dt \; d^2{\mathbf r} \; \left( \psi^\dagger \tau_3 \psi \right)^2.
\end{eqnarray}          

\begin{figure}[t!]
\subfigure[]{
\includegraphics[width=4.0cm, height=4.0cm]{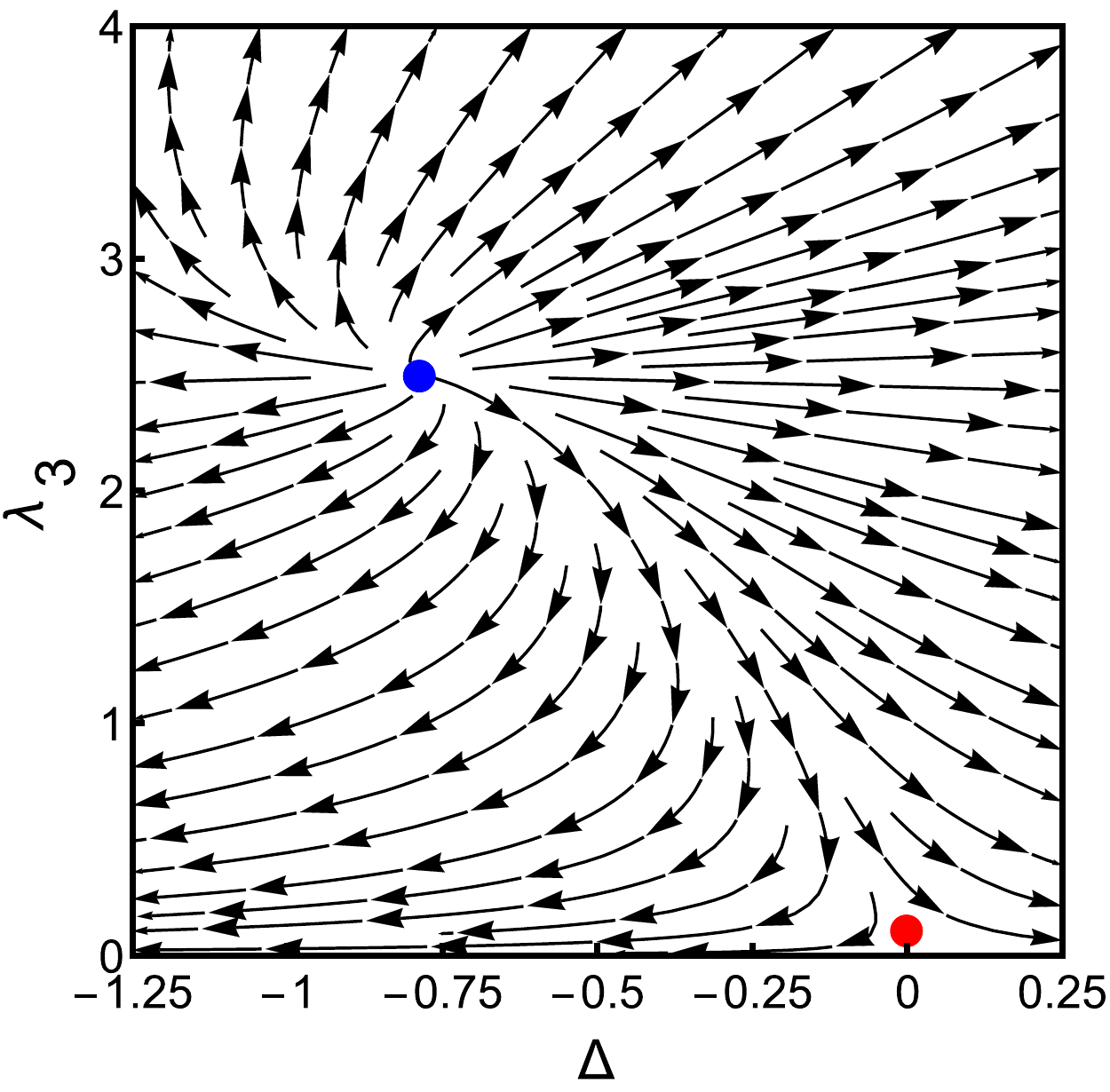}
\label{RG_Flow_Spinless}
}
\subfigure[]{
\includegraphics[width=4.0cm, height=4.0cm]{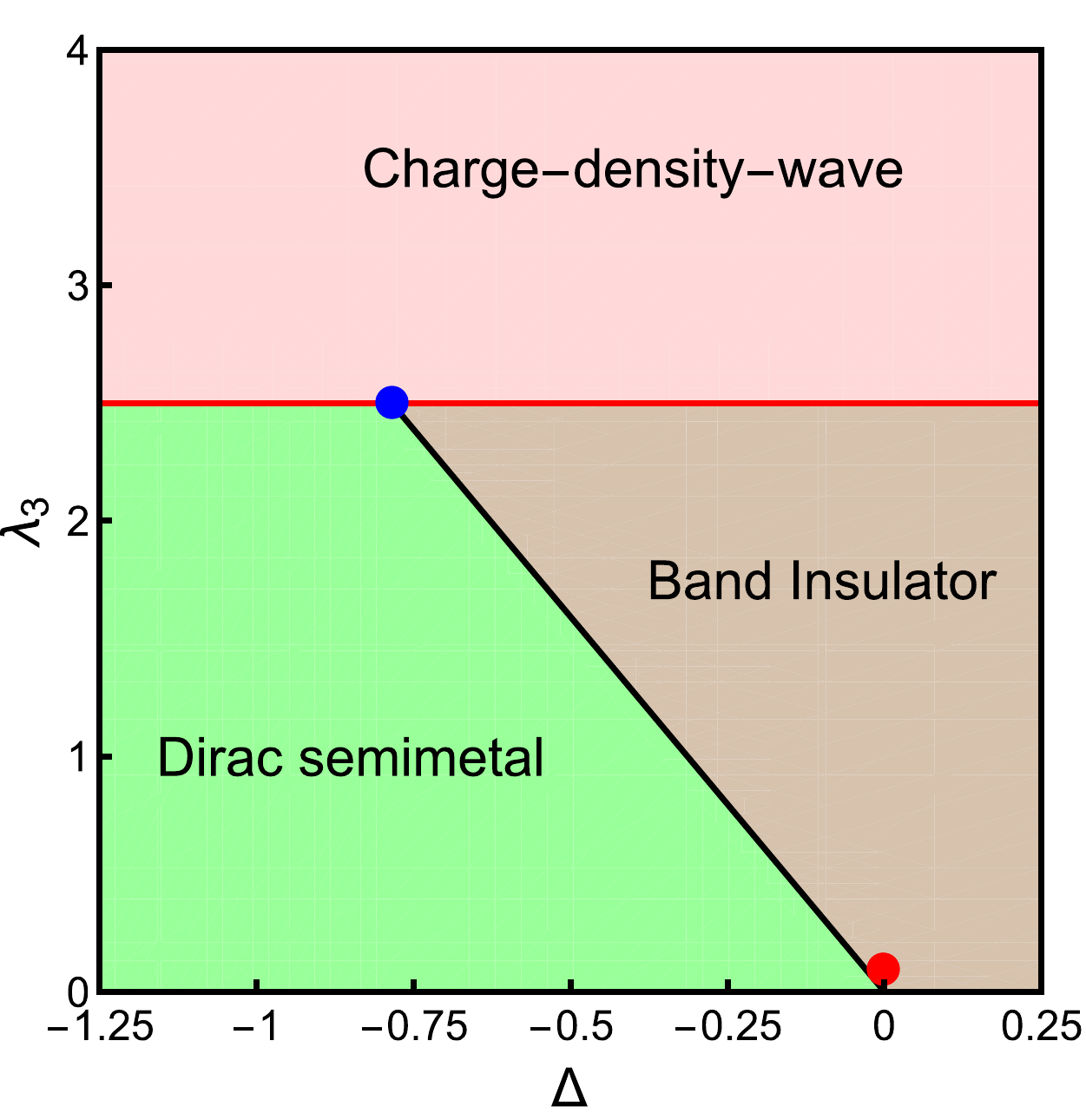}
\label{RG_PD_Spinless}
}
\caption{The renormalization group flow diagram obtained from Eq.~(\ref{RG-spinless})
for spinless fermions is shown in (a), and the corresponding phase diagram is shown in panel (b).
The red (blue) dot corresponds to the noninteracting critical
(interacting multicritical) point.
The (dimensionless) coupling constant $\lambda_3$ and band parameter $\hat{\Delta}$
in panels (a) and (b) are measured in units of $\epsilon$,
and in the RG flow equations [see Eq.~(\ref{RG-spinless})]
we substitute $n=10$ in $f_j(n)$.
}~\label{RG:spinless}
\end{figure}

After accounting for perturbative corrections to the quadratic order in $g_3$ (computing the one-loop diagrams shown in Fig.~\ref{Feynman}) and integrating out the fast Fourier modes within the Wilsonian shell $E_\Lambda e^{-l}	<	\sqrt{\omega^2+v^2k^2_x} <	E_\Lambda$ and 
$0<|k_y|<\infty$, we arrive at the following RG flow equations, 
\bsub\label{RG-spinless}
\begin{align}
	\frac{d\hat{\Delta}}{dl} 	=&\, \hat{\Delta} + \lambda_3 \left[ \hat{\Delta} \left[ 2 f_1(n) + f_2(n) \right] + f_3(n) \right], 
\\
	\frac{d \lambda_3}{dl}=&\,-\epsilon \lambda_3 + 4 \; \lambda^2_3 \; f_2 (n),
	\label{lambda3Flow--spinless}
\end{align} 
\esub 
where $\hat{\Delta}=\Delta/E_\Lambda$ and $\lambda_3= g_3 E^\epsilon_\Lambda/(8 \pi^2 v b^\epsilon)$ are dimensionless parameters. Here $E_\Lambda$ is the energy cutoff for $\omega$ and $v |k_x|$, and $\omega$ is the Matsubara frequency.	The functions appearing in the RG flow equations are (see Appendix~\ref{Append-loop-integral} for details)  
\allowdisplaybreaks[4]
\begin{eqnarray}~\label{three_functions}
	f_1(n) 	&=& \frac{\pi  (2 n-1) \csc \left(\frac{\pi }{2 n}\right)}{4 n^2}=	1-\frac{1}{2n}+ {\mathcal O} \left( n^{-2} \right), 
\nonumber \\	
  f_2(n) 	&=& \frac{\pi  \csc \left(\frac{\pi }{2 n}\right)}{2 n^2}	=	\frac{1}{n}+ {\mathcal O} \left( n^{-2} \right), 
\nonumber \\
	f_3(n) 	&=& \frac{\pi  \sec \left(\frac{\pi }{2 n}\right)}{n}	=	\frac{\pi}{n}+ {\mathcal O} \left( n^{-2} \right).
\end{eqnarray}
Therefore, as $n \to \infty$ only the contribution from $f_1(n)$ survives, while $f_2(n)$ and $f_3(n)$ capture subleading logarithmic divergences. 
The above coupled flow equations support only two fixed points: 

\begin{enumerate}

\item $\left( \hat{\Delta}, \lambda_3 \right)=(0,0)$ corresponds to the noninteracting system, the ASM critical point 
that connects the Dirac semimetal to the symmetry-preserving band insulator. 
 
\item $\left( \hat{\Delta}, \lambda_3 \right)=\left( -\frac{\pi}{4}, \frac{n}{4} \right) \epsilon$, which, on the other hand, stands as an interacting 
multicritical point
in the $\left( \hat{\Delta}, \lambda_3 \right)$ plane, where 
the
Dirac-semimetal, band-insulator,
ASM and CDW phases meet. Therefore, this 
point
controls a continuous 
quantum phase transition
of critical excitations residing at the phase boundary between the 
Dirac semimetal and band insulator
towards the formation of the CDW phase. The RG flow and the phase diagram are respectively shown in Fig.~\ref{RG_Flow_Spinless} and Fig.~\ref{RG_PD_Spinless}.      
\end{enumerate}

We now show how to compute the location of the interacting 
multicritical point
from Eq.~\ref{RG-spinless}. Respectively, $d\hat{\Delta}/dl=0$ and $d\lambda_3/dl=0$ leads to
\begin{eqnarray}~\label{solgeneral:Deltalambda}
\Delta_{\ast} &=&- \frac{\lambda_{3,\ast}}{1+ \lambda_{3,\ast} \left[ 2 f_1(n) + f_2(n) \right]} \; f_3(n) \nonumber \\
&=& - f_3 (n) \lambda_{3,\ast} \; \left( 1- \lambda_{3,\ast} \left[ 2 f_1(n) + f_2(n) \right] + \cdots \right), \nonumber \\
\lambda_{3,\ast} &=& \frac{\epsilon}{4 f_2(n)}=\frac{n}{4} \epsilon, 
\end{eqnarray}
where quantities with subscript ``$\ast$" denotes their fixed point values. Since, $\lambda_{3,\ast} \sim \epsilon$, for a systematic extraction of $\hat{\Delta}_\ast$ to the leading order in $\epsilon$ (interaction mediated shift of the band parameter), we only need to account for the first term in the above expression for $\hat{\Delta}_\ast$, yielding~\cite{chaikin-lubensky}
\begin{equation}
\hat{\Delta}_\ast =-f_3(n) \lambda_{3,\ast} + {\mathcal O}(\lambda^2_{3,\ast})
= -\frac{\pi}{4} \epsilon + {\mathcal O}(\epsilon^2),
\end{equation}
leading to the result announced above. 
The remaining terms in the expression of $\hat{\Delta}_\ast$ from Eq.~(\ref{solgeneral:Deltalambda}) yields contributions at least ${\mathcal O}(\epsilon^2)$.
Therefore, while determining the location of the 
multicritical point,
RG flow and phase diagram in the $\left( \hat{\Delta}, \lambda_3\right)$ plane, we only accounted for the terms proportional to $\hat{\Delta}$ and $\lambda_3$, and neglected the contribution proportional to $\hat{\Delta} \lambda_3$ in $d\hat{\Delta}/dl$. We follow the same procedure for spin-$1/2$ electrons discussed in the next section.

From the leading order in $\epsilon$- and $1/n$-expansions we can only estimate the correlation length exponent ($\nu$) for the ASM-CDW continuous 
transition, yielding
\begin{equation}~\label{nu_definition} 
	\nu^{-1}	= \frac{d}{d\lambda_3} \; \left[ \frac{d\lambda_3}{dl} \right] \bigg|_{\lambda_3=\lambda^\ast_3}=	\epsilon, 
\end{equation}
where 
$\lambda^\ast_3= \frac{n}{4} \epsilon$ represents the strength of four-fermion interactions at the 
multicritical point. 
Hence, for the physically relevant situation with $\epsilon=1/2$ and $n=2$, we obtain $\nu=2$. 
On the other hand, when the 
Dirac semimetal
undergoes a continuous 
transition
into the CDW phase 
$\nu=1$ (to one-loop order)~\cite{HJR}. Thus, the universality class of the critical excitation-CDW 
transition
is fundamentally a new one.

However, we note that the
multicritical point
describing the transition is not strictly controlled by a small parameter as $\lambda^\ast_3 \sim n \epsilon \sim 1$. This outcome is specific (and to certain extent pathological) for spinless two-component fermions. For spin-1/2 electrons, we show in the next section that all of the interacting QCPs and the associated critical properties are simultaneously controlled by two small parameters $\epsilon$ and $1/n$.

The spinless case gives an order one coupling strength at the 
multicritical point
because the correction in Eq.~(\ref{lambda3Flow--spinless}) vanishes in the $n \rightarrow \infty$ limit. This is true to all orders in $\lambda_3$, since the interaction in Eq.~(\ref{S--spinless}) is equivalent to a U(1) current-current (Thirring~\cite{zinn-justin}) perturbation. This is an exactly marginal Luttinger interaction in the 1D limit \cite{giamarchi,tsvelik}, which obtains for $\epsilon = 0$ and $n \rightarrow \infty$. Thus the ASM-CDW 
quantum phase transition
is solely captured by the subleading divergence arising from $f_2(n)$, which can directly be tested at least in quantum Monte Carlo simulation with only NN interaction 
in a uniaxially stressed honeycomb lattice~\cite{troyer-honeycomb, hong-yao-NN-honeycomb} [with $t_2=2 t_1$, see Fig.~\ref{lattice_hopping}]. By contrast, for isotropic two-component massless Dirac fermions, the RG flow equation vanishes to at least order $\lambda^3_3$~\cite{gracey}, although it is believed that there exists a possibly continuous semimetal-insulator 
transition
at finite interaction strength~\cite{rosa, wetterich}. Thus far this issue remains unsettled. On the other hand, our leading order RG calculation predicts a continuous anisotropic semimetal-CDW transition.

\subsection{First vs.\ second-order transitions }

We have identified two incompatible scenarios for the strong coupling physics
in the spinless case, wherein the ASM is either replaced by a first-order transition,
or the CDW phase. These conclusions are reached via calculations with different control parameters: 
the number of flavors $N$ for the first-order transition, and the curvature of the dispersion along $k_y$ (characterized by the $n$) for the 
intervening CDW phase. Since we must set $N = 1$ and $n = 2$ for the spinless model,
both results obtain at strong coupling and we cannot make a rigorous mathematical
argument that one scenario is more likely. However, microscopic considerations 
provide some physical intuition which can favor one or the other.

For a spinless ASM obtained from a microscopic strained honeycomb lattice
model that is free of frustration 
(e.g., a ``$t$-$V$'' model with nearest-neighbor hopping and density-density interactions only), 
the CDW order is expected to preempt the topological first-order transition between the 
Dirac semimetal and band insulator, since the CDW order completely gaps out critical excitations,
producing a uniform mass gap in the spectrum.
This can be tested in a quantum Monte Carlo simulation~\cite{troyer-honeycomb, hong-yao-NN-honeycomb}. 
However, we cannot completely rule out the first-order transition, 
since bandwidth renormalization and/or the suppression of the quasiparticle weight may become 
crucial to determine the ultimate fate of the strongly interacting ASM. 
These effects are not included in the one-loop RG. 

For spin-1/2 electrons generic local interactions live in a four-dimensional coupling constant space (see the next section). 
We can again gain some intuition by considering a particular microscopic model, e.g.\ the 
strained honeycomb lattice extended Hubbard model [see Fig.~\ref{UV_PD}]. 
Again this model is ``frustration free,'' and the fully gapped mass orders CDW, AFM, and $s$-wave pairing 
are expected to occur at strong coupling.  
Nucleation of these mass orders via a continuous phase transition is energetically superior over the first order transition
out of the Dirac semimetal, since the ordered phases fully gap the spectrum (maximal gain of condensation energy). 

We therefore conclude that if the extended Hubbard model can be realized in strained graphene
or 
an anisotropic honeycomb
optical lattice for ultracold
fermions, the first-order transition will be unlikely. For a material with different microscopics 
[e.g.\ the organic compound $\alpha$-(BEDT-TTF)$_2\text{I}_3$], 
the Dirac-semimetal to band-insulator
transition could occur at strong coupling in a corner of the four-dimensional interaction parameter space
where the first-order transition is actually preferred. 
It is a nontrivial task to identify a microscopic model that can exhibit 
the proposed first order transition in this high-dimensional coupling space. 
Finally, it must be emphasized that the first-order transition we discuss here 
between two symmetry-preserving phases (Dirac semimetal and band insulator) 
is fundamentally different from the one between two distinct broken symmetry phases. 
Therefore, such a fluctuation-driven first-order transition 
cannot be captured by an RG calculation and the computation of free-energy density 
(non-perturbative analysis in the large-$N$ limit) is necessary~\cite{first-order:old}.

\section{Spin-$1/2$ electrons}~\label{spinful-RG}

Next we perform the RG calculation to the leading order in $\epsilon$ and $1/n$ starting from the interacting $H_{int}$ [defined in Eq.~(\ref{Hint_spinful})] for spin-$1/2$ electrons. We made a judicious choice in selecting the linearly independent coupling constants in $H_{int}$. 
Note that we did not choose $g^s_2$ as one of the four independent couplings, since as we know when this coupling constant is strong enough the
Dirac-semimetal to band-insulator quantum phase transition 
becomes a fluctuation-driven first-order transition [following the analysis presented in Sec.~\ref{1storder} but now for spinful electrons, which can be accomplished by taking $N \to 2N$]. Nonetheless, as we will discuss in the next section, the source term $\Delta^s_2$ (corresponding to the anisotropy parameter) never displays the \emph{leading divergence}. Thus, we can proceed with the following RG analysis tailored to 
address 
continuous phase transitions into broken symmetry phases, leaving aside the possibility of a first-order transition as an alternative scenario. Following the same procedure described for spinless fermions in Sec.~\ref{Sec:spinless-RG} (see Fig.~\ref{Feynman} for the relevant Feynman diagrams) we arrive at the following RG equations
\begin{widetext}
\allowdisplaybreaks[4]
\begin{eqnarray}~\label{RG-spinful}
\frac{d\hat{\Delta}}{dl} &=& \hat{\Delta} + \left( \lambda^s_{1}+\lambda^s_{3}-3\lambda^t_{2}+3\lambda^t_{3} \right) \times \left( \left[ 2 f_1(n)+f_2(n)\right] \hat{\Delta} + f_3(n)\right), \nonumber\\
\frac{d\lambda^s_{1}}{dl} &=& -\epsilon \lambda^s_{1} + \frac{\lambda^s_{1}}{2} \left[ \lambda^s_{1} + \lambda^s_{3} + 3\lambda^t_{2} + 3 \lambda^t_{3}\right] f_2(n)-3\lambda^s_{1} \lambda^t_{2} f_2(n), \nonumber \\
\frac{d\lambda^s_{3}}{dl} &=& -\epsilon \lambda^s_{3} + \frac{\lambda^s_{3}}{2} \left[ \lambda^s_{1} + \lambda^s_{3} + 3\lambda^t_{2} - 3 \lambda^s_{3}\right] \times \left[ 2 f_1(n) + f_2(n) \right] + f_1(n) \big[-\lambda^s_{1} \lambda^s_{3}-3\lambda^s_{3}\lambda^t_{2}+3 \left(\lambda^t_{2} \right)^2 \nonumber \\
&-& 3\lambda^t_{2} \lambda^t_{3}+3 \left( \lambda^t_{3} \right)^2 \big] + f_2(n) \left[ 3 \lambda^s_{1} \lambda^t_{2} +  3 \lambda^t_{2} \lambda^t_{3} + \frac{1}{2} \left( \left[ \lambda^s_{1} \right]^2 + \left[ \lambda^s_{3} \right]^2 + 3 \left[ \lambda^t_{2} \right]^2 + 3 \left[ \lambda^t_{3} \right]^2 \right) \right], \nonumber \\
\frac{d\lambda^t_{2}}{dl} &=& -\epsilon \lambda^t_{2} + \frac{\lambda^t_{2}}{2} \left[ \lambda^s_{1} + \lambda^s_{3} + 3 \lambda^t_{2} - \lambda^t_{3} \right] \times \left[ 2 f_1(n) - f_2(n) \right] + f_1(n) \bigg[ \frac{1}{3} \lambda^s_{1} \lambda^s_{3}-\lambda^s_{1} \lambda^t_{3} + \lambda^s_{3} \lambda^t_{2}-\left[ \lambda^t_{2} \right]^2 \nonumber \\
&-& \lambda^t_{2} \lambda^t_{3} -\left[ \lambda^t_{2} \right]^2 \bigg] + f_2(n) \left[ \lambda^s_{3} \lambda^t_{3} -\frac{1}{6} \left( \left[ \lambda^s_{1} \right]^2 + \left[ \lambda^s_{3} \right]^2 + 3 \left[ \lambda^t_{2} \right]^2 + 3 \left[ \lambda^t_{3} \right]^2 \right) \right], \nonumber \\
\frac{d\lambda^t_{3}}{dl} &=& -\epsilon \lambda^t_{3} + \frac{\lambda^t_{3}}{2} \left[ \lambda^s_{1} - \lambda^s_{3} -\lambda^t_{2} + \lambda^t_{3} \right] \times \left[ 2 f_1(n) + f_2(n) \right] + f_1(n) \left[ -\frac{1}{3} \lambda^s_{1} \lambda^s_{3} - \lambda^s_{1} \lambda^t_{2} -2 \lambda^t_{2} \lambda^t_{3} \right] \nonumber \\
&+& f_2(n) \left[ \lambda^s_{1} \lambda^t_{2} + \lambda^s_{3} \lambda^t_{2} -2 \lambda^t_{2} \lambda^t_{3} + \frac{1}{6} \left( \left[ \lambda^s_{1} \right]^2 + \left[ \lambda^s_{3} \right]^2 + 3 \left[ \lambda^t_{2} \right]^2 + 3 \left[ \lambda^t_{3} \right]^2 \right) \right],
\end{eqnarray}  
where $\lambda^{a}_\mu= g^a_\mu E^\epsilon_\Lambda/(\pi^2 v b^\epsilon)$ is the dimensionless coupling constant and $\hat{\Delta}=\Delta/E_\Lambda$ is the dimensionless band parameter, and recall that here $n$ can only take even integer values. The above set of RG flow equations can also be expressed in terms of the coupling constants introduced in Eq.~(\ref{Hint_1D}) (namely $U_N$, $U_A$, $W$ and $X$) according to
\allowdisplaybreaks[4]
\begin{eqnarray}~\label{RG:1Dcoupling}
\frac{d\hat{\Delta}}{dl}&=&\hat{\Delta}+ \left[ \frac{1}{2} \left( U_A + 3 U_N\right) + W \right] \left( \left[ 2 f_1(n)+f_2(n)\right] \hat{\Delta} + f_3(n)\right), \:
\frac{dX}{dl}   = -\epsilon X + \frac{1}{4} X \bigg[ U_A + 3 U_N + 2 W \bigg] f_2(n), \nonumber \\
\frac{dU_N}{dl} &=& -\epsilon U_N + U^2_N f_1(n)+ \frac{1}{72} \bigg[ 8 U^2_A + 27 U^2_N -33 U_N W -4 W^2 -15 U_N U_A -23 U_A W +14 X U_A \nonumber \\
&-& 12 U_N X -10 W X + 32 X^2 \bigg] f_2(n), \nonumber \\ 
\frac{dU_A}{dl} &=& -\epsilon U_A + W^2 f_1(n) + \frac{1}{24} \bigg[ 4 U^2_A + 15 U_A U_N + 9 U^2_N + 11 U_A W -3 U_N W + 4 W^2 -14 U_A X \nonumber \\
&+& 12 U_N X + 10 X W - 8 X^2 \bigg] f_2(n),  \nonumber \\
\frac{dW}{dl}   &=& -\epsilon W + U_A W f_1(n)+ \frac{1}{24} \bigg[ -2 U^2_A-3 U_N U_A + 9 U^2_N + 5 U_A W + 15 U_N W \nonumber \\
&+& 16 W^2-14 X U_A + 12 X U_N + 10 W X -8 X^2 \bigg] f_2(n). 
\end{eqnarray}
\end{widetext}
In the last set of equations all coupling constants are dimensionless. Next we systematically analyze the above set of RG flow equations.
We addressed the effect of RG flow of $\hat{\Delta}$ in the phase diagram of the interacting ASM for the spinless case. \emph{Unless otherwise mentioned we will work in the hyperplane defined by $\hat{\Delta}(l)=0$, where $\hat{\Delta}(l)$ is the renormalized band parameter. Therefore, all interacting fixed points with $\lambda^{s,t}_j \neq 0$ [see Tables~\ref{fixedpoint1D} and ~\ref{table-1overn}, and Fig.~\ref{CFT-flowdiagram}] have one additional unstable direction, which is $\hat{\Delta}$, 
and all of them are truly multi-critical in nature.} The functions $f_{1}(n), f_{2}(n)$ and $f_{3}(n)$ have already been defined in Eq.~(\ref{three_functions}).

\subsection{Emergent one-dimensional system: spin-charge separation}

\begin{figure}[b!]
\includegraphics[width=4.25cm]{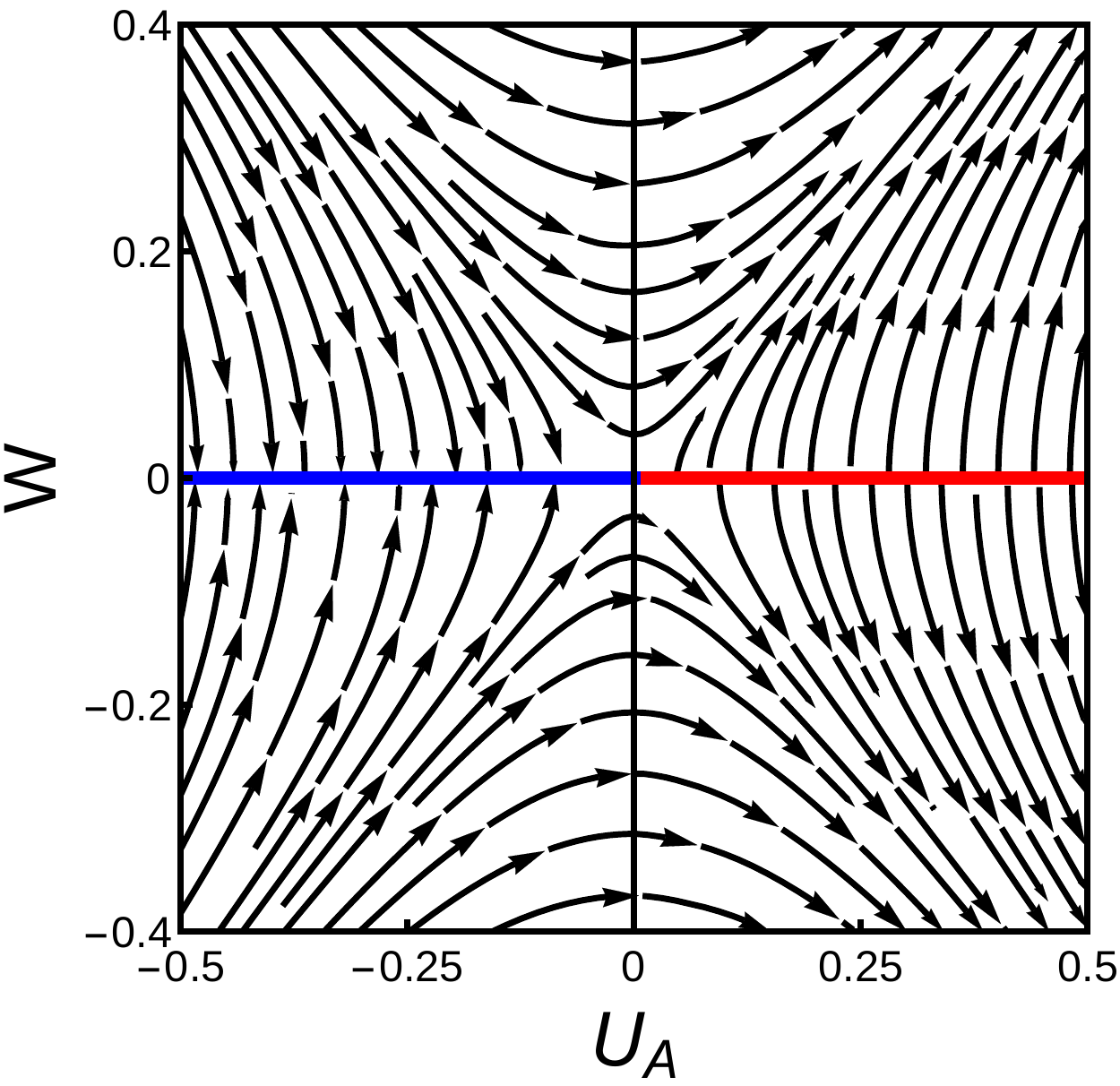}
\caption{RG flow diagram in the $\left(U_A ,W\right)$ plane in the strict one-dimensional limit 
[obtained by taking $n \to \infty$ in Eq.~(\ref{RG:1Dcoupling}) and setting $\epsilon=0$ in Eq.~(\ref{RG:1Dstrict})]. 
The stable (unstable) line of fixed points is indicated in blue (red).  
The flow determines the charge sector Kosterlitz-Thouless transition~\cite{giamarchi}.}~\label{KT}
\end{figure}

In the limit $n \rightarrow \infty$, Eq.~(\ref{RG:1Dcoupling}) reduces to 
\begin{align}~\label{RG:1Dstrict}
	\frac{dU_N}{dl} =&\, 	-\epsilon U_N + U^2_N, 	\quad\frac{dU_A}{dl}	= 	-\epsilon U_A + W^2, \nonumber\\
	\frac{dW}{dl} 	=&\,	-\epsilon W + U_A W, 	\quad	\frac{dX}{dl} 	= 	-\epsilon X.
\end{align}  
If we set the engineering dimension of all the four-fermion couplings $\epsilon=0$, then we recover the well-known RG flow equations for these coupling strengths in one dimension.

In this limit, the anisotropic semimetal with local interactions reduces to a decoupled collection of spin-1/2 one-dimensional ``wires''. Such systems exhibit spin-charge  separation \cite{giamarchi,tsvelik}, and this is reflected in Eq.~(\ref{RG:1Dstrict}). The only spin sector interaction $U_N$ couples to the SU(2)${}_1$ current-current perturbation [see Eq.~(\ref{1DOps--Def}) and the surrounding text for a precise definition]. For $\epsilon = 0$, $U_N$ resides on the inflowing (outflowing) part of the spin sector Kosterlitz-Thouless separatrix for $U_N < 0$ ($U_N > 0$) \cite{giamarchi,tsvelik}; spin SU(2) symmetry is preserved everywhere along the separatrix.

The remaining parameters couple to interaction operators that perturb the charge sector, as can be seen from the bosonization of the latter [Eq.~(\ref{Scharge})]. $U_A$ and $X$ couple to the $U(1)$ current-current and $U(1)$ stress tensor operators, respectively; these modify only the Luttinger parameter and charge velocity of the free boson description, as shown in Eq.~(\ref{vcKc}). By contrast, the umklapp interaction $W$ couples to a sine-Gordon perturbation. The flow equations for $U_A$ and $W$ with $\epsilon = 0$ are the charge sector Kosterlitz-Thouless equations. The RG flow in the $\left( U_A, W \right)$ plane is shown in Fig.~\ref{KT}.

Since spin and charge are independent, Eq.~(\ref{RG:1Dstrict}) with $\epsilon = 0$ implies that either sector can be a gapless Luttinger liquid or gapped Mott insulator. The spin sector becomes massive when $U_N \rightarrow +\infty$, while the charge sector becomes massive when $U_A \rightarrow +\infty$ and $W \rightarrow \pm \infty$. The product of these gives four different composite phases that must
be interpreted through a microscopic model. The one-dimensional extended Hubbard model described by the phase diagram in Fig.~\ref{UV_PD_1D} is an example; it is further reviewed in Sec.~\ref{extended-hubbard}, below.

The gapless charge sector Luttinger liquid phase is a stable line of fixed points, shown in Fig.~\ref{KT}. The tuning parameter is the interaction $U_A$ or equivalently, the charge sector Luttinger parameter $K_c$ [Eq.~(\ref{vcKc})]. Up to logarithmic corrections \cite{giamarchi} that we ignore, the spin Luttinger liquid is a fixed point with $U_N = 0$ and spin sector Luttinger parameter $K_s = 1$.

\begin{figure}
\subfigure[]{
\includegraphics[width=4.0cm, height=4.0cm]{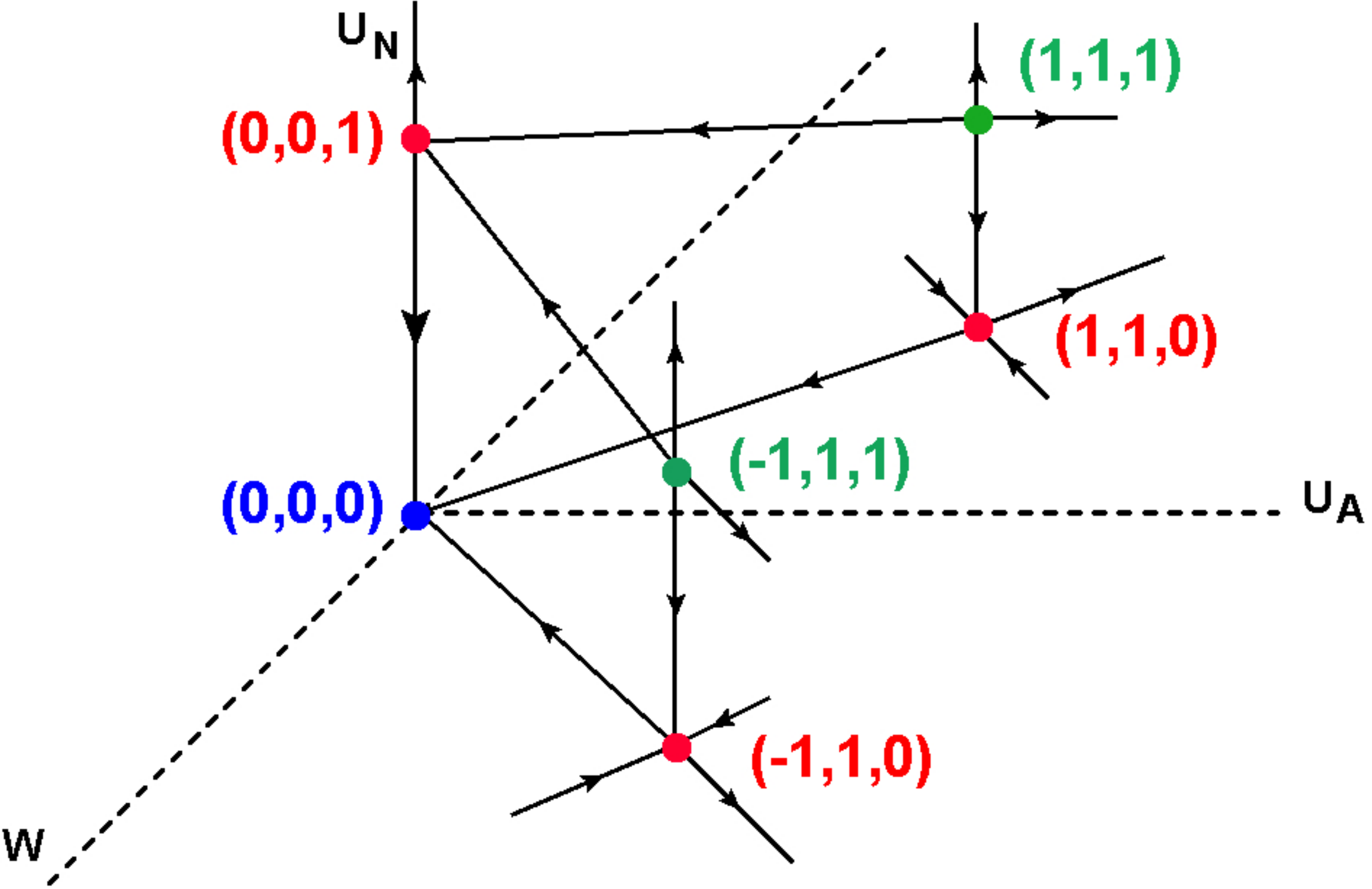}
\label{crit_point_1D}
}
\subfigure[]{
\includegraphics[width=4.0cm, height=4.25cm]{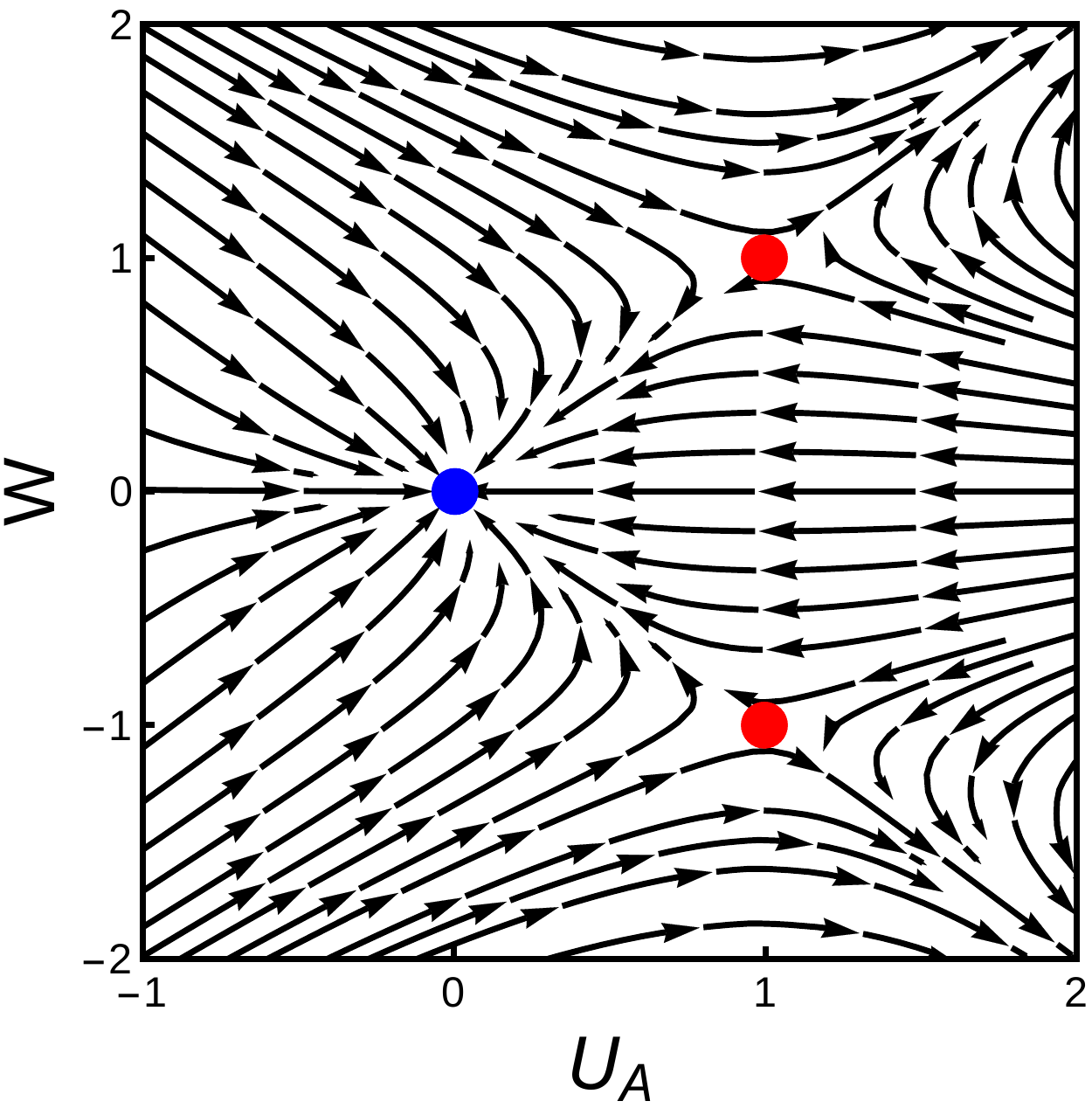}
\label{KT-lost}
}
\caption{ (a) A schematic representation of the renormalization group flow in three-dimensional coupling constant space $\left( W, U_A, U_N\right)$ for renormalized band parameter $\hat{\Delta}(l)=0$. The stable noninteracting anisotropic semimetal fixed point at $(0,0,0)$ is shown in blue, the quantum critical points at $(0,0,1)$ and $(\pm 1,1,0)$ are shown in red, while the remaining two bicritical points at $(\pm 1,1,1)$ are shown in green. The coupling constants are measured in units of $\epsilon$ [see also Table~\ref{fixedpoint1D}]. (b) Deformation of the flow diagram from Fig.~\ref{KT} in $\left( U_A, W \right)$ plane, upon incorporating the negative engineering dimension ($-\epsilon$) for the interaction couplings [see Eq.~(\ref{RG:1Dstrict})]. Recall that all the fixed points have one additional unstable direction along $\hat{\Delta}$, which controls the transition between a Dirac semimetal and band insulator.
}~\label{CFT-flowdiagram}
\end{figure}

\begin{table}
\begin{tabular}{|c|c|c|c|c|}
\hline
Fixed point & $\left( \lambda^s_1, \lambda^s_3, \lambda^t_2, \lambda^t_3 \right)$ & $\left( U_N, U_A, W, X \right)$ & $\hat{\Delta}$ & UDs \\
\hline \hline
FP$_1$ & $\left( 0,0,0,0 \right)$ & $\left( 0,0,0,0 \right)$ & $0$ & 0(1)\\
\hline
FP$_2$ & $\left( 0,\frac{3}{4},-\frac{1}{4}, 0 \right) \epsilon$ & $\left(1,0,0,0\right) \epsilon$ & $0$ & 1(2) \\
\hline
FP$_3$ & $\left( 0,\frac{1}{4},\frac{1}{4},0 \right) \epsilon$ & $\left(0,1,-1 ,0\right) \epsilon $ & $0$ & 1(2) \\
\hline 
FP$_4$ & $\left( 0,\frac{1}{4},-\frac{1}{12},\frac{1}{3} \right) \epsilon$ & $\left(0,1,1 ,0\right) \epsilon$ & $0$ & 1(2) \\
\hline
FP$_5$ & $\left( 0,1,0,0 \right)$ & $\left(1,1,-1 ,0\right) \epsilon$ & $0$ &  2(3)\\
\hline
FP$_6$ & $\left( 0,1,-\frac{1}{3}, \frac{1}{3} \right) \epsilon$ & $\left(1,1,1 ,0\right) \epsilon$ & $0$ &  2(3) \\
\hline \hline
\end{tabular}
\caption{Location of the fixed points obtained from the RG flow equations as we set $n \to \infty$ in Eq.~(\ref{RG-spinful}) and Eq.~(\ref{RG:1Dcoupling}), in the perturbative quadratic order corrections to the four-fermion coupling constants. The resulting RG flows are displayed in Eq.~(\ref{RG:1Dstrict}). The noninteracting semimetal is labeled FP$_1$ and has only one unstable direction; it controls the direct quantum phase transition between the Dirac semimetal and band insulator for sufficiently weak interactions. By contrast, each of FP$_{2,3,4}$ has two unstable directions and FP$_{5,6}$ has three unstable directions, and these correspond to interacting multicritical points. 
The penultimate rightmost column shows the value of the band parameter $\hat{\Delta}$ at each such fixed point. If we follow the RG trajectory along which the renormalized band parameter is kept fixed, i.e., $\hat{\Delta}=0$ then FP$_{2,3,4}$ represent interacting quantum critical points with only one unstable direction and FP$_{5,6}$ become bicritical points with two unstable directions. A schematic plot of these fixed points is shown in Fig.~\ref{CFT-flowdiagram}. The last column shows the number of unstable directions (UDs) at a given fixed point in the $\hat{\Delta}=0$ hyperplane as well as in the five dimensional coupling constant space (shown in parentheses).      
}~\label{fixedpoint1D}
\end{table}

\subsection{From 1D to $1 + \epsilon$}

Moving beyond the strict 1D limit, it is instructive to reinstate the dispersion in the $y$-direction in two steps. First, we consider the influence of the nonzero scaling dimension ($-\epsilon$) for all local four-fermion interactions in Eq.~(\ref{RG:1Dstrict}). 
With $\epsilon > 0$, these RG flow equations give birth to \emph{six} fixed points, tabulated in Table~\ref{fixedpoint1D} and schematically shown in Fig.~\ref{crit_point_1D}. All of them are located in a hyperplane defined by $X=0$ or $\lambda^s_1=0$. \emph{Note that inclusion of the nonzero scaling dimension does not destroy spin-charge separation}.

The gapless charge sector Luttinger liquid phase [stable fixed line in Fig.~\ref{KT}] gets replaced by a stable noninteracting fixed point and two QCPs, as shown in Fig.~\ref{KT-lost}. An additional quantum critical point appears in the spin sector at $U_N = \epsilon$.

Now we characterize all six fixed points in the hyperplane $\hat{\Delta}(l)=0$, cataloged in Table~\ref{fixedpoint1D}. The fixed point 
FP$_1$ represents the noninteracting ASM, stable against sufficiently weak generic short-range interactions. Each of FP$_2$, FP$_3$ and FP$_4$ is characterized by only one unstable direction (in addition to $\hat{\Delta}$). These three fixed points correspond to interacting QCPs and are analogous to the interacting 
multicritical point 
shown in Fig.~\ref{RG:spinless}. They describe continuous transitions from the ASM to 
broken symmetry phases.

To determine the actual nature of the 
broken symmetry
across the various QCPs in the ASM requires inclusion of $1/n$ corrections to the RG flow equations (\ref{RG-spinful}) or (\ref{RG:1Dcoupling}). These corrections are due to quantum fluctuations beyond one dimension and eliminate
certain special symmetries specific to 1D, such as spin-charge separation. In Sec.~\ref{susceptibility}, we will combine the RG with a scaling analysis of fermion bilinear susceptibilities in order to pin the pattern of symmetry breaking at strong coupling in an unbiased fashion.

The remaining two fixed points FP$_5$ and FP$_6$ each possess two unstable directions, as shown in Fig.~\ref{crit_point_1D}. In the $\hat{\Delta}(l)=0$ hyperplane they represent interacting \emph{bicritical points}, separating the basins of attraction for the various interacting QCPs. As shown in Fig.~\ref{crit_point_1D}, FP$_5$ separates the domain of attraction for FP$_2$ and FP$_3$, while FP$_6$ separates FP$_4$ and FP$_2$. The basins of attraction for FP$_3$ and FP$_4$ are separated by FP$_1$, which can be seen more transparently in Fig.~\ref{KT-lost}.

\begin{table*}
\begin{tabular}{|c|c|c|c|c|}
\hline
Fixed point & $\left( \lambda^s_1, \lambda^s_3, \lambda^t_2, \lambda^t_3 \right)$ & $\left( U_N, U_A, W, X \right)$ & $\hat{\Delta}$ & UDs \\
\hline \hline
FP$_1$ & $(0,0,0,0)$ & $(0,0,0,0)$ & $0$ & 0(1)\\
\hline 
FP$_2$ & $\left( 0,  \frac{3}{4}+\frac{0.25}{n} , -\frac{1}{4} - \frac{0.08334}{n} , \frac{0.15}{n} \right) \epsilon$ & $\left(1+\frac{0.18}{n}, \frac{0.46}{n}, \frac{0.44}{n}, 0 \right) \epsilon$ & $0$ & 1(2) \\
\hline
FP$_3$ & $\left( 0,  \frac{1}{4} + \frac{0.63}{n}, \frac{1}{4} + \frac{0.075}{n}, \frac{0.025}{n} \right) \epsilon$ & $\left(\frac{0.53}{n}, 1+\frac{0.93}{n}, -1-\frac{0.78}{n}, 0 \right) \epsilon$ & $\frac{\pi}{2 n} \epsilon$ & 1(2) \\
\hline 
FP$_4$ & $\left( 0, \frac{1}{4} -\frac{0.225}{n}, -\frac{1}{12} +\frac{0.0775}{n}, \frac{1}{3} -\frac{0.085}{n} \right) \epsilon$ & $\left(-\frac{0.2175}{n}, 1-\frac{0.2475}{n}, 1-\frac{0.2625}{n}, 0 \right) \epsilon$ & $ -\frac{\pi}{n} \epsilon$ & 1(2) \\
\hline
FP$_5$ & $\left(0,  1 -\frac{0.29}{n}, \frac{0.19}{n}, \frac{0.075}{n} \right) \epsilon$ & $\left(1-\frac{0.555}{n}, 1+\frac{0.505}{n}, -1-\frac{0.055}{n}, 0\right) \epsilon$ & $-\frac{\pi}{n} \epsilon$ & 2(3) \\
\hline 
FP$_6$ & $\left( 0,  1 +\frac{0.35}{n},  -\frac{1}{3} -\frac{0.13}{n}, \frac{1}{3} -\frac{0.5}{n} \right) \epsilon$ & $\left(1+\frac{0.98}{n}, 1-\frac{1.54}{n}, 1-\frac{1.46}{n}, 0 \right) \epsilon$ & $-\frac{3 \pi}{n} \epsilon$ & 2(3) \\
\hline
\end{tabular}
\caption{ 
Location of the six fixed points tabulated in Table~\ref{fixedpoint1D} after accounting for $1/n$ corrections in Eqs.~(\ref{RG-spinful}) and (\ref{RG:1Dcoupling}). The results obtain by utilizing the large-$n$ expansions for $f_1(n)$, $f_2(n)$ and $f_3(n)$ [Eq.~(\ref{three_functions})]. The inclusion of $1/n$ corrections does not change the number of fixed points (at least for large enough $n \geq 4$), but eliminates the spin-charge separation. Thus the $\epsilon$-expansion, augmented by $1/n$-expansion allows a controlled route to access 
some strong coupling phenomena (such as the quantum phase transition between the anisotropic semimetal and a broken symmetry phase) in two spatial dimensions. The coefficients of $1/n$ appearing in the locations of the fixed points are extracted numerically. The last column shows the number of unstable directions (UDs) at a given fixed point in the $\hat{\Delta}=0$ hyperplane as well as in the five dimensional coupling constant space (shown in parentheses).
}~\label{table-1overn}
\end{table*}

\subsection{RG analysis with $1/n$ corrections}

Upon incorporating $1/n$ corrections, the RG flow equations [see Eq.~(\ref{RG-spinful}) and Eq.~(\ref{RG:1Dcoupling})] still support six fixed points. Even though the number of fixed points is impervious to the choice of $n$ (as long as $n \geq 4$), their locations receive nontrivial corrections for finite $n$ from the ones reported in Table~\ref{fixedpoint1D} in the $n \to \infty$ limit. It is quite challenging to find the location of these fixed points analytically. Nevertheless, upon numerically locating the fixed points for various values of $n$, we can extract the functional dependence on $1/n$. At least for large $n$ (namely for $n \geq 8$) we find that the position of all the fixed points are well approximated functions of $1/n$, and they are tabulated in Table~\ref{table-1overn}. All fixed points are located in the hyperplane $\lambda^s_1=0$. Therefore, this coupling constant does not change any outcome qualitatively or quantitatively.

The location of the six fixed points in the four dimensional coupling constant space spanned by $\left( \lambda^s_1, \lambda^s_3, \lambda^t_2, \lambda^t_3 \right)$ or $\left( U_N, U_A, W, X\right)$ are presented in Table~\ref{table-1overn}. The nature of all the fixed points FP$_{j}$ for $j=1, \cdots, 6$ remains unchanged from that discussed in the previous section after $1/n$ corrections are included (see the last column of Table~\ref{table-1overn}). From the leading order in $\epsilon$ and $1/n$ expansion we can also allude to some emergent quantum critical phenomena at the interacting QCPs (namely FP$_{2,3,4}$), describing 
continuous 
transitions
from an ASM to various
broken symmetry phases.
For example, the correlation length exponent at the interacting QCPs is given by $\nu^{-1}=\epsilon$, and thus for $n=2$ (physical situation) $\nu=2$, 
similar to the situation discussed previously for spinless fermions [see Eq.~(\ref{nu_definition}) for definition of $\nu$]. The fact that correlation length exponent is same at all QCPs is, however, likely an artifact of the leading order calculation.

The residue of the quasiparticle pole of critical excitations (determined by fermionic anomalous dimension $\eta_\Psi$) is expected to vanish smoothly as these QCPs are approached from the ASM side of transition. (To one-loop order 
$\eta_\Psi=0$, so that a nontrivial fermionic anomalous dimension requires the computation of self-energy diagrams to two-loop order. We leave this exercise for future investigation.) On the other hand, if we subscribe to an appropriate order-parameter theory (also known as the Yukawa formalism), where the order-parameter or bosonic field is coupled with gapless fermions, a leading order calculation yields non-trivial anomalous dimensions for both fermionic and bosonic fields~\cite{ASM-Yuakawa}. The residue of the quasiparticle pole can serve as the 
order parameter
in the semimetallic phase. The interacting QCPs describe strongly coupled non-Fermi liquids where the notion of sharp quasiparticle excitations becomes moot.

\section{Susceptibility of source terms}~\label{susceptibility}

\begin{figure}
\includegraphics[width=8cm, height=3cm]{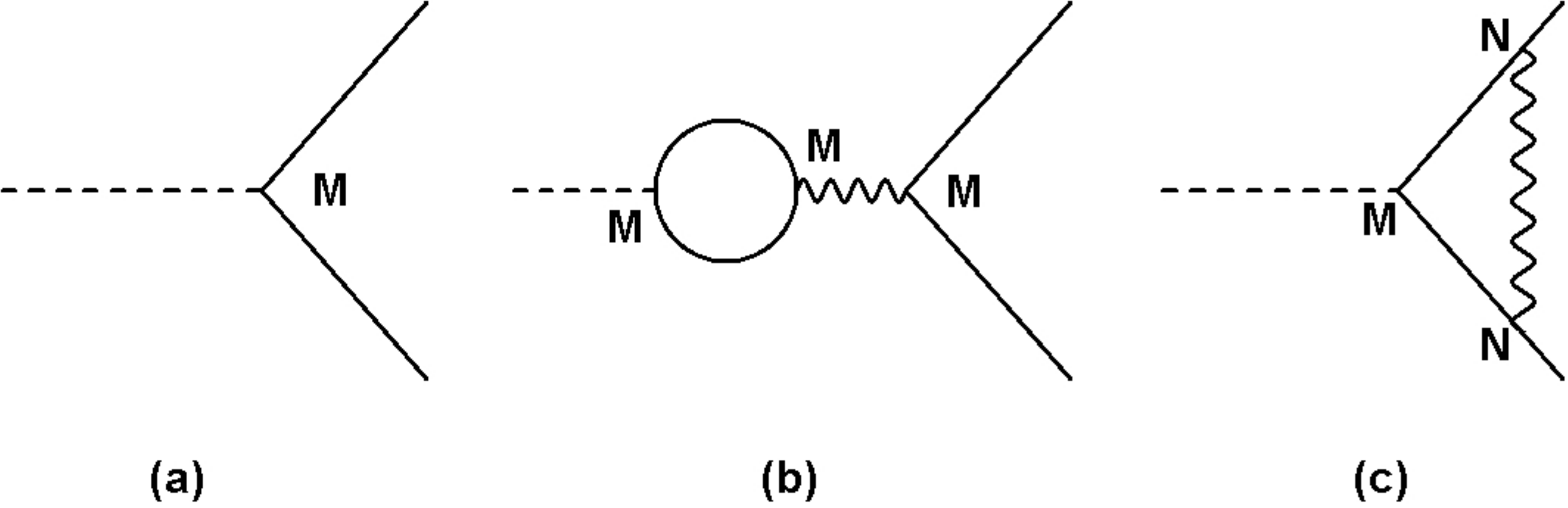}
\caption{(a) The bare vertex associated with a source term $\Psi^\dagger M \Psi$ (particle-hole channel) or $\Psi M \Psi$ (particle-particle channel). $M$ and $N$ are four-dimensional Hermitian matrices [see Eqs.~(\ref{source_Hamil}) and (\ref{source-pair})]. Renormalization of the bare vertex (a) to the leading order in interaction couplings arises from diagrams (b) and (c). Note there is no mixing of bilinear operators since all transform differently under the symmetries, see Table~\ref{order-parameters} and Appendix~\ref{symmetry-append}. 
}~\label{feynman-source}
\end{figure}

To pin the actual nature of the broken symmetry, 
we track the enhancement or suppression of two-point correlations amongst Hermitian fermion bilinears (order parameters). We do this by computing the flow equations for their corresponding source terms. The Hamiltonian including all of the source terms in the particle-hole or excitonic channel reads  
\begin{align}~\label{source_Hamil}
	H^{ph}_s = - \int d^2{\mathbf r} \, \sum_{\mu = 0}^3 \Psi^\dagger \left[ \Delta^s_{\mu} \sigma_0 \tau_\mu +	\Delta^t_{\mu} \vec{\sigma} \tau_\mu \right] \Psi. 
\end{align} 
The physical interpretation of all fermion bilinears was summarized in Table~\ref{order-parameters}. The leading order perturbative corrections to the source terms can be derived after computing the Feynman diagrams shown in Fig.~\ref{feynman-source}. The resulting RG flow equations are 
\begin{widetext}
\allowdisplaybreaks[4]
\begin{eqnarray}~\label{susceptibilities}
\frac{d\ln\Delta^s_{0}}{dl} &=& 1+ \frac{1}{4} \left( \lambda^s_{1} + \lambda^s_{3} + 3 \lambda^t_{2} + 3 \lambda^t_{3} \right)f_2(n) \equiv 1+ \frac{1}{4} \left(U_A + X \right) f_2(n), \nonumber \\ 
\frac{d\ln\Delta^s_{1}}{dl} &=& 1+ \frac{1}{4} \left( 3\lambda^s_{1} + \lambda^s_{3} + 3 \lambda^s_{2} + 3 \lambda^s_{3} \right)f_2(n) \equiv 1+ \frac{1}{4} \left(U_A - X \right) f_2(n), \nonumber \\
\frac{d\ln\Delta^s_{2}}{dl} &=& 1+ \frac{1}{2}\left( \lambda^s_{1} + \lambda^s_{3} - 3 \lambda^t_{2} + 3 \lambda^t_{3} \right) \left[ f_1(n) -\frac{f_2(n)}{2} \right] \equiv 1+ \frac{1}{4} \left( U_A+3 U_N +2 W \right) \left[ f_1(n) -\frac{f_2(n)}{2} \right] , \nonumber \\
\frac{d\ln\Delta^s_{3}}{dl} &=& 1+ \frac{1}{2}\left( \lambda^s_{1} + 3\lambda^s_{3} + 3 \lambda^t_{2} - 3 \lambda^t_{3} \right) \left[ f_1(n) +\frac{f_2(n)}{2} \right] \equiv 1+ \frac{1}{4} \left( U_A+3 U_N - 2 W  \right) \left[ f_1(n) +\frac{f_2(n)}{2} \right], \nonumber \\
\frac{d\ln\Delta^t_{0}}{dl} &=& 1+ \frac{1}{4}\left( \lambda^s_{1} + \lambda^s_{3} - \lambda^t_{2} - \lambda^t_{3} \right)f_2(n) \equiv 1+ \frac{1}{4} \left( U_N -X \right) f_2(n), \nonumber \\
\frac{d\ln\Delta^t_{1}}{dl} &=& 1+ \frac{1}{4}\left( -\lambda^s_{1} + \lambda^s_{3} - \lambda^t_{2} - \lambda^t_{3} \right)f_2(n) \equiv 1+ \frac{1}{4} \left( U_N + X \right) f_2(n), \nonumber \\
\frac{d\ln\Delta^t_{2}}{dl} &=& 1+ \frac{1}{2}\left(\lambda^s_{1} + \lambda^s_{3}+5 \lambda^t_{2} - \lambda^t_{3} \right)\left[ f_1(n) -\frac{f_2(n)}{2} \right] \equiv 1 + \frac{1}{4} \left( U_A-U_N-2W \right) \left[ f_1(n) -\frac{f_2(n)}{2} \right], \nonumber \\
\frac{d\ln\Delta^t_{3}}{dl} &=& 1+ \frac{1}{2}\left(\lambda^s_{1} - \lambda^s_{3}- \lambda^t_{2} +5 \lambda^t_{3} \right)\left[ f_1(n) -\frac{f_2(n)}{2} \right] \equiv 1+ \frac{1}{4} \left( U_A-U_N +2 W\right) \left[ f_1(n) +\frac{f_2(n)}{2} \right].  
\end{eqnarray}
\end{widetext}

\begin{table*}
\begin{tabular}{|c|c|c|c|c|c|c|}
\hline
Source 	& PS 	& Physical Meaning 	& Matrix representation & FP$_2$ 	&  FP$_3$ 	& FP$_4$ \\
term 	&	&			&			&		&		&	\\
\hline \hline
$\Delta^s_{0}$ &III	& Chemical potential or density & $\eta_3 \sigma_0 \tau_0$ & $1$ & $1+\frac{\epsilon}{4n}$ & $1+\frac{\epsilon}{4n}$  \\
\hline 
$\Delta^s_{1}$ &IV	& Abelian current in $x$ direction & $\eta_0 \sigma_0 \tau_1$ & $1$ & $1+\frac{\epsilon}{4n}$ & $1+\frac{\epsilon}{4n}$ \\
\hline 
$\Delta^s_{2}$ 	&	& Anisotropy parameter & $\eta_3 \sigma_0 \tau_2$ & {\color{blue}$1+\left[ \frac{3}{4}-\frac{0.275}{n} \right] \epsilon$} & $1+\left[ -\frac{1}{4}+\frac{0.5}{n} \right] \epsilon$ & {\color{blue}$1+\left[ \frac{3}{4}-\frac{0.85}{n} \right] \epsilon$}  \\
\hline 
$\Delta^s_{3}$ 	&I	& Charge density wave & $\eta_3 \sigma_0 \tau_3$& {\color{red}$1+\left[ \frac{3}{4}+\frac{0.025}{n} \right] \epsilon$} & {\color{red}$1+\left[ \frac{3}{4}+\frac{1.02}{n} \right] \epsilon$} & $1+\left[ -\frac{1}{4}-\frac{0.35}{n} \right] \epsilon $ \\
\hline \hline 
$\Delta^t_{0}$ && Magnetization & $\eta_0 \left(\sigma_1, \sigma_2, \sigma_3 \right) \tau_0$ & $1+\frac{\epsilon}{4n}$ & $1$ & $1$ \\
\hline 
$\Delta^t_{1}$ && Abelian spin current along $x$ & $\eta_3 \left(\sigma_1, \sigma_2, \sigma_3 \right) \tau_1$ & $1+\frac{\epsilon}{4n}$ & $1$ & $1$  \\
\hline 
$\Delta^t_{2}$ &II & Spin bond-density-wave & $\eta_0 \left(\sigma_1, \sigma_2, \sigma_3 \right) \tau_2$ & $1+\left[ -\frac{1}{4}+\frac{0.092}{n} \right] \epsilon$ & {\color{blue}$1+\left[ \frac{3}{4}-\frac{0.26}{n} \right] \epsilon$} & $1+\left[ -\frac{1}{4}+\frac{0.28}{n} \right] \epsilon$  \\
\hline 
$\Delta^t_{3}$ && Antiferromagnet or N\'{e}el & $\eta_0 \left(\sigma_1, \sigma_2, \sigma_3 \right) \tau_3$ & $1+\left[ -\frac{1}{4}+\frac{0.292}{n} \right] \epsilon$ & $1+\left[ -\frac{1}{4}-\frac{0.29}{n} \right] \epsilon$ & {\color{red}$1+\left[ \frac{3}{4}+\frac{0.286}{n} \right] \epsilon$}  \\
\hline \hline 
$\Delta_{s}$ &I & $s$-wave pairing & $\left( \eta_1, \eta_2 \right) \sigma_0 \tau_0$ & {\color{red}$1+ \left[ \frac{3}{4} + \frac{0.025}{n} \right] \epsilon$} & $1+ \left[-\frac{1}{4} + \frac{0.165}{n} \right] \epsilon$ & $1+ \left[-\frac{1}{4} - \frac{0.35}{n} \right] \epsilon$  \\
\hline 
$\Delta^{1}_{ch}$ &IV & Chiral singlet pairing$_1$ & $\left( \eta_1, \eta_2 \right) \sigma_0 \tau_2$ & $1$ & $1 -\frac{\epsilon}{4n}$ & $1 +\frac{\epsilon}{4n}$ \\
\hline 
$\Delta^{2}_{ch}$ &III & Chiral singlet pairing$_2$ & $\left( \eta_1, \eta_2 \right) \sigma_0 \tau_3$ & $1$ & $1 -\frac{\epsilon}{4n}$ & $1 +\frac{\epsilon}{4n}$  \\
\hline 
$\Delta_t$ &II & Triplet pairing & $\left( \eta_1, \eta_2 \right) \left(\sigma_1, \sigma_2, \sigma_3 \right) \tau_1$ & $1+ \left[-\frac{1}{4} +\frac{0.092}{n} \right] \epsilon$ & $1+ \left[-\frac{1}{4} -\frac{0.115}{n} \right] \epsilon$ & $1+ \left[-\frac{1}{4} +\frac{0.28}{n} \right] \epsilon$ \\
\hline \hline
\end{tabular}
\caption{The scaling dimensions (SDs) of the source term coupling constants conjugate to various fermionic bilinears at three interacting quantum critical points (QCPs); FP$_2$, FP$_3$ and FP$_4$, possessing only one unstable direction in the $\hat{\Delta}(l)=0$ hyperplane (see text for details). Note that at each QCP, we display the largest SD in red and subdominant SD in blue. As we take $n \to \infty$ (strict one-dimensional system) the red and blue channels possess exactly equal SDs and constitute an O(4) vector. However, such O(4) symmetry gets lifted once the $1/n$ corrections are systematically accounted for. Upon incorporating the $1/n$ corrections, the SD for one of the fully gapped (true mass) orders [charge-density-wave (CDW), anti-ferromagnet (AFM), $s$-wave pairing] for the two-dimensional anisotropic semimetal is largest at each QCP. We therefore anticipate that FP$_2$, FP$_3$ and FP$_4$ respectively nucleate CDW/$s$-wave pairing, CDW order, and N\'eel antiferromagnetism. 
Here all fermionic bilinears are presented in the Nambu-doubled spinor basis introduced in Eq.~(\ref{nambu-special}) [see also Appendix~\ref{symmetry-append}] and the Pauli matrices $\eta_\mu = \left\{ \eta_0, \eta_1, \eta_2, \eta_3 \right\}$ operate on the Nambu index. The column ``PS'' indicates bilinears that transform together under pseudospin SU(2) rotations; see Fig.~\ref{pseudospinrotation}. 
The other fermion bilinears transform as scalars under pseudospin rotation. 
 }~\label{anomalous-dim-table}
\end{table*}

We also track the RG flow of the source terms corresponding to various local (momentum independent or intra unit-cell) superconducting orders, presented in Table~\ref{order-parameters}. The Hamiltonian capturing all local pairings assumes the form
\begin{eqnarray}~\label{source-pair}
	H^{pp}_{s} &=& -\int d^2{\mathbf r} \bigg[ \Delta_s \Psi \sigma_2 \tau_3 \Psi  + \Delta^1_{ch}  \Psi \sigma_2 \tau_1 \Psi + \Delta^2_{ch} \Psi\sigma_2 \tau_0 \Psi  \nonumber \\
&+& \Delta_t \big[ \Psi \sigma_3 \tau_2 \Psi + \Psi \sigma_0 \tau_2 \Psi + \Psi \sigma_1 \tau_2 \Psi \big] \bigg] + \text{H.c.}. 
\end{eqnarray}
The leading RG flow equations for all the source terms associated with local pairings are given by 
\begin{widetext}
\allowdisplaybreaks[4]
\begin{eqnarray}~\label{susceptibilities:pairing}
\frac{d\ln\Delta_s}{dl} &=& 1-\frac{1}{2} \left( \lambda^s_{1} - \lambda^s_{3} + 3 \lambda^t_{2} + 3 \lambda^t_{3} \right) \left[f_1(n) + \frac{f_2(n)}{2}\right] \equiv 1 + \frac{1}{4} \left( -U_A +3 U_N \right) \left[f_1(n) + \frac{f_2(n)}{2}\right], \nonumber \\
\frac{d\ln\Delta^1_{ch}}{dl} &=& 1 + \frac{1}{4} \left( \lambda^s_{1}-\lambda^s_{3} -3 \lambda^t_{2} + 3 \lambda^t_{3} \right) f_2(n) \equiv 1+ \frac{1}{4} \left( W-X \right) f_2(n), \nonumber \\
\frac{d\ln\Delta^2_{ch}}{dl} &=& 1 -\frac{1}{4} \left( \lambda^s_{1} + \lambda^s_{3} + 3 \lambda^t_{2} -3 \lambda^t_{3} \right) f_2(n) \equiv 1+ \frac{1}{4} \left( W+ X\right) f_2(n), \nonumber \\
\frac{d\ln\Delta_t}{dl}&=& 1-\frac{1}{2} \left( \lambda^s_{1} + \lambda^s_{3} + \lambda^t_{2} + \lambda^t_{3} \right) \left[f_1(n) - \frac{f_2(n)}{2}\right]
\equiv 1- \frac{1}{4} \left( U_A + U_N \right) \left[f_1(n) - \frac{f_2(n)}{2}\right]. 
\end{eqnarray}
\end{widetext}

The quantities on the right hand side of Eqs.~(\ref{susceptibilities}) and (\ref{susceptibilities:pairing}) are the \emph{scaling dimensions} of the corresponding source terms. 
The scaling dimensions associated to all fermionic bilinear source terms at the three interacting QCPs (namely FP$_2$, FP$_3$ and FP$_4$) are displayed in Table~\ref{anomalous-dim-table}. If the source term has scaling dimension $\ys$, the anomalous dimension of the associated fermion bilinear $\yb$ is given by 
\[
\yb = 2 + 1/n - \ys, 
\]
using the fact that the operator and coupling constant scaling dimensions add to the scaling dimension of spacetime \cite{Cardy}. The latter is equal to $2 + 1/n$ according to Sec.~\ref{sec:HamLagAct}. Since $\yb$ controls the decay of correlations (in the $x$-direction) of the bilinear two-point function at the quantum critical point \cite{Cardy}, a larger source dimension $\ys$ means stronger correlations (smaller $\yb$, slower decay).

\subsection{Scaling dimensions of fermion bilinears: spin and pseudospin symmetries}~\label{anomalous-dim}

In Table~\ref{anomalous-dim-table}, the bilinears previously cataloged in Table~\ref{order-parameters} are identified with $8 \times 8$ matrices, obtained by combining physical spin ($\sigma$) and sublattice/orbital ($\tau$) with particle-hole ($\eta$) degrees of freedom. 
The matrix representations of the various fermionic bilinears in Table~\ref{anomalous-dim-table} are obtained in the eight-component Nambu doubled spinor basis  
\begin{eqnarray}~\label{nambu-special}
\Psi^\top_N &=& \bigg[ c^\dagger_{A,{\mathbf k}, \uparrow}, c^\dagger_{B,{\mathbf k}, \uparrow}, c^\dagger_{A,{\mathbf k}, \downarrow}, c^\dagger_{B,{\mathbf k}, \downarrow}, \nonumber \\
&& c_{A,-{\mathbf k}, \downarrow}, -c_{B,-{\mathbf k}, \downarrow}, -c_{A,-{\mathbf k}, \uparrow}, c_{B,-{\mathbf k}, \uparrow} \bigg].
\end{eqnarray}
In this basis the three generators of physical spin 
($\vec{S}$) 
and 
pseudospin ($\vec{PS}$)~\cite{SCZhang90,Auerbach94,Hermele07, roy-own-nambu} symmetry are given by 
\begin{equation}~\label{generators}
	\vec{S}= \eta_0 \left( \sigma_1, \sigma_2, \sigma_3 \right) \tau_0, \;
	\vec{PS}= \left(\eta_1 \sigma_0 \tau_3, \eta_2 \sigma_0 \tau_3, \eta_3 \sigma_0 \tau_0 \right).   
\end{equation}
Note that the pseudospin generators do not involve genuine spin degrees of freedom, and one of its generators, namely $PS_3$, is the charge density operator. The remaining two generators of pseudospin symmetry are the real and imaginary components of the 
chiral singlet pairing$_2$. See Appendix~\ref{symmetry-append} for symmetries in the Nambu basis.

We now discuss the results. First, we consider the 1D ($n \rightarrow \infty$) limit, as a check. Then Eq.~(\ref{three_functions}) implies that we should set $f_1(n) = 1$ and $f_2(n) = 0$ in Eqs.~(\ref{susceptibilities}) and (\ref{susceptibilities:pairing}). We see that the 
charge bond density wave/anisotropy, CDW, spin bond density wave, and antiferromagnet source fields $\Delta_{2,3}^{s,t}$ receive an anomalous correction $+U_A/4$, while the singlet and triplet SC terms $\Delta_{s,t}$ receive the anomalous correction $-U_A/4$, when $U_N = W = X = 0$. The remaining terms receive no corrections in this limit. This is consistent with the predictions of bosonization, summarized 
in Table~\ref{order-parameters-1D} and using the Luttinger parameter in Eq.~(\ref{vcKc}).

\begin{figure}
\includegraphics[width=7cm, height=7cm]{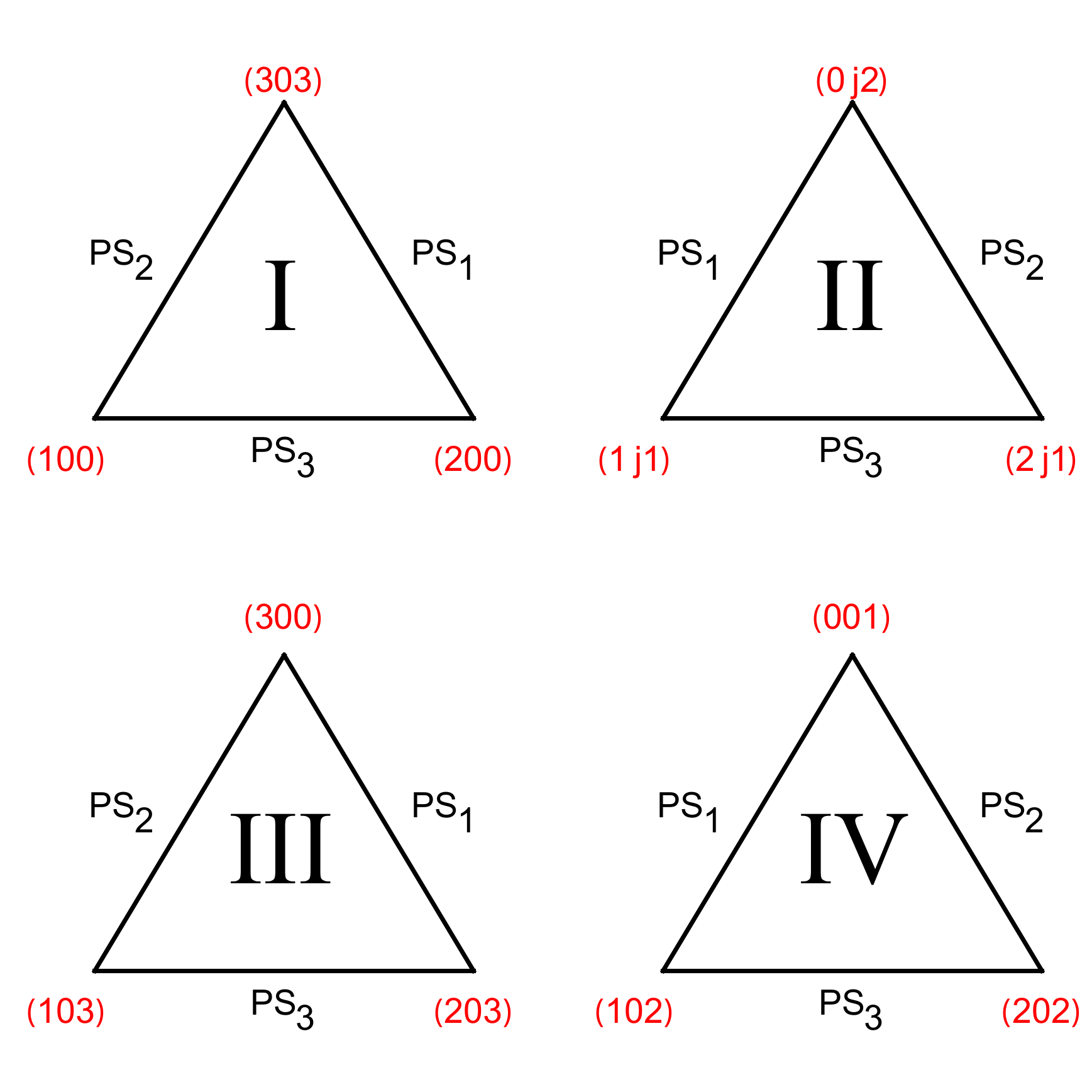}
\caption{Four sets of three mutually anticommuting fermionic bilinears, denoted by $(ikl) \equiv \eta_i \sigma_k \tau_l$, sitting at the vertices of four triangles [see Table~\ref{anomalous-dim-table}]. Each triangle demonstrates the SU(2) pseudospin rotation among 
(I) the charge density wave and two (real and imaginary) components of $s$-wave pairing, 
(II) the spin bond-density-wave and two components (real and imaginary) of triplet pairing with a specific spin orientation ($j=1,2,3$), 
(III) the charge density and the components (real and imaginary) of chiral singlet pairing$_2$, and 
(IV) the $x$-charge current and the components (real and imaginary) of chiral singlet pairing$_1$. 
The pseudospin generators (PS$_j$ with $j=1,2,3$) [Eq.~(\ref{generators})] correspond to (III). Any arm of a triangle represents the rotation between two bilinears, sitting at two corresponding vertices, by a specific generator. 
}~\label{pseudospinrotation}
\end{figure}

Next, note that at FP$_2$, the 
scaling dimensions
for CDW and $s$-wave pairing are \emph{largest} and exactly equal, reflecting the underlying pseudospin SU(2) symmetry among these two orders, see Fig.~\ref{pseudospinrotation}(I). Since the bipartite lattice Hubbard Hamiltonian possesses pseudospin SU(2) symmetry at half filling~\cite{SCZhang90, Auerbach94,Hermele07}, we expect that FP$_2$ can be accessed by tuning the strength of onsite attraction (see Sec.~\ref{extended-hubbard}). On the other hand, the 
scaling dimension
for CDW order is largest amongst all possible orders at the interacting QCP FP$_3$. Hence, it is natural to anticipate that repulsive NN interaction in the strained honeycomb model can induce the transition across this fixed point. By contrast, the AFM order possesses the largest 
scaling dimension
at the interacting QCP FP$_4$, indicating that the onset of AFM for strong repulsive Hubbard interaction is controlled by this fixed point. In the following section we will substantiate these observations from the phase diagram of an extended Hubbard model. The fact that a two-dimensional ASM supports only \emph{six} mass matrices representing CDW (1), AFM (3) and $s$-wave pairing (2), where the quantities in the parentheses denotes the number of matrices required to define a specific 
order parameter
[see Table~\ref{anomalous-dim-table}], can also be established from the Clifford algebra of real matrices, as shown in Appendix~\ref{cliffordalgebra-Append}.

The pseudospin symmetry between $s$-wave pairing and CDW order at FP$_2$ extends to three other bilinear ``triads''. Each triad contains
three bilinears that transform in the pseudospin triplet representation, as explained in Fig.~\ref{pseudospinrotation}. At FP$_2$ and FP$_4$, the bilinears within each triad share a common scaling dimension, indicating that pseudospin SU(2) symmetry is \emph{emergent} at both of these QCPs, see Figs.~\ref{pseudospinrotation}(II), (III), and (IV). This suggests that FP$_{2,4}$ are realized via attractive or
repulsive onsite Hubbard interactions in the strained honeycomb lattice model, as we confirm in Sec.~\ref{extended-hubbard}. By contrast, pseudospin symmetry is explicitly broken at FP$_{3}$. The perturbative restoration of pseudospin symmetry at interacting QCPs that can be accessed by tuning the Hubbard interaction serves as a good anchoring ground of the analysis~\cite{roy-multicriticality}.

\subsection{Identification of broken-symmetry phases \label{sec:BSP-find}}

The strategy for identifying the actual nature of the
broken symmetry
at strong coupling is the following. We simultaneously run the flow of the interaction coupling constants and of the source terms. When interactions are weak all of them flow back to zero under course graining, 
and none of the source terms diverges, indicating the stability of the ASM for weak enough interactions. 

\begin{figure*}
\subfigure[]{
\includegraphics[width=5.5cm, height=5.25cm]{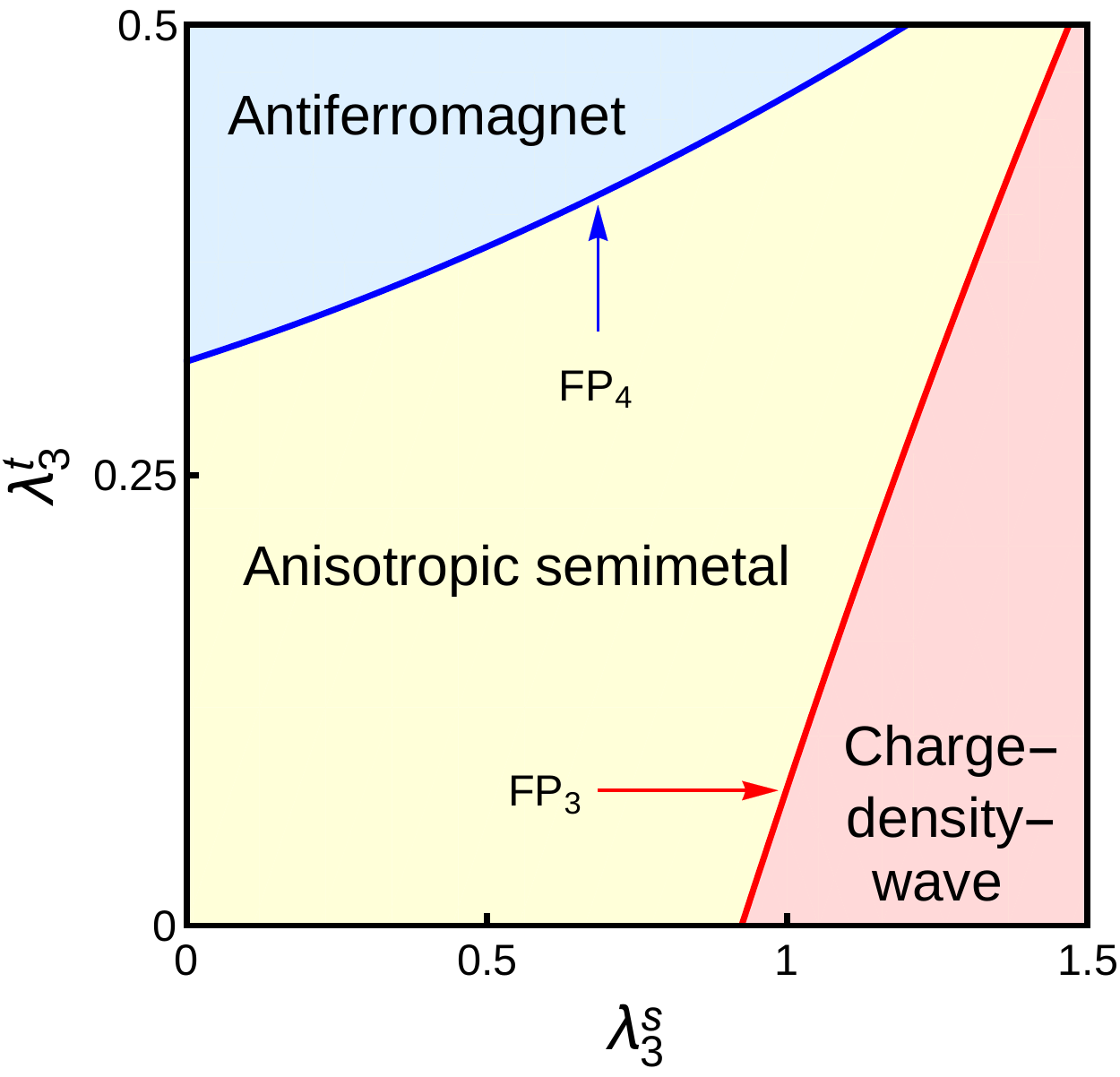}
}
\subfigure[]{
\includegraphics[width=5.5cm, height=5.25cm]{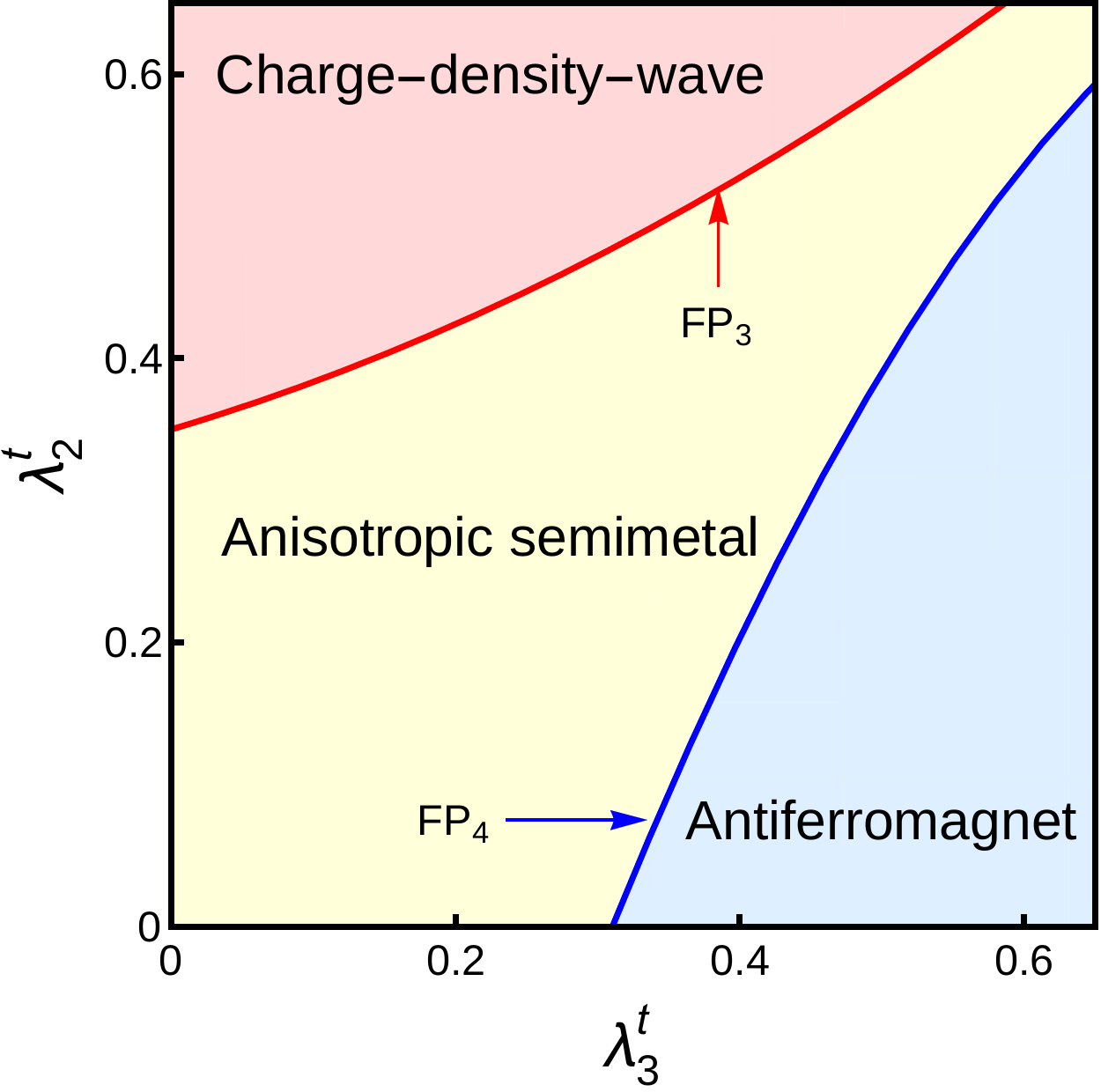}
}
\subfigure[]{
\includegraphics[width=5.5cm, height=5.25cm]{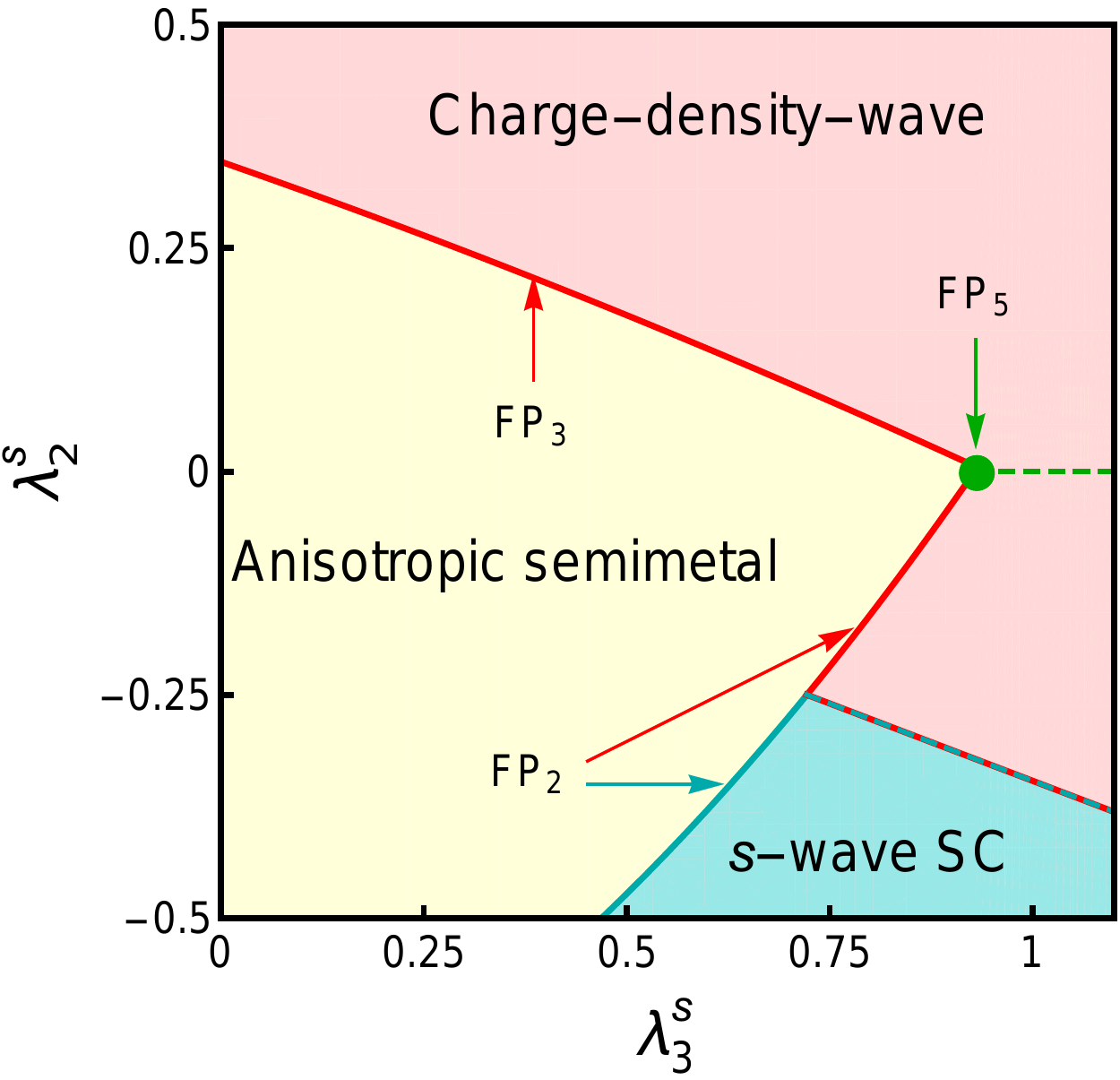}
}
\caption{ Various cuts of the phase diagram for the spin-1/2 interacting anisotropic semimetal (ASM), residing at the quantum critical point (QCP) between the two-dimensional Dirac semimetal and the band insulator. These phase diagrams are obtained for $n=10$, which parametrically keeps the quantum fluctuations beyond one dimension small. All coupling constants are dimensionless and measured in units of $\epsilon$. Various fixed points (FP$_j$s, see Table~\ref{table-1overn}) controlling the continuous quantum phase transition from the ASM to different broken-symmetry phases are displayed in each cut of the phase diagram. Note that FP$_{2,3,4}$ are interacting QCPs, while FP$_5$ in an interacting bicritical point [appearing at the point of inflection in the panel (c), the green dot]. Along the red/cyan dashed line in (c) 
charge density wave (CDW) and $s$-wave superconductor (SC) simultaneously display long range order, due to the pseudospin SU(2) symmetry. Below (above) this line only $s$-wave SC (CDW) order is realized as true long-ranged order. The green dashed line is a crossover boundary, across which the short-range correlations of $s$-wave pairing vanish smoothly from bottom to top. While extracting the phase diagram in each two-dimensional coupling constant space, we set the bare value of the other two couplings to be zero.    
}~\label{Phasediagram-finiten}
\end{figure*}

As we keep increasing the strength of interaction, beyond a critical strength at least one of the interaction coupling constants diverges. At the same time at least one of the source term diverges. The channel (say $\Delta_\mu$) that diverges fastest determines the pattern of spontaneous symmetry breaking and the 
resulting phase
is characterized with the order parameter $\Delta_\mu \neq 0$. Following this strategy we present various cuts of the phase diagram of the interacting ASM in Fig.~\ref{Phasediagram-finiten}. We follow the same strategy to arrive at the phase diagram of an extended Hubbard model, discussed in the next section.

\subsection{1D physics in 2D?}

The $1/n$ corrections to the interaction RG flow in Eq.~(\ref{RG:1Dcoupling}) proportional to $f_2(n)$ mix the spin ($U_N$) and charge ($U_A$, $W$, $X$) sector interaction coupling constants. This suggests that spin-charge separation is completely destroyed for any finite $n$; the one-dimensional physics does not survive once quantum fluctuations in 2D are incorporated. On the other hand, the pseudospin-SU(2)-symmetric fixed points FP$_{2}$ and FP$_{4}$ are reminiscent of the 1D spin-charge separated Luther-Emery liquid and SDW phases, respectively. Both the 1D phases and 2D fixed points sit on the precipice of long-range CDW/$s$-wave pairing or antiferromagnetic order, and both are expected to lack well-defined quasiparticles. It would therefore be extremely interesting to look for remnants of 1D physics at these fixed points upon incorporating higher order corrections (two loops and beyond).

\section{Extended Hubbard model}~\label{extended-hubbard}

We now focus on the phase diagram of the extended Hubbard model in 1D and for the ASM in two spatial dimensions. We discuss these two cases separately.  

\subsection{One dimension}~\label{Hubbard:1D}

The extended Hubbard interaction in 1D is 
\begin{eqnarray}~\label{hubbard:1D}
	H^{1D}_{UV} &=& U \sum_{j} c^\dagger_{j, \uparrow} c_{j, \uparrow} c^\dagger_{j, \downarrow} c_{j, \downarrow} \nonumber \\
	&+& V \sum_{j,\sigma=\uparrow, \downarrow} c^\dagger_{j, \sigma} c_{j, \sigma} c^\dagger_{j+1, \sigma} c_{j+1, \sigma}. 
\end{eqnarray}
Here $j$ is the 1D lattice site index. Since the single-band 1D system supports two Dirac points at momentum $k_x=\pm k_F$, we can expand the lattice fermion operator in terms of the low energy Fourier modes according to 
\begin{eqnarray}
	c_{\sigma}= e^{i k_F x} R_\sigma + e^{-i k_F x} L_\sigma, 
\end{eqnarray} 
where $R_\sigma$ and $L_\sigma$ are right- and left-mover operators for modes at $\pm k_F$, with spin projection $\sigma=\uparrow, \downarrow$. At half-filling the extended Hubbard interaction from Eq.~(\ref{hubbard:1D}) leads to the continuum form
\begin{align}
	H^{1D}_{UV} &=& \int dx \, 
	\left[ \begin{aligned}
	&\, U_N \, \mathcal{O}_N +	U_A \, \mathcal{O}_A \\&\,
	+	\frac{W}{2}\left(\mathcal{O}_W + \bar{\mathcal{O}}_W\right) +	X \, \mathcal{O}_X
	\end{aligned} \right],
\end{align} 
where the interaction operators are defined in Eq.~(\ref{1DOps--Def}). The initial or bare conditions (values of coupling constants at RG time $l=0$) for the 1D extended Hubbard model are
\begin{align}
\begin{gathered}
	U_N =	V - \ts{\frac{1}{2}} U, \quad U_A = \ts{\frac{1}{2}} U + 3 V,  \\
	W =	\ts{\frac{1}{2}} U-V, \quad	X = \ts{\frac{1}{2}} U+V. 
\end{gathered}
\end{align} 
The standard 1D RG flow equations in Eq.~(\ref{RG:1Dstrict}) (with $\epsilon=0$) then determine the phase diagram of the one-dimensional extended Hubbard model, displayed in Fig.~\ref{UV_PD_1D}~\cite{giamarchi}. The diagonal phase boundaries in Fig.~\ref{UV_PD_1D} are determined by $U=\pm 2 V$, based on the lowest order Kosterlitz-Thouless equations [Eq.~(\ref{RG:1Dstrict}) with $\epsilon=0$].

\subsection{Two-dimensional ASM}

\begin{figure*}[t!]
\subfigure[]{
\includegraphics[width=4cm,height=3.5cm]{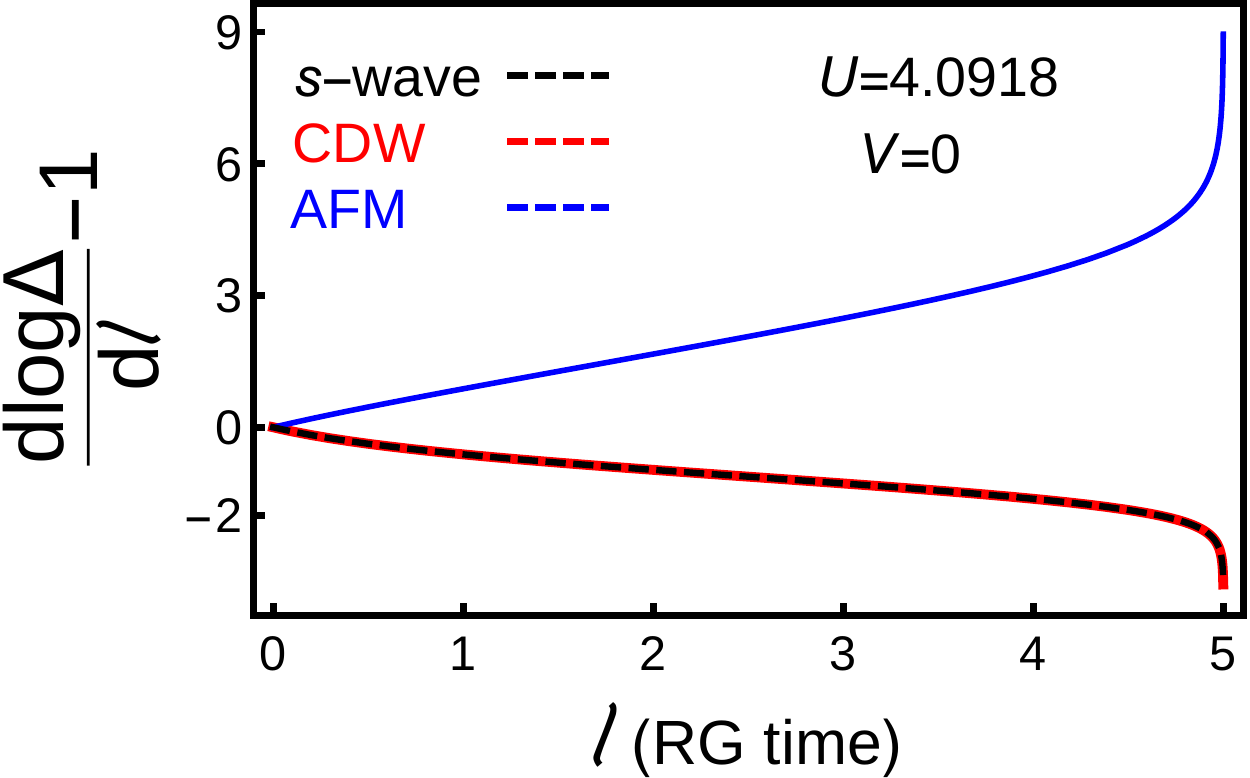}
~\label{repulsive-U}
}
\subfigure[]{
\includegraphics[width=4cm,height=3.5cm]{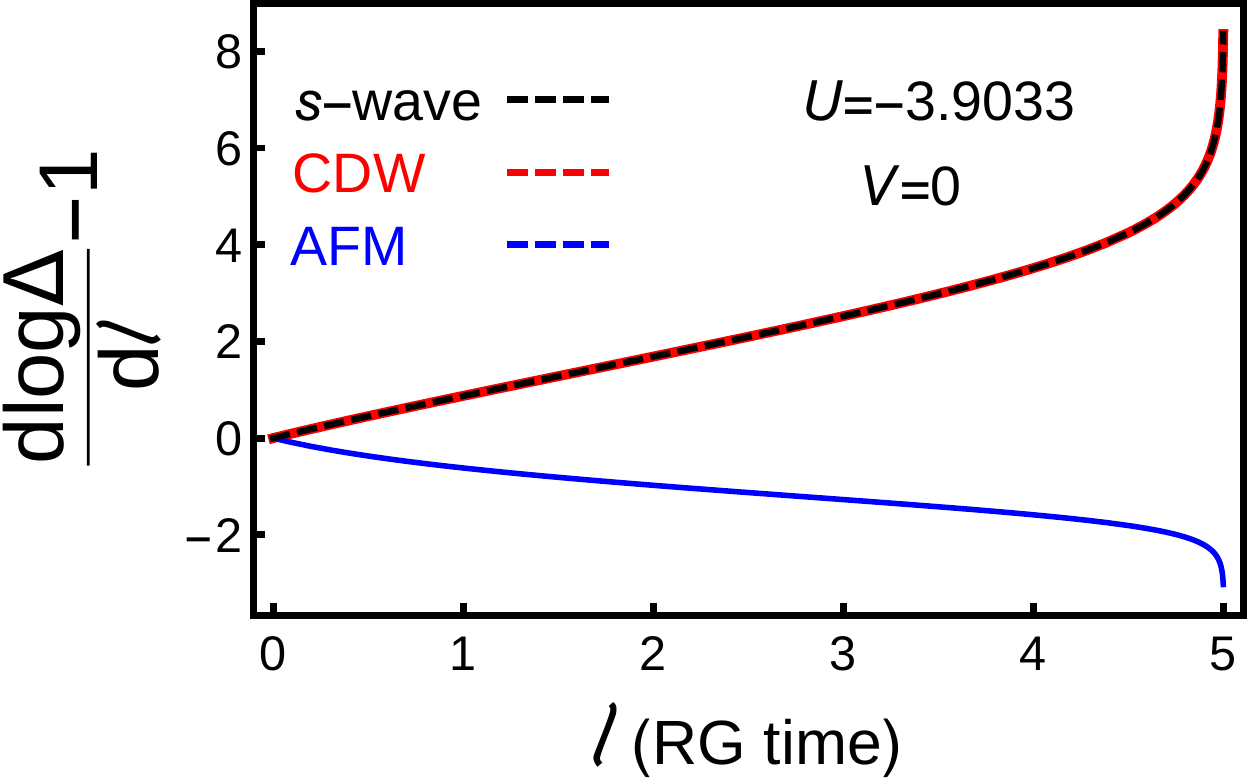}
\label{attractive-U}
}
\subfigure[]{
\includegraphics[width=4cm,height=3.5cm]{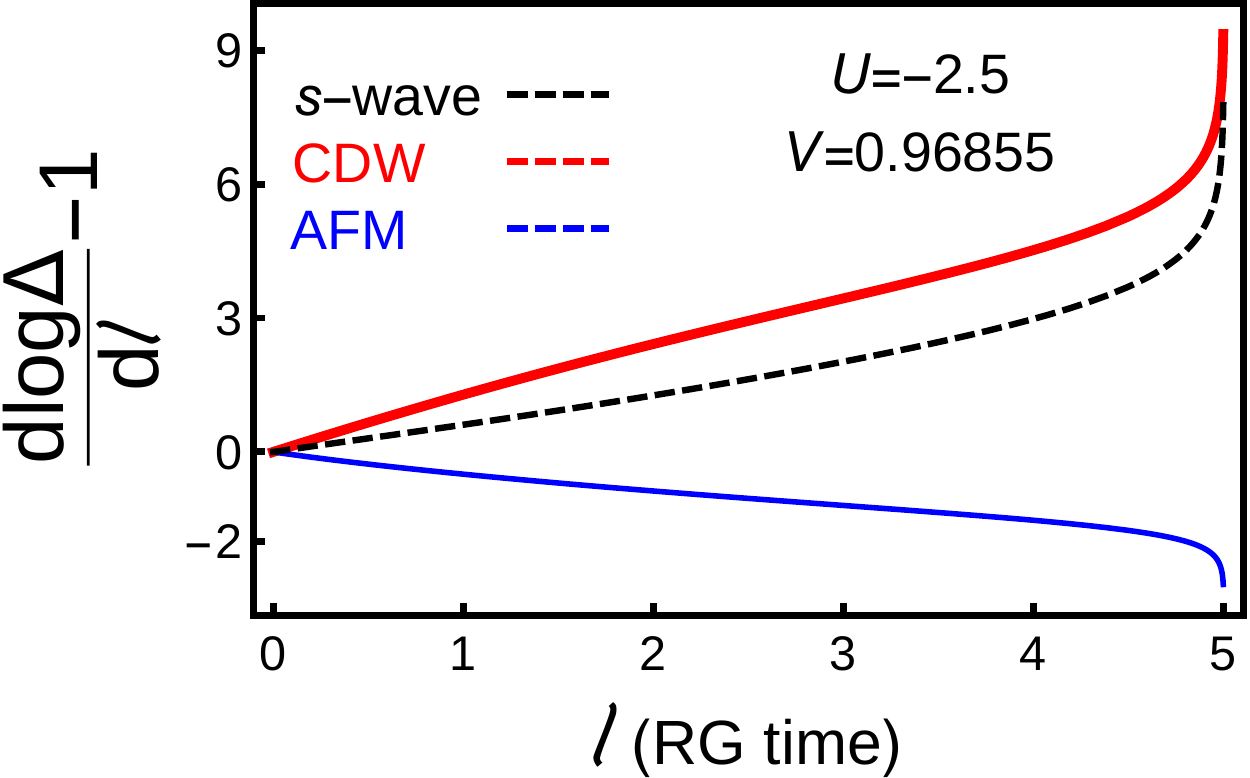}
\label{repulsive-V}
}
\subfigure[]{
\includegraphics[width=4cm,height=3.5cm]{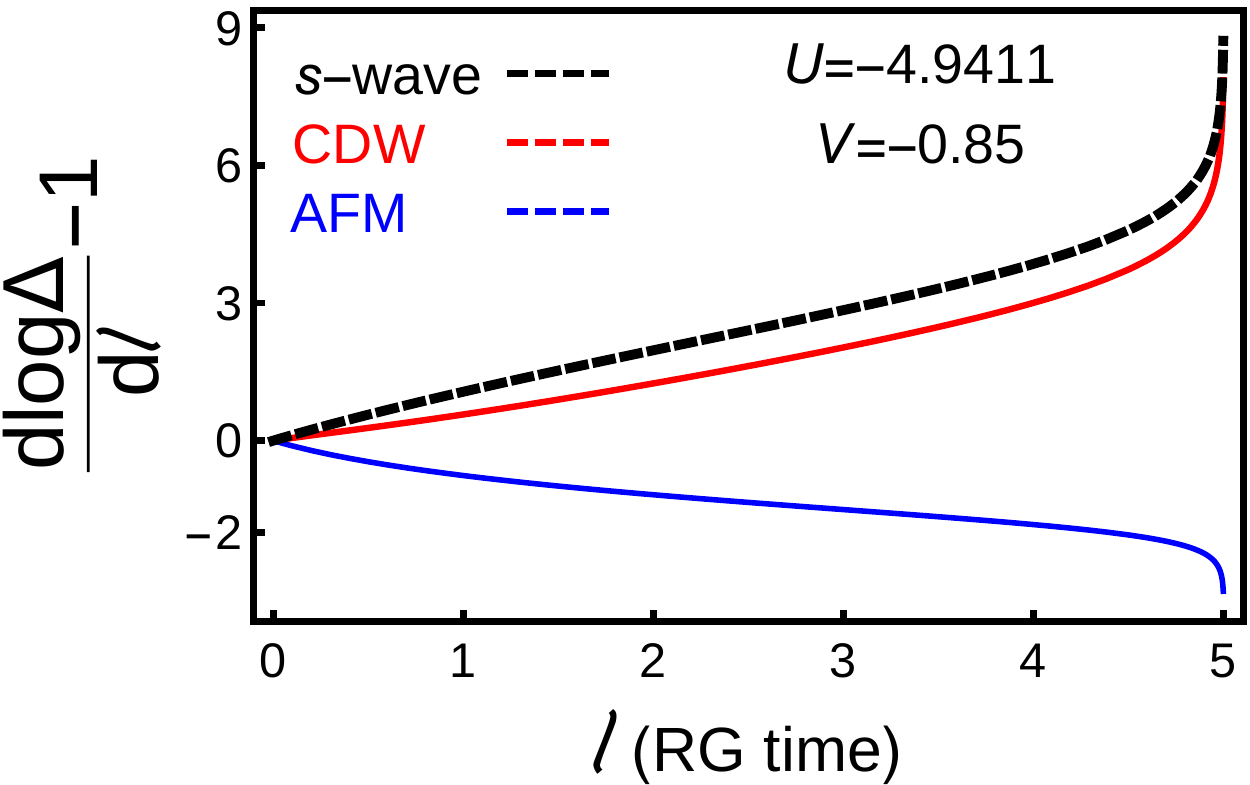}
\label{attractive-V}
}
\caption{Renormalization group flow of the source terms in the antiferromagnet (blue), charge density wave (red), and singlet $s$-wave pairing (black) channels for various choices for the bare strength (at $l=0$) of onsite ($U$) and nearest-neighbor ($V$) interactions in a uniaxially strained honeycomb lattice. The strength of both of the interactions are quoted in dimensionless units. Onset of an order-parameter is associated with the divergence ($\to + \infty$) of the corresponding source term. The flow of the source terms shows that strong repulsive $U > 0$ causes formation of antiferromagnetic order [see panel (a)]. By contrast, for a strong attractive $U < 0$ diverging susceptibilities in the charge density wave (CDW) and $s$-wave pairing channels are exactly degenerate, as guaranteed by the pseudospin SU(2) symmetry of the system with $V = 0$ [see Sec.~\ref{anomalous-dim}]. These two orders nucleate simultaneously in the strong coupling regime [see panel (b)]. The addition of the nearest-neighbor
  interaction $V$ breaks such exact degeneracy. In particular, $V > 0$ ($V < 0$) favors the nucleation of the CDW ($s$-wave pairing), as shown in panel (c) [(d)]. The resulting phase diagram in the $U$-$V$ plane is shown in Fig.~\ref{UV_PD}. For the clarity of presentation we here do not present the RG flow of the remaining channels. We have verified that in the entire phase diagram only one of these three channels displays the leading divergence at strong coupling. 
  }~\label{UV:susceptibility}
\end{figure*}

Since the 
Dirac-semimetal to band-insulator quantum phase transition
takes place when two Dirac points with opposite vorticity merge at a high symmetry point in the Brillouin zone, the ASM in two dimensions lacks any valley (chiral) degeneracy. Thus, we can simply identify the sublattice density operators with 
\begin{eqnarray}
	n_{\uparrow/\downarrow}(\vec{A}) &=& \frac{1}{4} \; \Psi^\dagger \left( \sigma_0 \pm \sigma_3 \right) \left( \tau_0 + \tau_3 \right) \Psi, \nonumber \\
	n_{\uparrow/\downarrow}(\vec{B}) &=& \frac{1}{4} \; \Psi^\dagger \left( \sigma_0 \pm \sigma_3 \right) \left( \tau_0 - \tau_3 \right) \Psi. 
\end{eqnarray}
Taking the continuum limit of the extended Hubbard Hamiltonian from Eq.~(\ref{hubbard}), we obtain the interaction Hamiltonian  $H_{int}$ in the form of Eq.~(\ref{Hint_spinful}). We thereby determine the initial conditions for the extended Hubbard model in the two-dimensional ASM,  
\begin{equation}
\begin{gathered}
	g^s_{1} =	- \ts{\frac{3}{8}} V - \ts{\frac{1}{4}}	U, \quad 
	g^s_{3}	=	\ts{\frac{3}{4}} V, \\ 
	g^t_{2} =	\ts{\frac{1}{8}} V + \ts{\frac{1}{12}} U, \quad 
	g^t_{3}	=	\ts{\frac{1}{4}} V + \ts{\frac{1}{6}}	U.
\end{gathered} 
\end{equation}				
In this notation $X>0$ corresponds to repulsive interaction, while attractive interaction is realized for $X<0$, where $X=U, V$.

We follow the strategy described in Sec.~\ref{sec:BSP-find} for various values of $U$ and $V$ (both attractive and repulsive interactions) to arrive at the phase diagram of the interacting ASM in the $U$-$V$ plane. Our result is shown in Fig.~\ref{UV_PD}. For strong enough onsite repulsion $U > 0$, the susceptibility in the AFM channel displays the leading divergence, as shown in Fig.~\ref{repulsive-U}, indicating the onset of collinear antiferromagnetic (N\'{e}el) order on the strained honeycomb lattice. On the other hand, for sufficiently strong attractive $U < 0$, the $s$-wave and CDW channels diverge in an exactly degenerate fashion, as shown in Fig.~\ref{attractive-U}, 
suggesting simultaneous nucleation of these two orders. Thus the onset of 
long-range order
for strong repulsive and attractive Hubbard interaction takes place through spontaneous breaking of SU(2) spin and pseudospin symmetries, respectively. However, the exact (microscopic) pseudospin SU(2) symmetry gets destroyed for any $V \neq 0$. In particular, for $V > 0$ ($V < 0$) only the susceptibility of the CDW ($s$-wave pairing) diverges, as shown in Fig.~\ref{repulsive-V} [Fig.~\ref{attractive-V}].

Finally, we discuss the role of various interacting QCPs, discussed in Sec.~\ref{spinful-RG}, on the phase diagram of the extended Hubbard model. The fixed point controlling the 
transition into the broken symmetry phase
can be identified by scrutinizing the RG flow trajectories of the interaction coupling constants. The continuous 
transition
between the ASM and AFM for repulsive $U > 0$ is controlled by the interacting QCP FP$_4$, while the 
transition
into a coexisting phase of $s$-wave pairing and CDW order ($U < 0$, $V = 0$) takes place through the interacting QCP FP$_2$. The ASM-CDW continuous 
transition
driven by NN repulsion is governed by the interacting QCP FP$_3$. Notice that for $V > 0$ as we keep increasing the strength of attractive onsite interaction, such that the 
broken symmetry phase
at strong coupling is always the CDW, the fixed point that controls the ASM-CDW continuous 
transition 
switches from FP$_3$ to FP$_2$ for an intermediate strength of negative Hubbard-$U$. At this point ASM-CDW phase boundary displays a \emph{cusp}, as shown
  with a \emph{dot} in Fig.~\ref{UV_PD}, and the ASM-CDW 
transition
at this point is controlled by the bicritical point FP$_5$. Since such a bicritical point is accessed by holding one of its two unstable directions fixed, 
the ASM-CDW 
transition 
is always continuous.

In the phase diagram of the extended Hubbard model for the two-dimensional ASM [see Fig.~\ref{UV_PD}] we show a green dashed line for repulsive NN but attractive onsite interaction. We now discuss the physical significance of this line. The scaling dimensions for $s$-wave pairing and CDW order source terms at the interacting fixed points FP$_2$, FP$_5$, FP$_3$ respectively read as
\allowdisplaybreaks[4]
\begin{align}
	\mathsf{y}_{{\stackrel[\text{SC}]{}{\text{$s$-wave}}} } =&\, 1 + 
	\left\{ 
	\stackrel[\ts{\text{FP}_2}]{}{\frac{3}{4} + \frac{0.025}{n}}, 
	\stackrel[\ts{\text{FP}_5}]{}{\frac{1}{2} - \frac{0.5245}{n}}, 
	\stackrel[\ts{\text{FP}_3}]{}{-\frac{1}{4} + \frac{0.165}{n}} 
	\right\} \epsilon 
\nonumber\\
	\mathsf{y}_{\text{CDW}}	=&\, 1 + 
	\left\{ 
	\stackrel[\ts{\text{FP}_2}]{}{\frac{3}{4} + \frac{0.025}{n}}, 
	\stackrel[\ts{\text{FP}_5}]{}{\frac{3}{2} - \frac{0.2625}{n}}, 
	\stackrel[\ts{\text{FP}_3}]{}{\frac{3}{4}+\frac{1.02}{n} }  
	\right\} \epsilon. 
\end{align}
While the
scaling dimensions
for these two orders are exactly equal at FP$_2$, at the two other fixed points the 
scaling dimension
for the CDW is always bigger than that for $s$-wave SC. 
Consequently, we determine that CDW order 
is favored
for strong attractive $U < 0$ and repulsive $V > 0$. Also note that, starting from $V = 0$ and attractive $U < 0$, if we increase $V$ along the ASM-CDW phase boundary [see the red line in Fig.~\ref{UV_PD}], the transition is first controlled by FP$_2$ (for weak $V > 0$), then FP$_5$ (for a specific strength of $V$ indicated by the green dot), 
and finally by FP$_3$ (stronger NN repulsion $V > 0$). The 
scaling dimension
of the $s$-wave pairing source term decreases monotonically from FP$_2$ to FP$_5$ to FP$_3$ for any $n \geq 2$. Therefore, in the entire regime of the CDW phase [the red shaded region in Fig.~\ref{UV_PD}], $s$-wave pairing can only be found as a short-range order, without any global long range order. Furthermore, the coherence length ($\xi$) of such local $s$-wave pairing decreases monotonically as we keep increasing $V > 0$ (keeping $U < 0$), and across the green dashed line $\xi$ vanishes smoothly. Therefore, above the green dashed line in Fig.~\ref{UV_PD}, local short-range $s$-wave pairing loses its support. It must be noted that there is no genuine phase transition across the green line, it only represents a crossover boundary.

Our RG analysis leads to a fairly complete phase diagram of the extended Hubbard model in the uniaxially strained honeycomb lattice, residing at or very close to the 
Dirac-semimetal to band-insulator QCP. We identified the role of each of the six fixed points (Table~\ref{table-1overn}) in the phase diagram of a simple microscopic interacting model. With recent progress in controlling the strength of electronic interactions in optical honeycomb lattices for ultracold fermionic atoms~\cite{coldatom-experiment-1, coldatom-experiment-2} and the realization of an ultracold-atom Fermi-Hubbard antiferromagnet~\cite{fermi-hubbard}, as well as quantum Monte Carlo simulation with onsite or NN interaction~\cite{sorella-1, herbut-assaad-1, herbut-assaad-2, sorella-2, kaul, Gremaud, troyer-honeycomb, hong-yao-NN-honeycomb}, we believe that the phase diagram of the extended Hubbard model can be established (at least partially) in numerical simulation and/or experiments.

The \emph{existence} of the three interacting critical points FP$_{2,3,4}$ 
can be argued to survive beyond perturbation theory. 
For strong repulsive Hubbard interactions on a bipartite lattice in 2D, 
SU(2) spin rotational symmetry is broken and the system enters into the AFM phase, 
while for strong attractive Hubbard interactions pseudospin SU(2) symmetry is lifted 
leading to simultaneous nucleation of $s$-wave pairing and CDW. 
By contrast, for strong repulsive nearest-neighbor interactions on a bipartite lattice, 
the system enters the CDW phase where discrete $Z_2$ sublattice symmetry is broken. 
The existence of these phases at sufficiently strong interaction can be anticipated for the
strained honeycomb lattice model, which is free of frustration for all of the above orderings. 
However, due to the vanishing DoS in the ASM ($\varrho(E) \sim \sqrt{E}$) all the broken-symmetry phases 
are realized only beyond a critical threshold of interaction, and thus through a quantum phase transition. 
We expect that the transitions are continuous in the strained honeycomb lattice extended Hubbard model, and thus controlled by 
QCPs. We note that quantum Monte Carlo simulation in the Dirac system (isotropic honeycomb lattice model) also 
indicate continuous transitions through various 
QCPs~\cite{sorella-1, herbut-assaad-1, herbut-assaad-2, sorella-2, kaul, Gremaud, troyer-honeycomb, hong-yao-NN-honeycomb}. 
Thus, future numerical works can shed light on the non-perturbative nature of these transitions, which are
accessed here from leading order $\epsilon$ and $1/n$ expansions.


\section{Discussion and conclusion}~\label{conclusion}

In this paper we discuss the effect of generic short-range interactions at the quantum critical point separating a two-dimensional topological Dirac semimetal and a symmetry-preserving band insulator. The critical excitations separating these two phases constitute an anisotropic semimetal. The quasiparticle spectra in the anisotropic semimetal display both linear and quadratic dependence on momenta along two mutually orthogonal directions. Consequently, the quantum critical regime in the noninteracting system displays peculiar power-law scaling of thermodynamic and transport quantities that are distinct from their counterparts in a Dirac semimetal or a band insulator, as shown in Table~\ref{Table:scaling-noninteracting}.

Due to the vanishing density of states ($\varrho(E) \sim \sqrt{E}$) the critical point separating the Dirac semimetal from the band insulator remains stable against sufficiently weak short range electron-electron interactions. However, at strong interaction this direct transition either $(i)$ becomes a fluctuation-driven first-order transition (see Sec.~\ref{1storder}), or 
$(ii)$ gets avoided by an intervening broken-symmetry phase (see Sec.~\ref{spinful-RG}). Using renormalization group (RG) analysis we identify charge density wave, antiferromagnet, and spin singlet $s$-wave pairing as the dominant channels of symmetry breaking for the interacting model containing generic short-range interactions that for example encompass the extended Hubbard model, see the phase diagrams in Fig.~\ref{UV_PD} and 
Fig.~\ref{Phasediagram-finiten}.

We demonstrate that (see Sec.~\ref{spinful-RG}) the RG can be controlled by expanding about one dimension, for which short-range interactions are marginal at tree level. We recover standard results for 1D systems such as the absence of corrections for a U(1) current-current (Luttinger) perturbation, as well as spin-charge separation and Kosterlitz-Thouless transitions for spin-$1/2$ electrons. From the leading order calculation we find that the correlation length exponent at all the interacting multi-critical points, describing continuous quantum phase transition from anisotropic semimetal to various broken symmetry phases is $\nu=2$, which is strikingly different from the ones across a 
Dirac-semimetal to broken-symmetry phase
transition for which $\nu \approx 1$, as predicted in field theoretic works~\cite{HJR, zinn-justin, gracey, rosa, wetterich, rosenstein, HJVafek, roy-multicriticality} and has also been established numerically~\cite{herbut-assaad-1, herbut-assaad-2, sorella-2, kaul, troyer-honeycomb, hong-yao-NN-honeycomb}.

Throughout we have neglected the long-range tail of the Coulomb interaction and focused only on its short-range pieces. Recently it has been proposed that infinitesimally weak long-range Coulomb interaction can cause an instability of the critical excitations and drive the system toward the formation of an \emph{infrared-stable non-Fermi liquid} phase~\cite{longrange-1, longrange-2}. In the future it will be interesting to study the interplay of non-Fermi liquid and broken-symmetry phases. However, in an optical honeycomb lattice (composed of neutral atoms) our study should provide a fairly complete picture of the phase diagram for the interacting uniaxially strained honeycomb lattice. Our results also apply to solid state compounds with screened Coulomb interactions (due e.g.\ to an external gate).

The RG framework developed here (combined $\epsilon$ and $1/n$ expansion) can be applied to many different systems where quasiparticle excitations possess anisotropy, such as a three-dimensional general Weyl semimetal constituted by monopoles and antimonopoles of strength ${\mathcal N}$~\cite{goswami-roy-juricic}, with ${\mathcal N}=1,2,3$ in a crystalline environment~\cite{Fang-HgCrSe, bergevig, nagaosa}. In addition, our formalism (with minor modifications) can also be subscribed to address the effects of quenched disorder at the Weyl semimetal-band insulator quantum critical point and possible onset of a metallic phase at strong disorder through a continuous quantum phase transition across a multi-critical point~\cite{roy-slager-juricic}. Furthermore, our formalism can also shed light on the effects of electronic interaction in a two-dimensional Dirac semimetal. At the cost of the in-plane rotational symmetry ($v_x \neq v_y$, where $v_j$ is the Fermi velocity of massless Dirac fermions in the $j$ direction), which can be achieved by applying a weak uniaxial strain in monolayer graphene (so that the system is sufficiently far from 
the Dirac-semimetal to band-insulator
quantum critical point), we can generalize the proposed $\epsilon$ and $1/n$ expansion to address competing orders, emergent quantum critical phenomena, and the global phase diagram of a 2D Dirac system (with the caveat that $n$ can only be \emph{odd} integer). The interacting theory for a \emph{slightly} anisotropic Dirac semimetal (with two valleys) is described by 4 (8) independent coupling constants for spinless (spinful) fermions. This exercise will allow us to investigate an intriguing competition among various ordered phases, such as anti-ferromagnet, valence-bond-solid, charge-density-wave, singlet and triplet superconductors as well as topological quantum anomalous/spin Hall insulators, in a controlled renormalization group framework~\cite{hou-chamon-mudry}. We leave this problem for future work~\cite{sharma}.

Finally, we point out that the dispersion of the ASM studied here closely resembles that of quasiparticles near \emph{hot-spots} of the 
Fermi surface in a two-dimensional square lattice system~\cite{sachdev}. Recently there has been a surge of theoretical works geared toward understanding the effects of strong forward scattering interactions and quantum critical phenomena within the hot-spot or patch model~\cite{polchinski, backscattering, sachdev-metlitski-1, sachdev-metlitski-2, sur-lee}. Prior works are often based on appropriate order parameter field (spin-density-wave or nematic, for example), coupled with gapless fermions, residing close to the hot-spots, through \emph{Yukawa coupling}. Our approach can provide an alternative route to investigate the effects of strong electronic interactions on the Fermi surface, 
the role of competition among various incipient orderings, emergent symmetry near quantum critical points, etc. Very recently it has also been argued that a topologically protected, multi-flavored 2D anisotropic semimetal can also be realized in Sr$_2$IrO$_4$~\cite{sr2iro4}. 
When electron-doped, Sr$_2$IrO$_4$ is believed to support high-$T_c$ $d$-wave superconductivity~\cite{dwave-1, dwave-2}. Therefore, one can (at least in principle) generalize the formalism outlined here to address the intriguing confluence of electronic correlations, exotic broken-symmetry phases, competing orders, quantum critical phenomena, the role of topology (such as in Weyl materials) in a wide variety of systems within a unified perturbatively controlled RG scheme.

\acknowledgements

This work was supported by the Welch Foundation Grant No.~C-1809 and by NSF CAREER Grant no.~DMR-1552327. Useful correspondence with Leon Balents and Fakher F. Assaad is thankfully acknowledged.

\appendix

\section{Optical conductivity in the anisotropic semimetal}~\label{append_conductivity}

Due to the distinct power-law dependence of the quasiparticle dispersion along the $x$- and $y$-directions, the critical excitations show 
distinct scaling of the dynamic (frequency-dependent ac) conductivity along these two directions (denoted by $\sigma_{xx}$ and $\sigma_{yy}$, respectively). We use the Kubo formula to compute the dynamic conductivity. The expression for the polarization bubble is
\begin{eqnarray}
\Pi_{jj}(i \omega_n) &=& N \; \frac{e^2}{\beta} \; \sum_{m} \; \int_{\mathbf k} {\mbox{\bf Tr}} \bigg[ \hat{J}_j \; G_0 \left( {\mathbf k}, i p_m \right) \; \hat{J}_j \nonumber \\ 
&\times&G_0 \left( {\mathbf k}, i p_m+i \omega_n  \right) \bigg],
\end{eqnarray}    
where $e$ is the electronic charge, $\beta$ is the inverse temperature, $\omega_n, p_m$ are fermionic Matsubara frequencies, $\hat{J}_x=v \tau_1$, $\hat{J}_y=2 b k_y \tau_2$, and $N$ counts the flavor degeneracy. Thus for spinful fermions $N=2$. 
The noninteracting Green's function reads as 
\begin{equation}
G_0 \left( {\mathbf k}, i \omega_n \right)=\frac{1}{i \omega_n+ \mu-H ({\mathbf k},0)}=\int^{\infty}_{-\infty} \frac{d\epsilon}{2 \pi} \: \frac{A(\mathbf k, \epsilon)}{i \omega_n-\epsilon},
\end{equation}  
where $\mu$ is the chemical potential and we have used the Lehmann representation in the final expression. The spectral function $A(\mathbf k, \omega)$ is given by 
\begin{eqnarray}
A(\mathbf k, \omega) &=& \pi \left( 1 + \sum^2_{a=1} \tau_a \hat{d}_a \right) \; \delta \left(\omega+\mu -E_{\mathbf k} \right) \nonumber \\
&+& \pi \left( 1 - \sum^2_{a=1} \tau_a \hat{d}_a \right) \; \delta \left(\omega+\mu +E_{\mathbf k} \right),
\end{eqnarray}   
where $E_{\mathbf k}=\sqrt{v^2 k^2_x + b^2 k^4_y}$, $\hat{d}_a=d_a/\sqrt{d^2_1+d^2_2}$ for $a=1,2$ and $d_1=v k_x, d_2=b k^2_y$. The conductivity is defined as $\sigma_{jj}=-\mbox{Im}[\Pi_{jj}(\omega)]/\omega$. The Drude conductivity (the contribution that comes only from the intra-band piece) evaluates to Eq.~(\ref{drude}),
where $T_0 = v^2 / (k_B b)$ and 
\begin{eqnarray}~\label{conductivity_parameters}
a_x &=& \frac{K\left(\frac{1}{2}\right)}{3 \pi^2} \approx 0.06262, 
\,
a_y = \frac{3 \Gamma \left(-\frac{1}{4}\right)^2}{20 \pi ^{5/2}} \approx 0.20602, \nonumber \\
F_{n}(x) &=& -\sum_{\sigma=\pm} \int^{\infty}_{0} dy \; y^{n-\frac{1}{2}} \; \sech^2\left( y+\sigma x\right). 
\end{eqnarray}
The scaling of $F_1(x)$ and $F_2(x)$ is shown in Fig.~\ref{conductivity}. The result for the inter-band component of the optical conductivity is given by Eq.~(\ref{OC:interband}), where $\omega_0=v^2/b$. Thus as the frequency $\omega \to 0$, the dynamic conductivity along the $y$- ($x$)-direction vanishes (diverges) as $\sqrt{\omega}$ ($1/\sqrt{\omega}$). Such distinct power-law dependence of the conductivity along different directions can be directly measured in experiment to pin the quasiparticle dispersion.

\section{Diamagnetic susceptibility}~\label{diamagnetic}

Next we will discuss the effects of external magnetic fields on critical excitations, and in particular we focus on the diamagnetic susceptibility. The Landau level (LL) spectrum of the ASM, described by the Hamiltonian $H ({\mathbf k},0)$, is $\pm E_n (B)$, where $n$ is the LL index~\footnote{The LL index $n$ in this Appendix should not be confused with the anisotropy parameter 
introduced in Eq.~(\ref{hamil-nonint})} 
and for $n=0,1,2, \cdots$ 
\begin{equation}
	E_n(B) = A \; \left( \frac{2 v^2}{b}\right)^{1/3} \left[ \omega_c \left( n+\frac{1}{2} \right) \right]^{2/3},
\end{equation}  
with $A \approx 1.173$, $\omega_c=2 e B b$ is the cyclotron frequency, and $B$ is the strength of the external magnetic field~\cite{montambauz-1}. The free energy at $T=0$ is defined as  
\allowdisplaybreaks[4]
\begin{eqnarray}
\Omega_0 (B) &=& -C B \sum^{\infty}_{n=0} E_n(B) =- C \alpha B^{5/3} \sum^{\infty}_{n=0} \left( n+\frac{1}{2}\right)^{2/3} \nonumber \\
&=& - C \alpha B^{5/3} \zeta\left(-\frac{2}{3}, \frac{1}{2} \right) \nonumber \\
&=&- C \alpha B^{5/3} \left( 2^{-2/3}-1\right) \zeta\left(-\frac{2}{3}\right).
\end{eqnarray}
Here $C=e/h$ and $CB$ counts the LL degeneracy. Note that $\Omega_0$ is a divergent quantity, which, however, can be regularized by using the zeta regulator as follows~\cite{zeta-book, goswami-ghosal, roy-zetaregulator}
\begin{equation}
\pi^s \zeta\left(1-s\right)=2^{1-s} \Gamma(s) \zeta(s) \cos \left(\frac{s \pi}{2}\right).
\end{equation}   
Upon using the above regularization scheme and the definition of the 
diamagnetic susceptibility
$\chi=\partial^2\Omega_0/\partial B^2$, we arrive at the expression 
for the two-dimensional ASM at $T=0$ 
\begin{equation}
\chi=-\frac{{\mathcal A}}{B^{1/3}} \; \frac{e}{h} \; \left( e v \sqrt{\frac{b}{2}}\right)^{2/3} \equiv \chi_0,
\end{equation} 
where 
\begin{equation}
{\mathcal A}=\frac{10}{9}\zeta\left( -\frac{2}{3}\right) \left( 2^{-2/3}-1\right) A \approx 0.075.
\end{equation}
Notice that the diamagnetic susceptibility diverges as $B^{-1/3}$ as $B \to 0$.

The free energy at finite temperature reads as $\Omega(T)=\Omega_0+\Omega_T$, where 
\begin{equation}
	\Omega_T=-\frac{k_B T}{ \pi \ell^2_B} \sum^{\infty}_{n=0} \log \left(1+e^{-\frac{E_n}{k_B T}} \right),
\end{equation}
and $\ell_B=\sqrt{\hbar/(eB)}$ is the magnetic length. After following the standard steps, highlighted above, we arrive at the following expression for the 
diamagnetic susceptibility
at finite temperature $\chi(T)=\chi_0 + \chi_T$, where $\chi_T=\chi_0 f\left( \lambda_{Th}/l_B \right)$ and 
\begin{eqnarray}~\label{DMS_finiteT}
f(x) &=&- \left(\frac{20}{9 \pi}\; \frac{A}{{\mathcal A}} \right) \times \left\{ g(x) + \frac{3}{10} \; \frac{\partial}{\partial x} g(x) \right\} \nonumber \\
& \approx & - (11.063) \; \left\{ g(x) + \frac{3}{10} \; \frac{\partial}{\partial x} g(x) \right\},
\end{eqnarray}
with
\begin{eqnarray}
\lambda_{Th} &=& \frac{\hbar \sqrt{v} (b/2)^{1/4}}{\left( k_B T\right)^{3/4}},\; x=\frac{\lambda_{Th}}{l_B} \sim \frac{\sqrt{B}}{T^{3/4}}, \nonumber\\
 g(x)&=& \sum^{\infty}_{n=0} \frac{\left( n+\frac{1}{2}\right)^{2/3}}{e^{x^{4/3} \left( n+\frac{1}{2}\right)^{2/3}} +1}.
\end{eqnarray}
Here $\lambda_{Th}$ is the thermal de Broglie wavelength for the critical excitations. Our proposed scaling of the 
diamagnetic susceptibility
at finite temperature is valid as long as $l_B<\lambda_{Th}<\xi_\Delta$, with $\xi_\Delta \sim 1/\Delta$. Note $f(x)$ is a universal function of dimensionless argument, and in Fig.~\ref{diamagnetism} we have shown the scaling of $\log [-f(x)]$ vs. $\log[x]$.

\section{Symmetries}~\label{symmetry-append}

\begin{table}[t!]
\begin{tabular}{|c|c|c|}
\hline
Bilinear 			& description &	PH			 	\\
\hline \hline
$\Psi^\dagger \sigma_0 \tau_0 \Psi$ & density & 		$\times$	 \\
\hline 
$\Psi^\dagger \sigma_0 \tau_1 \Psi$ & $x$-current & 		$\times$	 \\
\hline
$\Psi^\dagger \sigma_0 \tau_2 \Psi$ & AP & 			$\checkmark$	 \\
\hline
$\Psi^\dagger \sigma_0 \tau_3 \Psi$ & CDW & 			$\times$	 \\
\hline
$\Psi^\dagger \vec{\sigma} \tau_0 \Psi$ & ferromagnet & 	$\checkmark$	 \\
\hline 
$\Psi^\dagger \vec{\sigma} \tau_1 \Psi$ & $x$-spin-current & 	$\checkmark$	 \\
\hline
$\Psi^\dagger \vec{\sigma} \tau_2 \Psi$ & spin-BDW & 		$\times$	 \\
\hline
$\Psi^\dagger \vec{\sigma}\tau_3 \Psi$ & AFM & 			$\checkmark$	 \\
\hline
\end{tabular}
\caption{Symmetry of the particle-hole channel fermion bilinears listed in Table~\ref{order-parameters} under particle-hole (PH) transformation [Eqs.~(\ref{LP-Def}) and (\ref{PH-BdG})].  
}~\label{order-parameters-PH}
\end{table}
	
We summarize all symmetries of the noninteracting ASM defined via Eqs.~(\ref{hamil-nonint:2D}) and (\ref{hamil-nonint}). We use these to provide a more complete classification of the fermion bilinears in Table~\ref{order-parameters} and to connect to the Nambu
notation in the scaling dimension Table~\ref{anomalous-dim-table}.  

In terms of the four-component field $\Psi$ [Eq.~(\ref{S0})], sublattice symmetry (SLS) [Eq.~(\ref{SLS})] and time-reversal invariance ($\mathcal{T}$) are both encoded as \emph{antiunitary} transformations,
\bsub
\begin{align}
	\Psi(\vex{r}) \rightarrow&\, \sigma_0 \tau_3 \left[\Psi^\dagger(\vex{r})\right]^{\T}, \;\; i \rightarrow -i, & \text{SLS},
\\
	\Psi(\vex{r}) \rightarrow&\, i \sigma_2 \tau_3 \Psi(\vex{r}), \;\; i \rightarrow -i, & \mathcal{T},
\end{align}
\esub
where $\T$ denotes the transpose. We can also define the product of SLS and $\mathcal{T}$ as an effective particle-hole (PH) symmetry.
This is implemented by the unitary transformation
\begin{align}\label{LP-Def}
	\Psi(\vex{r}) \rightarrow&\, \sigma_2 \tau_0 \left[\Psi^\dagger(\vex{r})\right]^{\T}, & \text{PH}.
\end{align}
For the noninteracting problem, these imply the following conditions on the single-particle Hamiltonian in Eq.~(\ref{hamil-nonint}):
\begin{align}\label{HnSyms}
\begin{aligned}
	- \sigma_0 \tau_3 \, H_n({\mathbf k},\Delta) \, \sigma_0 \tau_3 =&\, H_n ({\mathbf k},\Delta),		& \text{SLS},	\\
	\sigma_2 \tau_3 \, H_n^*(-{\mathbf k},\Delta) \, \sigma_2 \tau_3 =&\, H_n ({\mathbf k},\Delta),		& \mathcal{T},	\\
	- \sigma_2 \tau_0 \, H_n^\T(-{\mathbf k},\Delta) \, \sigma_2 \tau_0 =&\, H_n ({\mathbf k},\Delta),	& \text{PH}.	
\end{aligned}
\end{align}

We switch to the Nambu basis introduced in Eq.~(\ref{nambu-special}). This equation can also be written as 
\[
	\Psi_N(\vex{r}) = 
	\begin{bmatrix}
	\Psi(\vex{r}) \\
	i \sigma_2 \tau_3 \left[\Psi^\dagger(\vex{r})\right]^{\T}
	\end{bmatrix},
\]
where the explicit decomposition is in the Nambu space, acted on by the Pauli matrices $\eta_\mu = \left\{ \eta_0, \eta_1, \eta_2, \eta_3 \right\}$. The Nambu spinor satisfies the reality condition
\begin{align}\label{MajCond}
	\Psi_N^\dagger = \Psi_N^\T \,	\eta_2 \sigma_2 \tau_3.
\end{align}
At the level of Bogoliubov-de Gennes (BdG) mean field theory, the Hamiltonian can be written as 
\begin{align}
	H_{\mathsf{BdG}} = \frac{1}{2} \Psi_N^\dagger h \Psi_N.
\end{align}
Here $h$ is the single-particle BdG Hamiltonian. Eq.~(\ref{MajCond}) implies that $h$ always satisfies the ``automatic'' 
particle-hole condition
\begin{align}\label{NPH}
	- \Mp \, h^T \, \Mp =&\, h, 	&\Mp =&\, \eta_2 \sigma_2 \tau_3,  	& \text{Nambu-PH}.
\end{align} 
The symmetries in Eq.~(\ref{HnSyms}) become in the Nambu basis
\bsub\label{STP}
\begin{align}
	- \Ms \, h \, \Ms =&\, h, 	&\Ms =&\, \eta_3 \sigma_0 \tau_3, 			& \text{SLS}, 	\label{SLS-BdG}\\ 
	\Mt \, h^* \, \Mt =&\, h, 	&\Mt =&\, \eta_0 \sigma_2 \tau_3 , 			& \mathcal{T},   \label{T-BdG}\\ 
	- \Mlp \, h^T \, \Mlp =&\, h, 	&\Mlp =&\, \eta_1 \sigma_0 \tau_3, 			& \text{PH}.	\label{PH-BdG} 
\end{align}	
\esub

We emphasize that there are two independent particle-hole symmetries in this language. The Nambu condition in [Eq.~(\ref{NPH})] is not really a symmetry; it is merely a consequence of Pauli exclusion. Nevertheless, this condition implies that there only 28 allowed independent fermion bilinears (without derivatives) in the Nambu basis, instead of 64. By contrast, the particle-hole condition in Eq.~(\ref{PH-BdG}) is instead derived from the microscopic sublattice symmetry; it is not automatic. Finally, it is useful to synthesize an alternative version of time-reversal invariance, by combining the Nambu-PH condition [Eq.~(\ref{NPH})] with Eq.~(\ref{T-BdG}). This yields the chiral condition 
\begin{align}\label{T-Alt}
	- \eta_2 \, h \, \eta_2 =&\, h, & (\text{Nambu-PH) $\otimes$ $\mathcal{T}$}.
\end{align}	
Again since the Nambu PH is automatic, Eq.~(\ref{T-Alt}) is an equivalent definition of time-reversal symmetry.

The full inventory of 28 Nambu-PH Hermitian fermion bilinears appears in Table~\ref{anomalous-dim-table}. The properties of these bilinears under SLS, $\mathcal{T}$, $x$-reflection ${\mathcal R}_\pi$, and spin SU(2) symmetry transformations were summarized in Table~\ref{order-parameters}. To this we add the properties under microscopic PH symmetry [Eqs.~(\ref{LP-Def}) and (\ref{PH-BdG})], shown in Table~\ref{order-parameters-PH}. We list only particle-hole channel bilinears.

\section{1D limit and review of spin-charge separation}~\label{1D:Coupling-Append}

We translate the anisotropic semimetal model into the standard notations for 1D physics. After a $\tau_2$-rotation that sends $\tau_1 \rightarrow \tau_3$, the field in Eq.~(\ref{S0}) is decomposed into right-$R$ and left-$L$ mover components via $\Psi^\T = \left[R, L\right]$, leading to 
\begin{align}\label{S0-Ch} 
	\!\!\!
	{\mathcal S}_0	=	\int	\! 	d^2 {\mathbf r} \, dt  \left[
	\begin{aligned}
	&\,	\bar{R}	\left(\partial_t - i v \partial_x\right) R + \bar{L}	\left(\partial_t + i v \partial_x\right) L \\&\,
	-	i b	\left( \bar{R} \, \partial_y^n \, L -	\bar{L} \, \partial_y^n \, R \right)
	\end{aligned}
	\right]\!.\!\!
\end{align} 
For the case of spin-1/2 electrons, the chiral components $R$ and $L$ are each two-component spinors.

The four four-fermion interaction operators appearing in Eq.~(\ref{Hint_1D}) are chosen to exploit spin-charge separation \cite{giamarchi,tsvelik} in the 1D ($n \rightarrow \infty$) limit. These are defined as
\begin{align}\label{1DOps--Def}
\begin{aligned}
	{\mathcal O}_N 
	\equiv&\,
	\left(\bar{R} \vec{\sigma} R\right)\cdot\left(\bar{L} \vec{\sigma} L\right),
\\
	{\mathcal O}_A 
	\equiv&\,
	\left(\bar{R} R\right) \left(\bar{L} L\right),
\\
	{\mathcal O}_W 
	\equiv&\,
	\left(\bar{R} \sigma_2 \bar{R}^{\T}\right)
	\left({L}^{\T} \sigma_2 {L}\right),
\\	
	\bar{\mathcal{O}}_W 
	\equiv&\,
	\left(\bar{L} \sigma_2 \bar{L}^{\T}\right)
	\left({R}^\T \sigma_2 {R}\right),
\\	
	{\mathcal O}_X
	\equiv&\,
	\left(\bar{R} R\right)^2 + \left(\bar{L} L\right)^2.
 \end{aligned}
 \end{align}
The operator ${\mathcal O}_N$ is an SU(2)${}_1$ current-current perturbation; the remaining operators couple only to the U(1) electric charge sector~\cite{tsvelik}.

\begin{table}[t!]
\begin{tabular}{|c|c|c|}
\hline
Bilinear & description & operator scaling dimension\\
\hline \hline
$\Psi^\dagger \sigma_0 \tau_0 \Psi$ & density & 		1 \\
\hline 
$\Psi^\dagger \sigma_0 \tau_1 \Psi$ & $x$-current & 		1 \\
\hline
$\Psi^\dagger \sigma_0 \tau_2 \Psi$ & AP & 			$(1/2) + ({K_c}/{2})$\\
\hline
$\Psi^\dagger \sigma_0 \tau_3 \Psi$ & CDW & 			$(1/2) + ({K_c}/{2})$\\
\hline
$\Psi^\dagger \vec{\sigma} \tau_0 \Psi$ & ferromagnet & 	1 \\
\hline 
$\Psi^\dagger \vec{\sigma} \tau_1 \Psi$ & $x$-spin-current & 	1 \\
\hline
$\Psi^\dagger \vec{\sigma} \tau_2 \Psi$ & spin-BDW & 		$(1/2) + ({K_c}/{2})$ \\
\hline
$\Psi^\dagger \vec{\sigma}\tau_3 \Psi$ & AFM & 			$(1/2) + ({K_c}/{2})$\\
\hline \hline
$\Psi \sigma_2 \tau_3 \Psi$ & $s$-wave SC & 			$(1/2) + ({1}/{2 K_c})$\\
\hline
$\Psi \sigma_2 \tau_1 \Psi$ & chiral SC$_1$ & 			$\left(K_c + 1/K_c\right)/2$\\
\hline
$\Psi \sigma_2 \tau_0 \Psi$ & chiral SC$_2$ & 			$\left(K_c + 1/K_c\right)/2$\\
\hline
$\Psi \sigma_{(0,1,3)} \tau_2 \Psi$ & triplet SC & 		$(1/2) + ({1}/{2 K_c})$\\
\hline
\end{tabular}
\caption{Scaling dimensions of all fermion bilinears in the 1D limit ($\epsilon = 0$, $n \rightarrow \infty$), in the stable gapless phase consisting independent spin and charge Luttinger liquids, c.f.\ Fig.~\ref{UV_PD_1D}. The common dimensions for the AP, CDW, spin-BDW, and AFM bilinears are the sum of the spin ($1/2$) and charge ($K_c/2$) sector contributions; the same applies to the $s$-wave SC and triplet SC. 
The chiral SC$_{1,2}$ bilinears reside entirely in the charge sector. We ignore logarithmic corrections to the spin sector due to nonzero, but marginally irrelevant $U_N < 0$ \cite{giamarchi}. 
}~\label{order-parameters-1D}
\end{table}

In the 1D limit, we can bosonize the U(1) charge and spin SU(2)${}_1$ sectors; up to irrelevant operators, these completely decouple \cite{tsvelik}. The imaginary time bosonic action for the charge sector can be written as 
\begin{align}\label{Scharge}
	S_c	= \int d^2\vex{r} \, d t \, \left[
	\begin{aligned}
	&\, i \partial_x \theta \, \partial_t \phi +	\frac{v_c}{2 K_c}	\left(\partial_x \theta\right)^2 + \frac{v_c K_c}{2}
	\left(\partial_x \phi\right)^2	\\&\,
	- \frac{2 W}{(2 \pi)^2} \cos\left(\sqrt{8 \pi} \theta\right)
	\end{aligned}
	\right]\!,\!\!
\end{align}
where $\phi$ ($\theta$) denotes the polar (axial) field, and the charge velocity $v_c$ and Luttinger parameter $K_c$ are given by 
\begin{align}\label{vcKc}
	v_c = v\left(1 + X\right),
	\quad
	K_c = 1 - U_A / 2. 
\end{align}
The U(1) current-current operator ${\mathcal O}_A$ and U(1) stress tensor operator ${\mathcal O}_X$ are completely absorbed into these parameters of the free boson, while the umklapp operator $\frac{1}{2}({\mathcal O}_W  + \bar{{\mathcal O}}_W)$ becomes a sine-Gordon perturbation. As usual \cite{giamarchi,tsvelik}, the expressions for $v_c$ and $K_c$ in terms of the microscopic coupling strengths $U_A$ and $X$ obtained via perturbative bosonization are valid only to first order.

When both charge and spin sectors are gapless, the theory asymptotes to a conformal fixed point in the infrared. This is a product of spin and charge Luttinger liquids [c.f.\ Fig.~\ref{UV_PD_1D}]. The SU(2) invariant fixed point has $U_N = W = 0$, with $U_A < 0$ ($K_c > 1$). The exact scaling dimensions of all fermion bilinears in Tables~\ref{order-parameters} and \ref{anomalous-dim-table} at this fixed point are summarized in Table~\ref{order-parameters-1D}.

\section{Table of integrals}~\label{Append-loop-integral}

To proceed with the perturbative RG analysis to the leading order in $\epsilon$ and $1/n$, 
we need to evaluate the following integrals 
\begin{eqnarray}
	I_1 &=& \int \frac{\omega^2}{\left( \omega^2 + v^2 k^2_x + b^2 k^{2n}_y\right)^2}, \nonumber \\
	I_2 &=& \int \frac{v^2 k^2_x}{\left( \omega^2 + v^2 k^2_x + b^2 k^{2n}_y\right)^2}, \nonumber \\
	I_3 &=& \int \frac{b^2 k^{2n}_y}{\left( \omega^2 + v^2 k^2_x + b^2 k^{2n}_y\right)^2}, \nonumber \\
	I_4 &=& \int \frac{b k^{n}_y}{\left( \omega^2 + v^2 k^2_x + b^2 k^{2n}_y\right)}.
\end{eqnarray} 
The relevant Feynman diagrams are shown in Figs.~\ref{Feynman} and ~\ref{feynman-source}. 
We introduce a new set of variables defined as 
\begin{eqnarray}
	\omega  =\rho \cos \theta, \:	v k_x =\rho \sin \theta, \: x=\frac{b^{1/n} k_y}{E^{1/n}_\Lambda},
\end{eqnarray}
where $\rho = \sqrt{\omega^2 +v^2 k^2_x}$, and $E_\Lambda$ is the ultraviolet energy cutoff. Within the framework of the Wilsonian RG, we integrate out the fast Fourier modes within the shell $E_\Lambda e^{-l} < \rho < E_\Lambda$, and subsequently integrate over $0 \leq \theta \leq 2 \pi$ and $0 \leq x \leq \infty$. The above integrals are then given by 
\begin{subequations}
\begin{align}
	I_1	=&\, I_2 = a_\epsilon l \int^{\infty}_0 dx \frac{1}{(1+x^{2n})^2} 
\nonumber\\
	=&\, a_\epsilon l \left[ \frac{\pi  (2 n-1) \csc \left(\frac{\pi }{2 n}\right)}{ 4 n^2} \right]
\nonumber\\
	=&\, a_\epsilon 
	\left(1-\frac{1}{2 n} \right) l +{\mathcal O} \left( \frac{1}{n^2}\right), \\
	I_3	=&\, a_\epsilon l \int^{\infty}_0 dx \frac{2 x^{2n}}{\left( 1+ x^{2n}\right)^2} 
\nonumber\\
	=&\, a_\epsilon l \left[\frac{\pi  \csc \left(\frac{\pi }{2 n}\right)}{2 n^2}\right] = a_\epsilon \left(\frac{1}{n}\right) l + {\mathcal O} \left( \frac{1}{n^2}\right),  \\
	I_4 =&\, E_\Lambda \; a_\epsilon l \int^{\infty}_0 dx \; \frac{2 x^n}{1+x^{2n}} \nonumber\\
	=&\, E_\Lambda \; a_\epsilon l \left[\frac{\pi  \sec \left(\frac{\pi }{2 n}\right)}{n}\right] 
\nonumber\\
	=&\,	E_\Lambda a_\epsilon \left( \frac{\pi}{n} \right) l +{\mathcal O} \left( \frac{1}{n^2}\right),
\end{align} 
\end{subequations}	
where $a_\epsilon \equiv	E^\epsilon_\Lambda/(8 \pi^2 v b^\epsilon)$.	Thus as $n \to \infty$ only the contributions from $I_1$ and $I_2$ survive. We note that all integrals over $x$ are convergent for any $n \geq 2$.

\section{Fierz identity}~\label{fierz}

In this Appendix we present the Fierz constraints among local four-fermion terms that allows us to 
extract the set of linearly independent quartic terms in an interacting model. 
We discuss this subject for spinless fermions and spin-$1/2$ electrons separately.

\subsection{Spinless fermions}~\label{fierz_spinless}

To find the number of independent coupling constants for spinless fermions, 
we define a four-dimensional vector as
\begin{equation}
X^\top_4= \left[ \left( \psi^\dagger \tau_0 \psi \right)^2, \left( \psi^\dagger \tau_1 \psi \right)^2, \left( \psi^\dagger \tau_2 \psi \right)^2, \left( \psi^\dagger \tau_3 \psi \right)^2 \right].
\end{equation}
The Pauli matrices $\tau_\mu =\left\{ \tau_0, \tau_1, \tau_2, \tau_3 \right\}$ 
constitute a complete basis for $2\times2$ matrices. The Fierz constraint for two component spinor ($\psi$) reads as
\begin{eqnarray}
\left[\psi^\dagger (x) \tau_a \psi(x) \right]\left[\psi^\dagger (y) \tau_b \psi(y) \right] = -\frac{1}{4} \mbox{Tr} \left[\tau_a \tau_c \tau_b \tau_d \right] \nonumber \\
\times \left[\psi^\dagger(x) \tau_c \psi(y) \right] \left[\psi^\dagger(y) \tau_d \psi(x) \right],
\end{eqnarray}
where $a,b,c,d=0,1,2,3$, and for local interaction $x=y$. 
The above Fierz constraint can then be compactly written as $F_{4 \times 4} X_4=0$, where
\begin{equation}
F_{4 \times 4}= \left[ \begin{array}{cccc}
3 & 1 & 1 & 1 \\
1 & 3 & -1 & -1 \\
1 & -1 & 3 & -1 \\
1 & -1 & -1 & 3
\end{array}
\right].
\end{equation}
The rank of $F_{4 \times 4}$ is $3$ and therefore the number of linearly independent coupling constants is $4-3=1$. Thus, we can choose $\left( \psi^\dagger \tau_3 \psi \right)^2$ as the independent interaction operator and the remaining quartic terms are related according to 
\begin{eqnarray}
-\left( \psi^\dagger \tau_0 \psi \right)^2=
\left( \psi^\dagger \tau_1 \psi \right)^2=\left( \psi^\dagger \tau_2 \psi \right)^2=\left( \psi^\dagger \tau_3 \psi \right)^2. \nonumber \\
\end{eqnarray}

\subsection{Spin-$1/2$ electrons}~\label{fierz_spinful}

To find the number of linearly independent coupling constants in the interacting Hamiltonian for spin-$1/2$ electrons we first define an eight component vector as 
\begin{eqnarray}
X^\top_8 &=& \bigg[ \left( \Psi^\dagger \sigma_0 \tau_0 \Psi \right)^2, \left( \Psi^\dagger \sigma_0  \tau_1 \Psi \right)^2, \left( \Psi^\dagger \sigma_0 \tau_2 \Psi \right)^2, \nonumber \\
&& \left( \Psi^\dagger \sigma_0 \tau_3 \Psi \right)^2, \left( \Psi^\dagger \vec{\sigma} \tau_0 \Psi \right)^2, \left( \Psi^\dagger \vec{\sigma}  \tau_1 \Psi \right)^2, \nonumber \\ 
&& \left( \Psi^\dagger \vec{\sigma} \tau_2 \Psi \right)^2, \left( \Psi^\dagger \vec{\sigma} \tau_3 \Psi \right)^2 \bigg].
\end{eqnarray}
The Fierz constraint reads as 
\begin{eqnarray}
\left[\Psi^\dagger (x) M \Psi(x) \right] \left[\Psi^\dagger (y) N \Psi(y) \right] = -\frac{1}{16} \mbox{Tr} \left[M \Gamma_a N \Gamma_b \right] \nonumber \\
 \times  \left[\Psi^\dagger(x) \Gamma_a \Psi(y) \right] \left[\Psi^\dagger(y) \Gamma_b \Psi(x) \right],
\end{eqnarray}
which we apply here for contact interaction $x=y$. 
Here $M$ and $N$ are $4 \times 4$ Hermitian matrices, 
and $\Gamma_a$ close the basis for all $4\times4$ Hermitian matrices, $a,b=1, \cdots, 16$. 
Thus the Fierz identity allows us to write each quartic term as a linear combination of rest. 
The above constraint can then be compactly written as $F_{8 \times 8} X_8=0$, where 
\begin{equation}
F_{8 \times 8}=\left[ \begin{array}{cccccccc}
5 & 1 & 1 & 1 & 1 & 1 & 1 & 1 \\
1 & 5 & -1 & -1 & 1 & 1 & -1 & -1 \\
1 & -1 & 5 & -1 & 1 & -1 & 1 & -1 \\
1 & -1 & -1 & 5 & 1 & -1 & -1 & 1 \\
3 & 3 & 3 & 3 & 3 & -1 & -1 & -1 \\
3 & 3 & -3 & -3 & -1 & 3 & 1 & 1 \\
3 & -3 & 3 & -3 & -1 & 1 & 3 & 1 \\
3 & -3 & -3 & 3 & -1 & 1 & 1 & 3
\end{array}
\right].
\end{equation}
The rank of the Fierz matrix $F_{8 \times 8}$ is $4$ and therefore the number of linearly independent four-fermion local interactions is $8-4=4$. We chose $\left( \Psi^\dagger \sigma_0  \tau_1 \Psi \right)^2$, $\left( \Psi^\dagger \sigma_0 \tau_3 \Psi \right)^2$, $\left( \Psi^\dagger \vec{\sigma} \tau_2 \Psi \right)^2$ and $\left( \Psi^\dagger \vec{\sigma} \tau_3 \Psi \right)^2$ as linearly independent quartic terms in $H_{int}$, as shown in Eq.~(\ref{Hint_spinful}). The rest of the quartic terms can then be expressed as linear combination of above four according to 
\begin{align}
	\left( \Psi^\dagger \sigma_0  \tau_0 \Psi \right)^2 	=&\, \left( \Psi^\dagger \sigma_0 \tau_1 \Psi \right)^2 - \left( \Psi^\dagger \sigma_0 \tau_3 \Psi \right)^2  \nonumber\\
	&\,	-	\ts{\frac{1}{3}} \left( \Psi^\dagger \vec{\sigma} \tau_2 \Psi \right)^2 -	\ts{\frac{2}{3}} \left( \Psi^\dagger \vec{\sigma} \tau_3 \Psi \right)^2, 
\nonumber \\
	\left( \Psi^\dagger \sigma_0 \tau_2 \Psi \right)^2 =&\, \left( \Psi^\dagger \sigma_0 \tau_3 \Psi \right)^2 - \ts{\frac{1}{3}} 
	\left( \Psi^\dagger \vec{\sigma} \tau_2 \Psi \right)^2 \nonumber\\
	&\,	+	\ts{\frac{1}{3}} \left( \Psi^\dagger \vec{\sigma} \tau_3 \Psi \right)^2,  
\nonumber \\
	\left( \Psi^\dagger \vec{\sigma} \tau_0 \Psi \right)^2 =&\, -	3 \left( \Psi^\dagger \sigma_0 \tau_1 \Psi \right)^2	+ \left( \Psi^\dagger \vec{\sigma}  \tau_2 \Psi \right)^2 \nonumber\\ 
	&\, +	\left( \Psi^\dagger \vec{\sigma} \tau_3 \Psi \right)^2,  \nonumber \\
	\left( \Psi^\dagger \vec{\sigma} \tau_1 \Psi \right)^2  =&\, -3 \left( \Psi^\dagger \sigma_0 \tau_1 \Psi \right)^2 + 3 \left( \Psi^\dagger \sigma_0  \tau_3 \Psi \right)^2 \nonumber\\
	&\,	+	\left( \Psi^\dagger \vec{\sigma}  \tau_3 \Psi \right)^2.
\end{align}
In the RG calculation whenever we generate any one of these terms, it is rewritten back in terms of the four quartic terms appearing in $H_{int}$ [Eq.~(\ref{Hint_spinful})].

\section{Clifford algebra of mass matrices in the 2D ASM}~\label{cliffordalgebra-Append}

We here provide an algebraic derivation of the internal algebra among various mass orders in the two-dimensional ASM (such as the CDW, N\'{e}el AFM and $s$-wave pairing). This analysis is purely based on the Clifford algebra of real matrices and does not rely on specific microscopic details of the model. Let us define a Nambu-doubled spinor as $\Psi=\left( \Psi_p, \Psi_h \right)^\top$ so that the effective Hamiltonian describing low energy excitations in an ASM assumes the form
\begin{equation}
H_{ASM} (\mathbf k)=H_0 (\mathbf k) \oplus \left[- H^\top_0 (-\mathbf k) \right],
\end{equation}
where $H_0 (\mathbf k)=\sum_{j=1,2} \alpha_j \; d_j (\mathbf k)$, with
\begin{equation}
d_1 (\mathbf k)=v k_x, \: d_2 (\mathbf k)=b k^2_y.
\end{equation}
Here $\alpha_1$ and $\alpha_2$ are two mutually anticommuting four component Hermitian matrices. Next we wish to find all the Hermitian eight-dimensional matrices ($M$s) that anticommute with $H_{ASM} (\mathbf k)$, such that $m=\langle \Psi^\dagger M \Psi \rangle \neq 0$. 
Therefore, $M$s represent mass matrices. Such condition is satisfied only when
\begin{equation}
M=- \left(\sigma_1 \otimes \mathbb I_4 \right) \; M^\top \; \left(\sigma_1 \otimes \mathbb I_4 \right),
\end{equation}
where $\mathbb I_4$ is a four-dimensional identity matrix. There exists a unitary matrix $U=U_2 \otimes \mathbb I_4$ such that $\tilde{M}=-\tilde{M}^\top$, where $\tilde{M}=U M U^\dagger$ and
\begin{equation}
U_2 = \sqrt{\pm i} \; e^{i \frac{\pi}{4} \sigma_3} \; e^{i \frac{\pi}{4} \sigma_2} \; e^{i \left( \frac{\pi}{4}-\phi\right) \sigma_3}.
\end{equation}
Therefore, after the unitary transformation all mass matrices are purely \emph{imaginary}~\cite{herbut-isospin, roy-linenode}.

Let us now write two mutually anticommuting four-dimensional matrices, appearing in $H_0(\mathbf k)$ as
\begin{equation}
\alpha_j = \Re(\alpha_j) + i \; \Im(\alpha_j),
\end{equation}
for $j=1,2$. After such decomposition the kinetic energy of the ASM assumes the form
\begin{eqnarray}
H_{ASM}(\mathbf k) &=& \left[ \sigma_0 \otimes \Re(\alpha_1) + i \sigma_3 \otimes \Im(\alpha_1) \right] d_1(\mathbf k) \nonumber \\
&+&  \left[ \sigma_3 \otimes \Re(\alpha_2) + i \sigma_0 \otimes \Im(\alpha_2) \right] d_2(\mathbf k).
\end{eqnarray}
Since $U_2 \sigma_3 U^\dagger_2=\sigma_2$, after the unitary transformation with $U$ the kinetic energy becomes
\begin{eqnarray}
\tilde{H}_{ASM}(\mathbf k) &=& \big[ \sigma_0 \otimes \Re(\alpha_1)+ i \sigma_2 \otimes \Im(\alpha_1) \big] d_1(\mathbf k) \nonumber \\
&+&  \left[ \sigma_2 \otimes \Re(\alpha_2) + i \sigma_0 \otimes \Im(\alpha_2) \right] d_2(\mathbf k) \nonumber \\
&=& \tilde{\Gamma}_1 d_1(\mathbf k) + \tilde{\Gamma}_2 d_2(\mathbf k).
\end{eqnarray}
Thus, $\tilde{\Gamma}_1$ and $\tilde{\Gamma}_2$ are respectively mutually anticommuting, but purely real and imaginary eight-dimensional Hermitian matrices. We are after all purely imaginary matrices ($\tilde{M}$) that anticommute with $\tilde{\Gamma}_1$ and $\tilde{\Gamma}_2$. Since, $i \tilde{\Gamma}_2$ and $i \tilde{M}$ are purely real and square to $-1$, while $\tilde{\Gamma}_1$ is purely real and squares to $+1$, we seek to know the maximal number of $q$ so that for $p \geq 1$, the dimensionality of real matrices is eight and together they close the $C(p,q)$ algebra. Notice that the $C(p,q)$ Clifford algebra is defined by a set of $p+q$ mutually anticommuting matrices, among which $p$ of them squares to $+1$, while $q$ of them squares to $-1$. The answer is $q=4$ and $p=1$~\cite{herbut-isospin, okubo}. Thus the maximal number of mutually anticommuting matrices is $5$, and we define a set of such five matrices as $\left\{R, I_1, I_2, I_3, I_4 \right\}$, where $R$ is real and the $I_j$s are imaginary for $j=1, \cdots, 4$. Two of those five matrices, say $R$ and $I_1$, can be used to define the kinetic energy. Therefore, $I_2, I_3$ and $I_4$ represents three mass matrices that together close a $Cl(3)$ algebra, defining a set of three mutually anticommuting matrices. However, we have not exhausted all possible mass matrices for a 2D ASM yet.

Note the Clifford algebra $C(1,4)$ follows the \emph{quaternionic} representation that supports \emph{three} real Casimir operators, $K_j$ for $j=1,2,3$, besides the standard identity matrix. The Casimir operators satisfy the \emph{quaternionic algebra}
\begin{equation}
K_i K_j=-\delta_{ij} + \epsilon_{ijk} K_k,
\end{equation}
where $\delta_{ij}$ is the Kronecker delta function. From the Casimir operators we can define imaginary matrices $E_j = i K_j$ for $j=1,2,3$ and $E_j$s satisfy the SU(2) algebra. Notice that $E_j$s are purely imaginary Hermitian matrices that commute with $H_{ASM}(\mathbf k)$. From these three imaginary matrices we can define yet another set of three mutually anticommuting matrices $M^\prime_j=i E_j R I_1$ for $j=1,2,3$ that together close a $Cl(3)$ algebra and also anticommute with $H_{ASM}(\mathbf k)$. Hence, there are all together six purely imaginary matrices (thus mass matrices) that anticommute with $H_{ASM}(\mathbf k)$ and they can be arranged into two sets according to
\begin{equation}
\left\{ I_2, I_3, I_4 \right\} \quad \mbox{and} \quad \left\{ M^\prime_1, M^\prime_2, M^\prime_3 \right\}. \nonumber
\end{equation}
Together these two sets close a $Cl(3) \times Cl(3)$ algebra of mass matrices, as we announced in the main part of the paper. One set we can identify with the three components of N\'{e}el AFM order, while the other set is constituted by the CDW and two components (real and imaginary) of $s$-wave pairing. We note that Clifford algebra of mass matrices in a 2D ASM is identical to that for a three-dimensional line-node semimetal~\cite{roy-linenode}.


\end{document}